\documentclass[twocolumn,showpacs,preprintnumbers,amsmath,amssymb,superscriptaddress,prl]{revtex4}
\usepackage{epsfig}
\usepackage{graphicx}
\usepackage{color}
\usepackage{bm}
\usepackage{dsfont}

\begin{document}

\title{Microscopic analysis of the valence band and impurity band theories of (Ga,Mn)As}

\author{J.~Ma\v{s}ek}
\affiliation{Institute of Physics ASCR, v.v.i., Na Slovance 2, 182 21 Praha
8, Czech Republic}

\author{F.~M\'aca}
\affiliation{Institute of Physics ASCR, v.v.i., Na Slovance 2, 182 21 Praha
8, Czech Republic}

\author{J.~Kudrnovsk\'y}
\affiliation{Institute of Physics ASCR, v.v.i., Na Slovance 2, 182 21 Praha
8, Czech Republic}

\author{O.~Makarovsky}
\affiliation{School of Physics and Astronomy, University of
Nottingham, Nottingham NG7 2RD, United Kingdom}

\author{L.~Eaves}
\affiliation{School of Physics and Astronomy, University of
Nottingham, Nottingham NG7 2RD, United Kingdom}

\author{R.~P.~Campion}
\affiliation{School of Physics and Astronomy, University of
Nottingham, Nottingham NG7 2RD, United Kingdom}

\author{K.~W.~Edmonds}
\affiliation{School of Physics and Astronomy, University of
Nottingham, Nottingham NG7 2RD, United Kingdom}

\author{A.~W.~Rushforth}
\affiliation{School of Physics and Astronomy, University of
Nottingham, Nottingham NG7 2RD, United Kingdom}

\author{C.~T.~Foxon}
\affiliation{School of Physics and Astronomy, University of
Nottingham, Nottingham NG7 2RD, United Kingdom}

\author{B.~L.~Gallagher}
\affiliation{School of Physics and Astronomy, University of
Nottingham, Nottingham NG7 2RD, United Kingdom}

\author{V.~Nov\'ak}
\affiliation{Institute of Physics ASCR, v.v.i., Cukrovarnick\'a 10, 162 53 Praha
6, Czech Republic}

\author{Jairo~Sinova}
\affiliation{Department of Physics, Texas A\&M University, College
Station, TX 77843-4242, USA}
\affiliation{Institute of Physics ASCR, v.v.i., Cukrovarnick\'a 10, 162 53 Praha
6, Czech Republic}

\author{T.~Jungwirth}
\affiliation{Institute of Physics ASCR, v.v.i., Cukrovarnick\'a 10, 162 53 Praha
6, Czech Republic} \affiliation{School of Physics and Astronomy,
University of Nottingham,
  Nottingham NG7 2RD, United Kingdom}

\begin{abstract}
We analyze microscopically the valence and impurity band models  of ferromagnetic (Ga,Mn)As. We find that the tight-binding Anderson approach with conventional parameterization and the full potential LDA+U calculations give a very similar picture of states near the Fermi energy which reside in an exchange-split {\em sp-d} hybridized valence band with dominant orbital character of the host semiconductor; this microscopic spectral character is consistent with the physical premise of the $\bm{k}\cdot\bm{p}$ kinetic-exchange model. On the other hand, the various models with a band structure comprising an impurity band detached from the valence band assume mutually incompatible microscopic spectral character. By adapting the tight-binding Anderson calculations individually to each of the impurity band pictures in the single Mn impurity limit and then by exploring the entire doping range we find that a detached impurity band does not persist in any of these models in ferromagnetic (Ga,Mn)As.
\end{abstract}

\pacs{75.10.Lp,75.30.Hx,75.50.Pp}

\maketitle

Over more than four decades,
(Ga,Mn)As has evolved from a pioneering direct gap p-doped semiconductor \cite{Chapman:1967_a} into an archetypical degenerate semiconductor with hole-mediated ferromagnetism \cite{Ohno:1998_a,Jungwirth:2006_a,Dietl:2008_b}. Paramagnetic insulating Ga$_{1-x}$Mn$_x$As materials prepared in the 1970's by melt growth  showed valence band (VB) to impurity band (IB) activation, with a non-systematic filamentary metallic conduction being observed at the highest studied dopings of $x\sim 0.1\%$ and ascribed to sample inhomogeneities \cite{Woodbury:1973_a}. The degenerate semiconductor regime was not reached in these materials. A comprehensive experimental assessment of basic doping only trends  became possible in the late 1990's with the development of epitaxial (Ga,Mn)As films \cite{Ohno:1998_a,Jungwirth:2007_a,Novak:2008_a,Wang:2008_e} which can be doped well beyond the equilibrium Mn solubility limit while avoiding phase segregation and maintaining a high degree of uniformity.  Transport measurements  on such films confirmed the insulating characteristics and the presence of the IB for $x\lesssim0.1\%$. For higher concentrations, $0.5\lesssim x\lesssim1.5\%$, no clear signatures of activation from the VB to the IB have been detected in the dc transport, suggesting that the bands start to overlap and mix, yet the materials remain insulating. At $x\sim 1.5\%$,  the low-temperature conductivity of the films  increase abruptly by several orders of magnitude and the material becomes a  bulk
degenerate semiconductor. The onset of ferromagnetism occurs on the insulating side of the transition at $x\sim 1\%$ and the Curie temperature gradually increases with increasing doping, reaching  $\sim$190~K at  the accessible substitutional Mn$_{\rm Ga}$ doping of $x\sim 8\%$.

One physical scenario for ferromagnetic (Ga,Mn)As, termed the VB picture, has an exchange-split band structure comprising the impurity band merged into the valence band. The states at the Fermi energy, $E_F$, retain the predominant orbital character of the host semiconductor and are moderately hybridized with the localized Mn $d$-electrons. This description, quantified by a variety of theoretical methods, has been a fruitful basis for analyzing and predicting a whole range of thermodynamic, magnetic, transport, and optical properties of ferromagnetic (Ga,Mn)As \cite{Jungwirth:2006_a,Dietl:2008_b}. Recently, several experimental observations have been interpreted using alternative models of an IB which is detached from the VB  \cite{Burch:2006_a,Stone:2008_a,Ando:2008_a,Tang:2008_a,Burch:2008_a}. However, it has been argued that a detailed analysis of the data in combination with transport experiments  is also consistent with the VB picture \cite{Jungwirth:2007_a}. The postulated IB models have not been previously derived from a microscopic theory considering all relevant orbital states in the mixed crystal. In order to help resolve the debate on these alternative interpretations, we examine here  the IB models by recreating them using microscopic modeling techniques and studying their band structure characteristics over the entire doping range.  These calculations (i) firmly establish the microscopic basis and internal consistency of the VB picture, (ii) demonstrate the mutual inconsistency of the various postulated IB models, and (iii) demonstrate that a detached IB does not persist in any of the models' band structures at dopings corresponding to ferromagnetic (Ga,Mn)As. Our theoretical analysis is based on the tight-binding Anderson (TBA) approach which includes all spectral components in the band structure forming the states near $E_F$, accounts for the Mn $d$-orbital electron-electron interaction effects using the self-consistent unrestricted Hartree-Fock method, and can be adopted to  realize microscopically the diverse proposed IB models. Additional physical insight is provided by comparisons to full-potential LDA and LDA+U calculations. More details on the techniques and more extensive numerical results can be found in the Supplementary material \cite{suppl}.

The perturbation of the crystal potential of GaAs due to a single Mn impurity has three components. (i) The first is the long-range hydrogenic-like potential
of a single acceptor in GaAs which produces a bound state at about 30 meV above the VB \cite{Marder:1999_a}.
(ii) The  second contribution is a short-range central-cell potential. It is specific to a given impurity and reflects the difference in the electro-negativity of the impurity and the host atom \cite{Harrison:1980_a}. For a conventional non-magnetic acceptor Zn$_{\rm Ga}$, which is the 1st nearest neighbor of Ga in the periodic table, the atomic {\em p}-levels are shifted  by $\sim$0.25~eV which increases the binding energy by $\sim 5$~meV. For Mn, the 6th nearest neighbor of Ga, the {\em p}-level shift is $\sim$1.5~eV which when compared to Zn$_{\rm Ga}$ implies
the  central-cell contribution   to the acceptor level of Mn$_{\rm Ga}$  $\sim 30$~meV \cite{Bhattacharjee:2000_a}.
(iii) The remaining part of the Mn$_{\rm Ga}$ binding energy is due to the spin-dependent hybridization of Mn $d$-states  with neighboring As $p$-states. Its contribution, which has been directly inferred from spectroscopic measurements of uncoupled Mn$_{\rm Ga}$ impurities \cite{Linnarsson:1997_a,Bhattacharjee:2000_a}, is again comparable to the binding energy of the hydrogenic single-acceptor potential. Combining  (i)-(iii) accounts for the experimental binding energy of the Mn$_{\rm Ga}$ acceptor of 0.1~eV. An important caveat to these elementary considerations, further quantified by our microscopic calculations \cite{suppl}, is that the short-range potentials alone of strengths inferred in (ii) and (iii)  would not produce a bound-state above the top of the VB but only a broad region of scattering states inside the VB.

The VB picture of ferromagnetic (Ga,Mn)As builds on the above conventional semiconductor description of the Mn$_{\rm Ga}$ acceptor in which the presence of the long-range hydrogenic-like impurity potential is  essential for creating a bound state in the band gap. With increasing doping, the impurity level broadens and for a sufficiently screened hydrogenic potential the impurity states must merge into the VB within this picture. The premise of the models with a persistent detached IB in ferromagnetic (Ga,Mn)As is distinct and can be reconciled by ascribing the main role in binding to the short-range potentials and a minor role to screening and impurity level broadening.

We now provide microscopic analysis of these scenarios by performing  TBA band-structure calculations. Disorder is treated in the coherent potential approximation (CPA) which allows us to scan the entire range of dopings from the single Mn-impurity limit to MnAs. Our results are consistent with available corresponding spectra obtained using the super-cell method \cite{Turek:2008_a} which justifies the validity of the CPA \cite{suppl} to represent the one-particle, orbital resolved, density of states (DOS).  We first take the conventional parameterization of atomic levels and overlap integrals \cite{Harrison:1980_a,suppl}. On-site electron-electron interactions on the Mn {\em d}-states are described using the Hubbard parameter $U=3.5$~eV and the Heisenberg parameter $J_H=0.6$~eV, which also correspond to a conventional parametrization of {\em d}-orbital correlations in atomic Mn or Mn in II-VI semiconductors, and are consistent with values of $U$ and $J_H$ inferred from photoemission experiments in (Ga,Mn)As \cite{Okabayashi:1998_a}. Since we are primarily interested in the ferromagnetic behavior which occurs at relatively high dopings ($\gtrsim1\%$) and is governed by the spin-dependent {\em p-d} hybridization potential, we omit the long-range Coulomb potential which is non-magnetic and largely screened at the relevant hole concentrations \cite{note2}.
\begin{figure}[ht]
\vspace*{-0.5cm}
\hspace*{-0.cm}\includegraphics[height=0.95\columnwidth,angle=0]{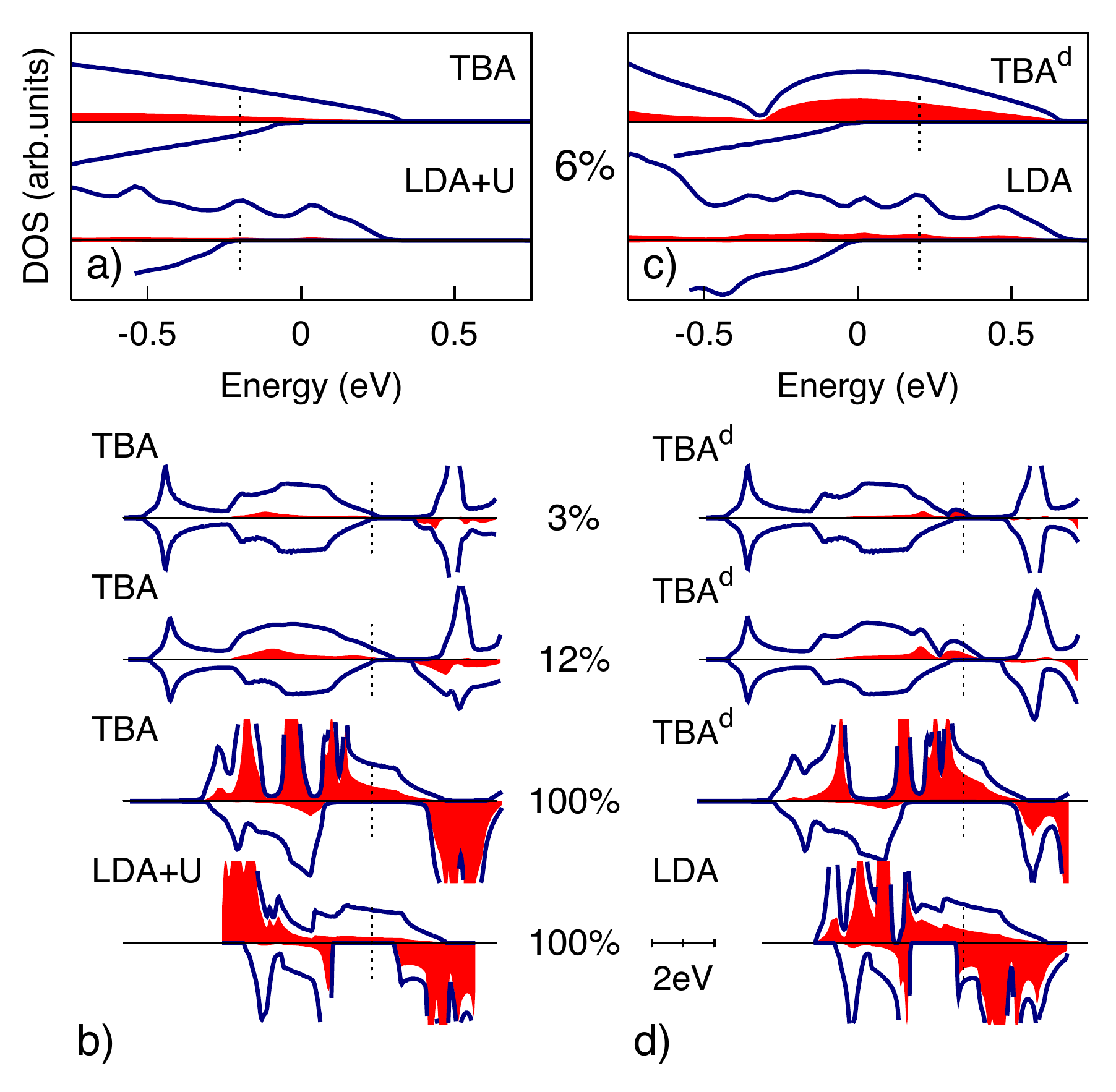}
\vspace*{-0.3cm} \caption{(a),(b) Total (solid line) and partial Mn $d$-orbital (red filled)  DOSs calculated using the TBA (CPA) and compared with the LDA+U (super-cell) results for Mn$_{\rm Ga}$ dopings of 3,6,12, and 100\%. (c),(d) Same for the TBA$^d$ compared with the LDA results. $E_F$ is indicated by the dashed  vertical line.}
\label{Figure1}
\end{figure}

Using the conventional values of the TBA parameters \cite{Harrison:1980_a,suppl} we first determined  occupation numbers on  Mn $d$-orbitals and corresponding on-site energies using the self-consistent unrestricted Hartree-Fock description of the Anderson Mn impurity embedded  in the semiconductor environment \cite{suppl}. The important result of these calculations is that we did not find any tendency to symmetry breaking in these occupation numbers, i.e., the three $t_{2g}$-orbitals (and similarly the two weakly hybridized $e_g$-orbitals) remain degenerate and strongly localized. After determining the Mn $d$-orbital on-site energies we proceed to calculate the microscopic DOS of (Ga,Mn)As over the entire doping range. In  Figs.~1(a),(b) we plot examples of both the total (black line) and the Mn $d$-orbital resolved (red filled) DOSs for $x=$3, 6, and 12\% together with the results for $x=$100\%, i.e. for the zinc-blende MnAs. The Mn {\em d} spectral weight is peaked at approximately 4~eV bellow the top of the VB, in agreement with photoemission data \cite{Okabayashi:1998_a}, and is significantly smaller near $E_F$ as further highlighted in Fig.~2(b). The Fermi level states at the top of the VB have a dominant As(Ga) $p$-orbital character; the stronger As $p$-component is plotted in Figs.~2(a).  The {\em p-d} coupling strength, $N_0\beta=\Delta/(Sx)$ \cite{Jungwirth:2006_a}, determined from the calculated VB exchange splitting $\Delta$ (and  taking $S=5/2$) is close to the upper bound of the reported experimental range of $N_0\beta\sim 1-3$~eV \cite{Matsukura:1998_a,Okabayashi:1998_a,Szczytko:1999_a,Bhattacharjee:2000_a,Omiya:2000_a}, as shown in Fig.~2(c). This is regarded as a moderately weak {\em p-d} coupling because the corresponding Fermi level states of the (Ga,Mn)As have a similar orbital character as the states in the host GaAs VB. The spectral features shown in Figs.~1(a),(b) and 2(a)-(c) are among the key characteristics of the VB picture. Note that the $\bm k\cdot\bm p$ kinetic-exchange (Zener) model calculations assume a value of $N_0\beta$ also within the range of 1-3~eV (typically closer to the lower experimental bound) \cite{Jungwirth:2006_a}. It is this moderate {\em p-d} hybridization that allows it to be treated perturbatively and to perform the Schrieffer-Wolff transformation  to effective valence band states experiencing a spin-dependent kinetic-exchange field \cite{Jungwirth:2006_a}. Hence, the effective kinetic-exchange model and the microscopic TBA theory provide a consistent physical picture of ferromagnetic (Ga,Mn)As.

\begin{figure}[ht]
\vspace*{-0cm}
\hspace*{-0.cm}\includegraphics[height=0.85\columnwidth,angle=0]{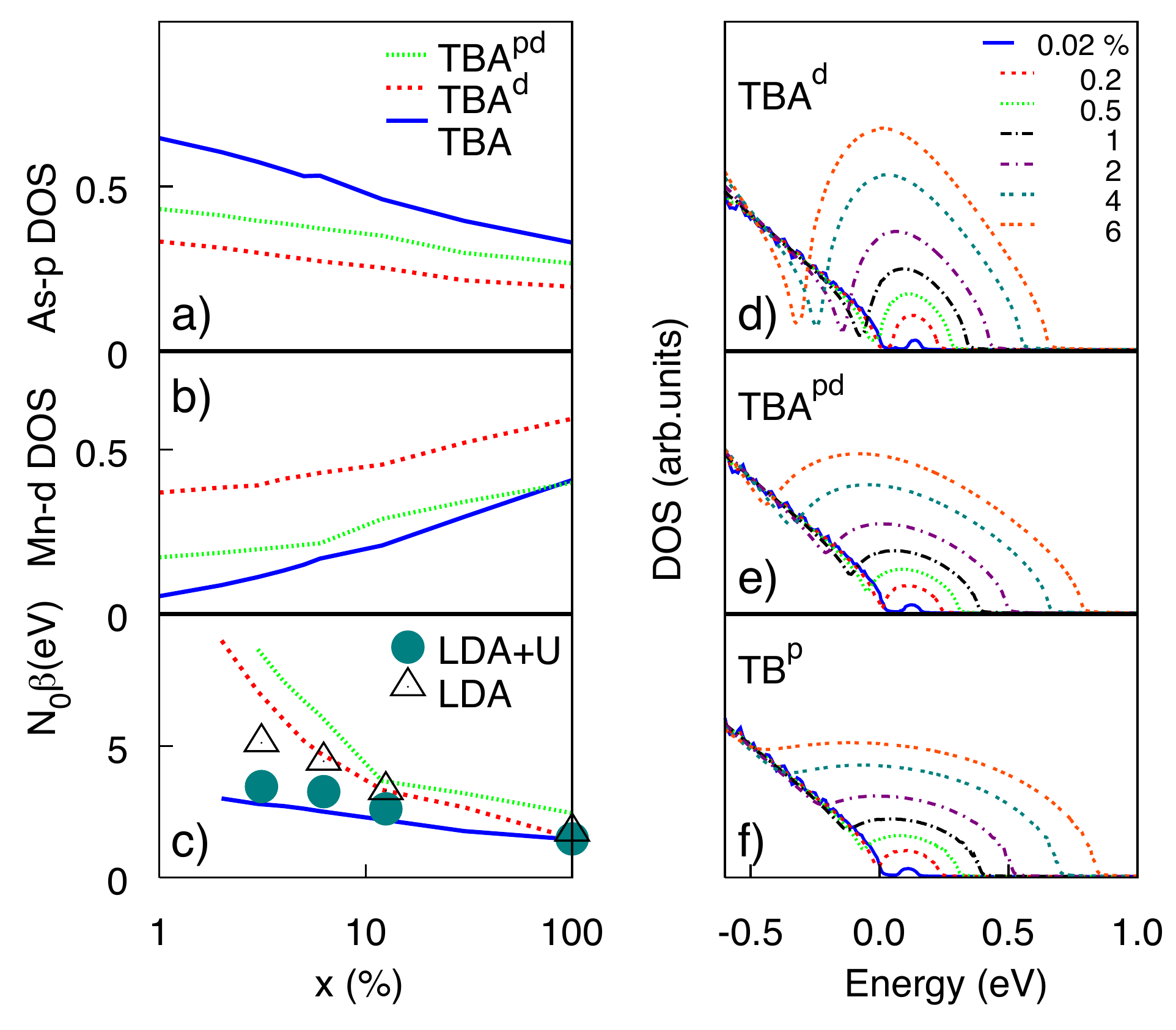}
\vspace*{-0.5cm} 
\caption{ Partial As $p$-orbital (a) and Mn $d$-orbital (b) DOSs at $E_F$ in the depicted TBA models. The remaining contribution to the total DOS  is primarily due to Ga and Mn $p$-orbitals with their relative weights given nearly precisely by the Ga/Mn ratio \protect\cite{suppl}. (c) The p-d coupling strength in the TBA models compared with the LDA (triangles) and LDA+U (circles) results. (d)-(f) Total DOSs showing the merging of the IB into the VB in the depicted tight-binding models.}
\label{Figure2}
\end{figure}

We next attempt to recreate the IB models by considering that the bound state at the single Mn  is formed by the short range impurity potentials. As noted above, this is not obtained from the conventionally parameterized TBA and the values of the atomic levels or overlap integrals have to be adjusted {\em ad hoc} to match the 0.1~eV binding energy \cite{suppl}. We first search for a bound state due to the central-cell potential by treating the Mn {\em p}-level  as a free parameter. We find that binding the hole by the central-cell potential alone requires physically incomprehensible {\em p}-level shifts of several tens of eV \cite{Tang:2004_a}. The reason for this is the short-range nature of the potential and the orbital composition of the top VB from which the bound state forms. The VB near its maximum is dominated by  {\em p}-orbitals of the As  not Ga sublattice.

A more favorable scenario to create a bound state through the short-range potentials is tuning the strength of the {\em p-d} hybridization. This term is less local as it affects four As neighbors of the Mn$_{\rm Ga}$ impurity and
acts on the As {\em p}-orbitals which form the top of the host VB. To tune the hybridization strength we can treat as a free parameter the atomic Mn {\em d}-level or the {\em p-d} hopping \cite{suppl}. The corresponding models are labelled as TBA$^d$ and as TBA$^{pd}$, respectively. For the TBA$^d$ model we obtain the 0.1~eV  bound-state when shifting the {\em d}-level by 1.5~eV. We
now fix this parameter and calculate the corresponding DOSs over the entire doping range, as shown in Figs.~1(c),(d).
More detailed characteristics of the corresponding spectra are summarized in Figs.~2(a)-(d). The key observation is that for dopings above $\sim 0.1$\% the band structure cannot be recast in a model with Fermi level states residing in a narrow IB (of width not exceeding the single impurity binding energy) which is detached from the VB. The TBA$^{pd}$ model yields the same general conclusion, as shown in Figs.~2(d),(e). A detached IB model for the ferromagnetic (Ga,Mn)As materials is therefore microscopically incompatible with the 0.1~eV acceptor level even if the binding of the hole to the Mn impurity was entirely due to short-range potentials.
We remark that $N_0\beta$ in the TBA$^{d(pd)}$ parameterization is about a factor of 2-3 stronger than in the conventionally parameterized TBA model, i.e., much larger than the upper experimental bound for the {\em p-d} coupling strength. This discrepancy is due to the omission of the long-range Coulomb potential when fitting the experimental single Mn$_{\rm Ga}$ acceptor state. Note also that the dip in the TBA$^{d(pd)}$ DOS, which persists and shifts deep in the band at high dopings, is another consequence of the {\em ad hoc} increased {\em p-d} coupling.

We next associate the IB models postulated in literature with corresponding microscopic TBA calculations. One proposed phenomenology assumes a dominant  Mn {\em d}-orbital nature of the detached  IB and allows for some hybridization with the host VB \cite{Burch:2006_a,Burch:2008_a}. The TBA$^d$ theory is the closest microscopic realization of this model. It shows that apart from the absence of the detached IB itself at dopings corresponding to ferromagnetic (Ga,Mn)As, the formation of the 0.1~eV acceptor state by shifting the Mn $d$-level does not yield a dominant Mn {\em d} spectral weight near $E_F$. From this perspective we regard the 0.1~eV  acceptor level as moderately shallow.

An orthogonal IB model, in terms of the assumed  orbital character of the IB, elaborates on a {\em sp}-tight-binding Hamiltonian with shifted {\em p}-levels on  four As neighbors of the Mn$_{\rm Ga}$ \cite{Tang:2008_a}. We label this model as TB$^p$. The shifts are introduced to effectively account for the microscopic {\em p-d} hybridization and again to obtain the 0.1~eV single impurity state without the hydrogenic long-range Coulomb potential. The model has a merit in the very dilute regime \cite{Tang:2004_a} as the extent of the bound state wavefunction (the exponential tail) is determined by the value of the binding energy and is insensitive to the specific choice of the confining potential. It also captures, by its design, the symmetries of the As {\em p}-orbital dominated bound state. The model can be associated with our microscopic TBA$^{pd}$ calculations and indeed the corresponding DOSs show very similar doping trends, as shown in Fig.~2 (e),(f). Again, no detached IB persists in the TB$^p$  DOS to dopings above $\sim 0.1$\%.

Another phenomenological proposal assumes that states in the IB have  Mn {\em p}-orbital character \cite{Stone:2008_a}. This corresponds to our first attempt to obtain the Mn$_{\rm Ga}$ acceptor level in short-range potentials by considering the central-cell component only. As discussed above, such a model would require an unphysical large shift of the Mn $p$-levels.
If we omit the microscopic justification of this IB model the approach has a merit as a phenomenological effective model describing the Mn acceptor level in the band-gap of the host semiconductor.  Since the impurity states are added in this type of effective modeling {\em ad hoc} to the spectrum, the model does not conserve the total number of states. (In a microscopic language it describes rather an interstitial than a substitutional impurity.) The applicability of the approach is therefore limited to small Mn concentrations and the model is not suitable for exploring trends with changing  Mn doping.  

Finally we show in Figs.~1(a)-(d) examples of the comparisons of the TBA$^d$ and TBA calculations with results of the LDA and LDA+U full-potential {\em ab initio} theories \cite{LDA,suppl}. We find a very good agreement between the TBA and LDA+U results \cite{note_lda}. The LDA+U, the TBA, and the kinetic-exchange Zener theories therefore all provide a compatible picture of the band structure of ferromagnetic (Ga,Mn)As. We also find a clear correspondence between the TBA$^d$ and LDA results. The large exchange splitting of the VB obtained in the LDA reflects the general deficiency of the LDA to account for localized states within an itinerant band. Mn {\em d}-states are more delocalized  and move closer to the VB edge in the LDA, which enhances the hybridization. Hence, the exchange splitting of the top of the VB is increased to values comparable to those of the TBA$^d$ Hamiltonian. 

To conclude, at the doping levels for which (Ga,Mn)As is ferromagnetic none of the postulated one-particle DOS models with a detached IB arising from the 0.1 eV Mn acceptor level in GaAs is microscopically justified. The Fermi level states in ferromagnetic (Ga,Mn)As can be regarded as residing in a modified VB of the host semiconductor due to disorder, exchange splitting, and admixture of the impurity orbitals. The corresponding one-particle band structure can be described by methods ranging from full-potential density-functional theory to multi-orbital tight-binding-Anderson or  envelope-function approaches which are all mutually consistent. We also emphasize nevertheless, that due to the vicinity of the metal-insulator transition and correlation phenomena these effective one-particle VB band models can only represent a proxy to the complex electronic structure of ferromagnetic (Ga,Mn)As materials. 

We acknowledge support from EU Grants FP7-215368 SemiSpinNet, FP7-214499 NAMASTE, Czech Republic Grants  GACR 202/07/0456, AV0Z10100520, AV0Z10100521, KAN400100652, LC510, Preamium Academiae, and U.S. Grants  ONR-N000140610122,  DMR-0547875, SWAN-NRI. JS is a Cottrell Scholar of Research Corporation.


\newpage
\onecolumngrid
\begin{center}
{\Large\bf SUPPLEMENTARY MATERIAL}
\end{center}
\vspace*{3.3cm}  
\section{Tight-binding Anderson description of ${\rm\bf (Ga,Mn)As}$}
\addtocounter{figure}{-2}
In this section we review our implementation of the tight-binding Anderson formalism for calculating the band structure of (Ga,Mn)As. First we introduce the Hamiltonian describing the semiconductor host GaAs. We use the Slater-Koster tight-binding approach \cite{Slater:1954_a,Harrison:1980_a2} which is a variant of the linear-combination-of-atomic-orbitals (LCAO) method implemented using the Bloch's theorem. Most of the calculations in this paper are done in the second-nearest-neighbor (nearest Ga-As, As-As, and Ga-Ga bonds) $sp^3$ parametrization \cite{Talwar:1982_a} of the tight-binding Hamiltonian. We have checked that the alternative nearest-neighbor (only nearest Ga-As bonds) $sp^3$ parametrization \cite{Chadi:1977_a} and the nearest-neighbor $sp^3s^{\ast}$ parametrization \cite{Vogl:1983_a} give very similar results for the physics discussed in this paper. Note that these parametrizations are obtained by fitting to the band structure calculations with pseudopotentials obtained from experiment. Anticipating the introduction of substitutional Mn$_{\rm Ga}$ impurities we formally included  Ga $d$ orbitals in the GaAs tight-binding Hamiltonian. There energy level is, however, far above the bottom of the conduction band and has no effect on the part of the band structure we are interested in. For the description of a single Mn impurity we use a combination of the Bloch's theorem based Slater-Koster approach in which Mn does not introduce new orbitals, i.e., it does not change the dimension of the Hilbert space, electron-electron interactions on the Mn $d$ shell are treated in the Anderson-impurity spirit, and the effect of the periodic crystal environment of the host GaAs on Mn impurity states is accounted using Green's function formalism.

\subsection{Slater-Koster tight-binding description of GaAs in the second-nearest-neighbor $sp^3$ parametrization}

For describing the valence and conduction bands of the host GaAs, the basis functions are written in the form of Bloch sums,

\begin{eqnarray}\label{bloch}
\Phi_{{\bm k} a\alpha}&=&N^{-1/2}e^{i{\bm k}\cdot{\bm p}_a}\sum_{n=0}^{N-1}e^{i{\bm k}\cdot{\bm R}_n}\phi_{a\alpha}({\bm r}-{\bm R_n}-{\bm p}_a)=\left(N^{-1/2}e^{i{\bm k}\cdot{\bm p}_a}e^{i{\bm k}\cdot{\bm r}}\sum_{n=0}^{N-1}e^{-i{\bm k}\cdot({\bm r}-{\bm R}_n)}\phi_{a\alpha}({\bm r}-{\bm R_n}-{\bm p}_a)\right)\nonumber \\
&=&N^{-1/2}\sum_{n=0}^{N-1}e^{i{\bm k}\cdot{\bm\rho}_{na}}\phi_{a\alpha}({\bm r}-{\bm\rho}_{na})\;\;\;{\rm with}\;\;{\bm\rho}_{na}={\bm R_n}+{\bm p}_a\;,
\end{eqnarray}
where $a$ is the atom index in the unit cell, $\alpha$ is the atomic orbital quantum number, $N$ is the number of atoms $a$ (unit cells), $n$ is the unit cell index, ${\bm R}_n$ is the unit cell vector, ${\bm p}_a$ is the position vector of the atom $a$ in the unit cell, and the overall phase factor $e^{i{\bm k}\cdot{\bm p}_a}$ is added to obtain more symmetric expressions for the Hamiltonian matrix elements derived below. The wavefunction $\phi_{a\alpha}({\bm r}-{\bm\rho}_{na})$ are centered around the atom $a$ in unit cell $n$ and are orthonormal, i.e., assumed to be constructed from the atomic orbitals $\psi_{a\alpha}$ following L\"owdin's orthonormalization procedure,

\begin{equation}\label{lowdin}
\phi_{a\alpha}({\bm r}-{\bm\rho}_{na})=\psi_{a\alpha}({\bm r}-{\bm\rho}_{na})-\frac12\sum_m\psi_{a\alpha}({\bm r}-{\bm\rho}_{ma})S_{mn}+\ldots\;,
\end{equation}
where $S_{mn}$ are the overlap integrals,
\begin{equation}\label{overlap}
S_{mn}=\int d{\bm r}\psi_{a\alpha}^{\ast}({\bm r}-{\bm\rho}_{ma})\psi_{a\alpha}({\bm r}-{\bm\rho}_{na})
\end{equation}
For the zinc-blende GaAs lattice we set,
\begin{eqnarray}
{\bm p}_a&=&(0,0,0)\; {\rm for}\; {\rm Ga} \nonumber\\
&=&\frac{a_{lc}}{4}(1,1,1)\; {\rm for}\; {\rm As}\;
\end{eqnarray}
where $a_{lc}$ is the cube lattice constant of GaAs.

The number of $k$ points in the first Brillouin zone is the same is the number of unit cells but since
the Hamiltonian has the same periodicity as the basis function in Eq.~(\ref{bloch}) it is diagonal in the ${\bm k}$-vector.
\begin{eqnarray}
\langle\Phi_{{\bm k^{\prime}} a^{\prime}\alpha^{\prime}}|H|\Phi_{{\bm k} a\alpha}\rangle&=&\frac{1}{N}\sum_{n}\sum_{n^{\prime}}\int d{\bm r}\phi^{\ast}_{a^{\prime}\alpha^{\prime}}({\bm r}-{\bm\rho}_{n^{\prime}a^{\prime}})H\phi_{a\alpha}({\bm r}-{\bm\rho}_{na})\nonumber \\
&=&\frac{1}{N}\sum_{n}\int d{\bm r}\phi^{\ast}_{a\alpha}({\bm r}-\Delta{\bm\rho}_{na})H\phi_{a^{\prime}\alpha^{\prime}}({\bm r})e^{i{\bm k}\cdot\Delta{\bm\rho}_{na}}\sum_{n^{\prime}}e^{-i{\bm k^{\prime}}\cdot{\bm\rho}_{n^{\prime}a^{\prime}}}e^{i{\bm k}\cdot{\bm\rho}_{n^{\prime}a^{\prime}}}\nonumber \\
&=&\delta_{kk^{\prime}}\sum_{n}e^{i{\bm k}\cdot{\bm\rho}_{n{\alpha}}}\int d{\bm r}\phi^{\ast}_{{\alpha}\alpha}({\bm r}-{\bm\rho}_{n{\alpha}})H\phi_{{\alpha^{\prime}}\beta}({\bm r})
\label{hopping}
\end{eqnarray}

In this Koster-Slater implementation of the tight-binding model the problem is simplified as we only diagonalize a Hamiltonian matrix of a dimension given by the number of atoms in the unit cell and valence orbitals considered for each atom.
Since every atom in GaAs has the same environment (there's one Ga and one As per unit cell),  As-Ga Hamiltonian matrix elements, keeping only the nearest-neighbor (4 atoms) hopping terms in the sum, can be written for all ${\bm k}$-vectors within the first Brillouin zone as,
\begin{equation}
H_{{\rm As}\alpha,{\rm Ga}\beta}({\bm k})=\sum_{n=0}^{3\;{\rm n.n.}}e^{i{\bm k}\cdot{\bm\rho}_{n{\rm As}}}\int d{\bm r}\phi^{\ast}_{{\rm As}\alpha}({\bm r}-{\bm\rho}_{n{\rm As}})H\phi_{{\rm Ga}\beta}({\bm r})\;.
\label{hoppingAsGa}
\end{equation}
The unit cell lattice vectors are ${\bm a}_1=\frac{a_{lc}}{2}(1,1,0)$, ${\bm a}_2=\frac{a_{lc}}{2}(1,0,1)$, and ${\bm a}_3=\frac{a_{lc}}{2}(0,1,1)$. For the Ga in the 0th unit cell considered above, ${\bm R}_0=(0,0,0)$, the 4 nearest As neighbors are in unit cells ${\bm R}_0$, ${\bm R}_1=-{\bm a}_1$, ${\bm R}_2=-{\bm a}_2$, and ${\bm R}_3=-{\bm a}_3$  at positions,
\begin{eqnarray}
{\bm\rho}_{0{\rm As}}&=&{\bm p}_{\rm As}+{\bm R}_0=\frac{a_{lc}}{4}(1,1,1)\nonumber\\
{\bm\rho}_{1{\rm As}}&=&{\bm p}_{\rm As}+{\bm R}_1=\frac{a_{lc}}{4}(-1,-1,1)\nonumber\\
{\bm\rho}_{2{\rm As}}&=&{\bm p}_{\rm As}+{\bm R}_2=\frac{a_{lc}}{4}(-1,1,-1)\nonumber\\
{\bm\rho}_{3{\rm As}}&=&{\bm p}_{\rm As}+{\bm R}_3=\frac{a_{lc}}{4}(1,-1,-1)
\end{eqnarray}
The Ga-Ga matrix elements, keeping the diagonal term and only the nearest Ga (12 atoms) hopping terms in the sum, read
\begin{equation}
H_{{\rm Ga}\alpha,{\rm Ga}\beta}({\bm k})=\sum_{n=0}^{12\;{\rm n.n.}}e^{i{\bm k}\cdot{\bm R}_{n}}\int d{\bm r}\phi^{\ast}_{{\rm Ga}\alpha}({\bm r}-{\bm R}_{n})H\phi_{{\rm Ga}\beta}({\bm r})\;,
\label{hoppingGaGa}
\end{equation}
with the positions of the 12 As neighbors given by,
\begin{eqnarray}
{\bm R}_1&=&-{\bm a}_1=\frac{a_{lc}}{2}(-1,-1,0)\nonumber\\
{\bm R}_2&=&-{\bm a}_2=\frac{a_{lc}}{2}(-1,0,-1)\nonumber\\
{\bm R}_3&=&-{\bm a}_3=\frac{a_{lc}}{2}(0,-1,-1)\nonumber\\
{\bm R}_4&=&{\bm a}_1=\frac{a_{lc}}{2}(1,1,0)\nonumber\\
{\bm R}_5&=&{\bm a}_2=\frac{a_{lc}}{2}(1,0,1)\nonumber\\
{\bm R}_6&=&{\bm a}_3=\frac{a_{lc}}{2}(0,1,1)\nonumber\\
{\bm R}_7&=&{\bm a}_2-{\bm a}_3=\frac{a_{lc}}{2}(1,-1,0)\nonumber\\
{\bm R}_8&=&{\bm a}_3-{\bm a}_2=\frac{a_{lc}}{2}(-1,1,0)\nonumber\\
{\bm R}_9&=&{\bm a}_1-{\bm a}_3=\frac{a_{lc}}{2}(1,0,-1)\nonumber\\
{\bm R}_{10}&=&{\bm a}_3-{\bm a}_1=\frac{a_{lc}}{2}(-1,0,1)\nonumber\\
{\bm R}_{11}&=&{\bm a}_1-{\bm a}_2=\frac{a_{lc}}{2}(0,1,-1)\nonumber\\
{\bm R}_{12}&=&{\bm a}_2-{\bm a}_1=\frac{a_{lc}}{2}(0,-1,1)\nonumber\\
\end{eqnarray}
Similarly the As-As elements read,
\begin{equation}
H_{{\rm As}\alpha,{\rm As}\beta}({\bm k})=\sum_{n=1}^{12\;{\rm n.n.}}e^{i{\bm k}\cdot{\bm R}_{n}}\int d{\bm r}\phi^{\ast}_{{\rm As}\alpha}({\bm r}-{\bm R}_{n})H\phi_{{\rm As}\beta}({\bm r})\;.
\label{hoppingAsAs}
\end{equation}
The diagonal in the atom index on site integrals are denoted as,
\begin{eqnarray}
\epsilon^{{\rm Ga}}_{\alpha}&=&\int d{\bm r}\phi^{\ast}_{{\rm Ga}\alpha}({\bm r})H\phi_{{\rm Ga}\alpha}({\bm r})\nonumber \\
\epsilon^{{\rm As}}_{\alpha}&=&\int d{\bm r}\phi^{\ast}_{{\rm As}\alpha}({\bm r})H\phi_{{\rm As}\alpha}({\bm r})\nonumber \\
\end{eqnarray}
The diagonal in the atom index hopping integrals are parametrized as,
\begin{eqnarray}
W^{{\rm Ga}}_{\alpha\beta}({\bm R}_{n})&=&\int d{\bm r}\phi^{\ast}_{{\rm Ga}\alpha}({\bm r}-{\bm R}_{n})H\phi_{{\rm Ga}\beta}({\bm r})\nonumber \\
W^{{\rm As}}_{\alpha\beta}({\bm R}_{n})&=&\int d{\bm r}\phi^{\ast}_{{\rm As}\alpha}({\bm r}-{\bm R}_{n})H\phi_{{\rm As}\beta}({\bm r})\;.
\end{eqnarray}
The off-diagonal in the atom index hopping integrals are parametrized by
\begin{equation}
W^{{\rm As},{\rm Ga}}_{\alpha\beta}({\bm\rho}_{n{\rm As}})=\int d{\bm r}\phi^{\ast}_{{\rm As}\alpha}({\bm r}-{\bm\rho}_{n{\rm As}})H\phi_{{\rm Ga}\beta}({\bm r})\;.
\end{equation}
Finally we rewrite the hopping integrals in the form of $\sigma$ and $\pi$  bonds between $s$, $p$, and $d$ orbitals as,
\begin{eqnarray}
W^{{\rm Ga}}_{ss}({\bm R}_{n})&=&V^{{\rm Ga}}_{ss\sigma}(|{\bm R}_{n}|)\nonumber\\
W^{{\rm Ga}}_{sp_i}({\bm R}_{n})&=&V^{{\rm Ga}}_{sp\sigma}(|{\bm R}_{n}|)\hat{R}_{n,i}\;\;\;, i=x,y,z\nonumber\\
W^{{\rm Ga}}_{p_ip_j}({\bm R}_{n})&=&V^{{\rm Ga}}_{pp\sigma}(|{\bm R}_{n}|)\hat{R}_{n,i}\hat{R}_{n,j}+V^{{\rm Ga}}_{p_ip_j\pi}(|{\bm R}_{n}|)(\delta_{ij}-\hat{R}_{n,i}\hat{R}_{n,j})\;\;\;, i,j=x,y,z\nonumber\\
W^{{\rm As}}_{ss}({\bm R}_{n})&=&V^{{\rm As}}_{ss\sigma}(|{\bm R}_{n}|)\nonumber\\
W^{{\rm As}}_{sp_i}({\bm R}_{n})&=&V^{{\rm As}}_{sp\sigma}(|{\bm R}_{n}|)\hat{R}_{n,i}\;\;\;, i=x,y,z\nonumber\\
W^{{\rm As}}_{p_ip_j}({\bm R}_{n})&=&V^{{\rm As}}_{pp\sigma}(|{\bm R}_{n}|)\hat{R}_{n,i}\hat{R}_{n,j}+V^{{\rm As}}_{p_ip_j\pi}(|{\bm R}_{n}|)(\delta_{ij}-\hat{R}_{n,i}\hat{R}_{n,j})\;\;\;, i,j=x,y,z\nonumber\\
W^{{\rm As},{\rm Ga}}_{ss}({\bm\rho}_{n{\rm As}})&=&V^{{\rm As},{\rm Ga}}_{ss\sigma}(|{\bm\rho}_{n{\rm As}}|)\nonumber\\
W^{{\rm As},{\rm Ga}}_{sp_i}({\bm\rho}_{n{\rm As}})&=&V^{{\rm As},{\rm Ga}}_{sp\sigma}(|{\bm\rho}_{n{\rm As}}|)\hat{\rho}_{n{\rm As},i}\;\;\;, i=x,y,z\nonumber\\
W^{{\rm As},{\rm Ga}}_{p_is}({\bm\rho}_{n{\rm As}})&=&V^{{\rm As},{\rm Ga}}_{ps\sigma}(|{\bm\rho}_{n{\rm As}}|)\hat{\rho}_{n{\rm As},i}\;\;\;, i=x,y,z\nonumber\\
W^{{\rm As},{\rm Ga}}_{p_ip_j}({\bm\rho}_{n{\rm As}})&=&V^{{\rm As},{\rm Ga}}_{pp\sigma}(|{\bm\rho}_{n{\rm As}}|)\hat{\rho}_{n{\rm As},i}\hat{\rho}_{n{\rm As},j}+V^{{\rm As},{\rm Ga}}_{pp\pi}(|{\bm\rho}_{n{\rm As}}|)(\delta_{ij}-\hat{\rho}_{n{\rm As},i}\hat{\rho}_{n{\rm As},j})\;\;\;, i,j=x,y,z \nonumber\\
W^{{\rm As},{\rm Ga}}_{sd_{ij}}({\bm\rho}_{n{\rm As}})&=&\sqrt{3}V^{{\rm As},{\rm Ga}}_{sd\sigma}(|{\bm\rho}_{n{\rm As}}|)\hat{\rho}_{n{\rm As},i}\hat{\rho}_{n{\rm As},j}\;\;\;, i,j=x,y,z \nonumber\\
W^{{\rm As},{\rm Ga}}_{sd_{x^2-y^2}}({\bm\rho}_{n{\rm As}})&=&\frac{\sqrt{3}}{2}V^{{\rm As},{\rm Ga}}_{sd\sigma}(|{\bm\rho}_{n{\rm As}}|)\left(\hat{\rho}_{n{\rm As},x}^2-\hat{\rho}_{n{\rm As},y}^2\right)\nonumber\\
W^{{\rm As},{\rm Ga}}_{sd_{3z^2-r^2}}({\bm\rho}_{n{\rm As}})&=&V^{{\rm As},{\rm Ga}}_{sd\sigma}(|{\bm\rho}_{n{\rm As}}|)\left[\hat{\rho}_{n{\rm As},z}^2-\frac12\left(\hat{\rho}_{n{\rm As},x}^2+\hat{\rho}_{n{\rm As},y}^2\right)\right]\nonumber\\
W^{{\rm As},{\rm Ga}}_{p_id_{ij}}({\bm\rho}_{n{\rm As}})&=&\sqrt{3}V^{{\rm As},{\rm Ga}}_{pd\sigma}(|{\bm\rho}_{n{\rm As}}|)\hat{\rho}_{n{\rm As},i}^2\hat{\rho}_{n{\rm As},j}+V^{{\rm As},{\rm Ga}}_{pd\pi}(|{\bm\rho}_{n{\rm As}}|)\hat{\rho}_{n{\rm As},j}\left(1-2\hat{\rho}_{n{\rm As},i}^2\right)\;\;\;,  \nonumber\\
W^{{\rm As},{\rm Ga}}_{p_id_{jk}}({\bm\rho}_{n{\rm As}})&=&\sqrt{3}V^{{\rm As},{\rm Ga}}_{pd\sigma}(|{\bm\rho}_{n{\rm As}}|)\hat{\rho}_{n{\rm As},i}\hat{\rho}_{n{\rm As},j}\hat{\rho}_{n{\rm As},k}-2V^{{\rm As},{\rm Ga}}_{pd\pi}(|{\bm\rho}_{n{\rm As}}|)\hat{\rho}_{n{\rm As},i}\hat{\rho}_{n{\rm As},j}\hat{\rho}_{n{\rm As},k}\;\;\;, i\neq j\neq k \nonumber\\
W^{{\rm As},{\rm Ga}}_{p_xd_{x^2-y^2}}({\bm\rho}_{n{\rm As}})&=&\frac{\sqrt{3}}{2}V^{{\rm As},{\rm Ga}}_{pd\sigma}(|{\bm\rho}_{n{\rm As}}|)\hat{\rho}_{n{\rm As},x}\left(\hat{\rho}^2_{n{\rm As},x}-\hat{\rho}^2_{n{\rm As},y}\right)+V^{{\rm As},{\rm Ga}}_{pd\pi}(|{\bm\rho}_{n{\rm As}}|)\hat{\rho}_{n{\rm As},x}\left(1-\hat{\rho}^2_{n{\rm As},x}+\hat{\rho}^2_{n{\rm As},y}\right)\nonumber\\
W^{{\rm As},{\rm Ga}}_{p_yd_{x^2-y^2}}({\bm\rho}_{n{\rm As}})&=&\frac{\sqrt{3}}{2}V^{{\rm As},{\rm Ga}}_{pd\sigma}(|{\bm\rho}_{n{\rm As}}|)\hat{\rho}_{n{\rm As},y}\left(\hat{\rho}^2_{n{\rm As},x}-\hat{\rho}^2_{n{\rm As},y}\right)-V^{{\rm As},{\rm Ga}}_{pd\pi}(|{\bm\rho}_{n{\rm As}}|)\hat{\rho}_{n{\rm As},y}\left(1+\hat{\rho}^2_{n{\rm As},x}-\hat{\rho}^2_{n{\rm As},y}\right)\nonumber\\
W^{{\rm As},{\rm Ga}}_{p_zd_{x^2-y^2}}({\bm\rho}_{n{\rm As}})&=&\frac{\sqrt{3}}{2}V^{{\rm As},{\rm Ga}}_{pd\sigma}(|{\bm\rho}_{n{\rm As}}|)\hat{\rho}_{n{\rm As},z}\left(\hat{\rho}^2_{n{\rm As},x}-\hat{\rho}^2_{n{\rm As},y}\right)-V^{{\rm As},{\rm Ga}}_{pd\pi}(|{\bm\rho}_{n{\rm As}}|)\hat{\rho}_{n{\rm As},z}\left(\hat{\rho}^2_{n{\rm As},x}-\hat{\rho}^2_{n{\rm As},y}\right)\nonumber\\
W^{{\rm As},{\rm Ga}}_{p_id_{3z^2-r^2}}({\bm\rho}_{n{\rm As}})&=&V^{{\rm As},{\rm Ga}}_{pd\sigma}(|{\bm\rho}_{n{\rm As}}|)\hat{\rho}_{n{\rm As},i}\left[{\hat{\rho}^2_{n{\rm As},z}}-\frac12\left(\hat{\rho}^2_{n{\rm As},x}+\hat{\rho}^2_{n{\rm As},y}\right)\right]-\sqrt{3}V^{{\rm As},{\rm Ga}}_{pd\pi}(|{\bm\rho}_{n{\rm As}}|)\hat{\rho}_{n{\rm As},i}\hat{\rho}^2_{n{\rm As},z}\;\;\;, i=x,y \nonumber\\
W^{{\rm As},{\rm Ga}}_{p_zd_{3z^2-r^2}}({\bm\rho}_{n{\rm As}})&=&V^{{\rm As},{\rm Ga}}_{pd\sigma}(|{\bm\rho}_{n{\rm As}}|)\hat{\rho}_{n{\rm As},z}\left[{\hat{\rho}^2_{n{\rm As},z}}-\frac12\left(\hat{\rho}^2_{n{\rm As},x}+\hat{\rho}^2_{n{\rm As},y}\right)\right]+\sqrt{3}V^{{\rm As},{\rm Ga}}_{pd\pi}(|{\bm\rho}_{n{\rm As}}|)\hat{\rho}_{n{\rm As},z}\left(\hat{\rho}^2_{n{\rm As},x}+\hat{\rho}^2_{n{\rm As},y}\right) \nonumber \\
&{}&
\end{eqnarray}
Here $\hat{\rho}_{n{\rm As},i}$ ($\hat{R}_{n,i}$)is the $i$-th component of the unit vector along ${\bm\rho}_{n{\rm As}}$ (${\bm R}_{n}$). Note that the diagonal in the atom index hopping integrals are parametrized within the three-center approximation, i.e., considering also the term in the periodic potential, $V({\bm r})=\sum_{na}V_a({\bm r}-{\bm R}_{n}-{\bm p}_a)$ centered between the wavefunctions on site corresponding to the other sublattice. In this approximation we need to keep spatial index ($x,y,z$) dependent parameters $V^{{\rm As}({\rm Ga})}_{p_ip_i\pi}(|{\bm R}_{n}|)$. For the near-neighbor As-Ga hopping we only consider the two-center integrals in which case all $V$'s including the $V^{{\rm As},{\rm Ga}}_{pp\pi}(|{\bm\rho}_{n{\rm As}}|)$ term depend only on the bond length. Values for the on-site energies and all the above $\sigma$ and $\pi$-bond hopping energies for $s$ and $p$ orbitals are taken from Ref.~\onlinecite{Talwar:1982_a}. For the Ga $d$ orbitals, which are unimportant for GaAs but are included anticipating the Mn substitution, we put the on-site energy to a very high value. The hopping energies $V^{{\rm As},{\rm Ga}}_{sd\sigma}=V^{{\rm As},{\rm Mn}}_{sd\sigma}$, $V^{{\rm As},{\rm Ga}}_{pd\sigma}=V^{{\rm As},{\rm Mn}}_{pd\sigma}$ and
$V^{{\rm As},{\rm Ga}}_{pd\pi}=V^{{\rm As},{\rm Mn}}_{pd\pi}$ are discussed in the following section.

We have tested that spin-orbit coupling, when included in the diagonal Ga and As $p$-orbital terms in the form of $\xi_{\rm Ga(As)}{\bm l}\cdot{\bm s}$, has no significant effect on the physics discussed in this paper. (Note that in the $sp^3s^{\ast}$ parametrization \cite{Vogl:1983_a}, spin-orbit is included in the fitting of the tight-binding parameters to the {\em ab initio} band structure. The second-nearest-neighbor $sp^3$ parametrization \cite{Talwar:1982_a} was done in the original paper without including spin-orbit coupling; in our implementation, $\xi_{\rm Ga(As)}$ were taken from Ref.~\onlinecite{Vogl:1983_a} and the on-site energies were then rigidly shifted to recover the value of the GaAs band gap.)

\subsection{Tight-binding parametrization of a single substitutional Mn$_{\rm Ga}$ impurtity in GaAs}
As mentioned above, the GaAs tight-binding Hamiltonian parameters \cite{Chadi:1977_a,Talwar:1982_a,Vogl:1983_a} are obtained by fitting to the empirical pseudopotential band structure calculations. For the nearest-neighbor $s$ and $p$-orbital As-Mn hopping energies we take the same values as for the As-Ga bonds. The Mn $d$-orbitals is included in our theory in the spirit of Harrison's rules \cite{Harrison:1980_a2}. The hopping
integrals involving Mn $d$-orbitals are parametrized as,
\begin{eqnarray}
V^{{\rm As},{\rm Mn}}_{sd\sigma}&=&\eta_{sd\sigma}\frac{\hbar^2}{m_ed^2}\left(\frac{r_d}{d}\right)^{3/2}\nonumber\\
V^{{\rm As},{\rm Mn}}_{pd\sigma}&=&\eta_{pd\sigma}\frac{\hbar^2}{m_ed^2}\left(\frac{r_d}{d}\right)^{3/2}\nonumber\\
V^{{\rm As},{\rm Mn}}_{pd\pi}&=&\eta_{pd\pi}\frac{\hbar^2}{m_ed^2}\left(\frac{r_d}{d}\right)^{3/2}\;,
\label{eta}
\end{eqnarray}
where we took for the dimensionless hopping parameters for the covalent Mn-As bonds,  $\eta_{sd\sigma}=-4.9964$, $\eta_{pd\sigma}=-4.6644$, and $\eta_{pd\pi}=2.1503$, the Mn-As distance $d=\sqrt{3/4}\,a_{lc}$, and for the effective Mn $d$-orbital radius $r_d=0.86\AA$ (see Ref.~\onlinecite{Harrison:1980_a2}).
The change of the crystal potential due to the substitutional Mn is represented by  replacing Ga on-site energies with on-site Mn energies,
$\epsilon^{\rm Mn}_{s}$, $\epsilon^{\rm Mn}_{p}$, and $\epsilon^{\rm Mn}_{d_{m},s}$ at the
site where the substitution takes place. This is a natural
tight-binding representation of the central cell correction
\cite{Koster:1954_a,Ralph:1975_a}. For the on-site Mn $d$ energies, the $t_{2g}$ ($m=xy,xz,yz$) and the $e_g$ ($m=x^2-y^2,3z^2-r^2$) orbitals are split by  the crystal field, $\Delta_{cf}$. The other reason why we keep explicitly the $d$-orbital index $m$ and the spin index $s=\pm$ is that the levels can be further spit by electron-electron interaction. This effect can be modeled by the multi-orbital Anderson many-body Hamiltonian solved in the unrestricted Hartree-Fock approximation. The direct and exchange Coulomb terms are parametrized by
\begin{eqnarray}
U_{mm^{\prime}}=\int d{\bm r}\int d{\bm r}^{\prime}\psi^{\ast}_{ms}({\bm r})\psi^{\ast}_{m^{\prime}s^{\prime}}({\bm r}^{\prime})
V(|{\bm r}-{\bm r}^{\prime}|)\psi_{ms}({\bm r})\psi_{m^{\prime}s^{\prime}}({\bm r}^{\prime})\nonumber \\
J_{mm^{\prime}}=\int d{\bm r}\int d{\bm r}^{\prime}\psi^{\ast}_{ms}({\bm r})\psi^{\ast}_{m^{\prime}s}({\bm r}^{\prime})
V(|{\bm r}-{\bm r}^{\prime}|)\psi_{ms}({\bm r}^{\prime})\psi_{m^{\prime}s}({\bm r})\;.
\label{ummjmm}
\end{eqnarray}
In the unrestricted Hartree-Fock theory the occupation numbers, $\langle n_{m,s}\rangle$, of the $d$-orbitals hybridized with the host GaAs are not {\em a priori} know and have to be calculated self-consistently. This procedure will be explained below; here we just assume for the moment that they are known. Using the above parametrization, the Hartree-Fock equations,
\begin{eqnarray}
\epsilon^{\rm Mn}_{d_{m},s}\psi_{ms}({\bm r})&=&\left(E_d+\Delta_m\right)\psi_{ms}({\bm r})+\sum_{m^{\prime},s^{\prime}}\int d{\bm r}^{\prime}\left|\psi_{m^{\prime}s^{\prime}}({\bm r}^{\prime})\right|^2V(|{\bm r}-{\bm r}^{\prime}|)\psi_{ms}({\bm r})\nonumber \\
&-&\sum_{m^{\prime}}\int d{\bm r}^{\prime}\psi^{\ast}_{m^{\prime}s}({\bm r}^{\prime})\psi_{ms}({\bm r}^{\prime})V(|{\bm r}-{\bm r}^{\prime}|)\psi_{m^{\prime}s}({\bm r})\;,
\end{eqnarray}
yield
\begin{equation}
\epsilon^{\rm Mn}_{d_{m},s}=E_d+\Delta_m+\sum_{m^{\prime},s^{\prime}}U_{mm^{\prime}}\left(\langle n_{m^{\prime},s^{\prime}}\rangle-f\right)
-\sum_{m^{\prime}}J_{mm^{\prime}}\left(\langle n_{m^{\prime},s}\rangle-f\right)\;.
\end{equation}
Here $E_d$ is the $d$-level of an isolated atomic Mn  calculated from {\em ab initio} assuming unpolarized Mn atom, i.e., $\langle n_{m,s}\rangle=f=1/2$, $\Delta_{m} =2/5 \Delta_{cf}$ for the three $t_{2g}$ orbitals and
$\Delta_{m} = -3/5 \Delta_{cf}$ for the two $e_g$ orbitals, $f=1/2$ is subtracted from the occupation numbers in the Hatree-Fock terms to recover $E_d$ in the unpolarized configuration of an isolated Mn atom.

Anderson \cite{Anderson:1961_a} reduced the number of direct and exchange interaction parameters by setting $U_{mm^{\prime}}\equiv U$ for all $m,m^{\prime}$ and $J_{mm^{\prime}}\equiv J$ for all $m\neq m^{\prime}$. The interaction Hamiltonian in this parametrization is, however, not rotationally invariant in the spin space which can be corrected \cite{Dworin:1970_a,Parmenter:1973_a}  by considering $U_{mm}\neq U_{mm^{\prime}}$, namely $U_{mm^{\prime}}=U$ for $m\neq m^{\prime}$ and $U_{mm}=(U+J)$. (Note that $U_{mm}=J_{mm}$ (see Eq.~(\ref{ummjmm}) which removes the self-interaction.) We use this parametrization in which the on-site Mn $d$ energies read,
\begin{equation}
\epsilon^{\rm Mn}_{d_{m},s}=E_d+\Delta_m+(U+J)\left(\langle n_{m,-s}\rangle-f\right)
+(U-J)\sum_{m^{\prime}\neq m}\left(\langle n_{m^{\prime},s}\rangle-f\right)
+U\sum_{m^{\prime}\neq m}\left(\langle n_{m^{\prime},-s}\rangle-f\right)\;.
\label{epsilonMn}
\end{equation}
The values of the relevant parameters are summarized in Table I.

The self-consistent occupation numbers are determined by employing the local Green's function formalism. The formalism allows us to evaluate local and orbital resolved density of states on the Mn in the environment of the GaAs host.
We remark that the same formalism can be used to evaluate the Mn d-orbital occupation numbers in the environment of (Ga,Mn)As with the Green's functions describing the (Ga,Mn)As environment obtained from the coherent potential approximation (discussed below). The important result of these occupation number calculations is that they remain very similar in the GaAs and (Ga,Mn)As environment and in both cases not very far from the polarized isolated Mn configuration. More details on this are given later in these supplementary notes section; in the following section we recap the formalism of the calculations.

\begin{table}[floatfix]
\begin{center}
\begin{tabular}{|l|l|l|l|}
 \hline
 $\epsilon^{\rm As}_{s}$ & -6.724   & $\epsilon^{\rm As}_{p}$ & 0.641 \\
 $\epsilon^{\rm Ga}_{s}$ & -3.978 & $\epsilon^{\rm Ga}_{p}$ & 2.874 \\
 $\epsilon^{\rm Mn}_{s}$ & -0.200   & $\epsilon^{\rm Mn}_{p}$ & 4.874 \\
 \hline
 $E_{d}$ & -0.100   & $V_{sd\sigma}$ & 6.770\\
 $\Delta_{cf}$ & 0.500 & $V_{pd\sigma}$ & -6.320\\
 $U$ & 3.500   & $V_{pd\pi}$ & 2.913\\
 $J$ & 0.600 & & \\
 \hline
$\epsilon^{\rm Mn}_{d_{t2g},+}$ & -2.220 &
$\epsilon^{\rm Mn}_{d_{t2g},-}$ & 2.634\\
$\epsilon^{\rm Mn}_{d_{eg},+}$ & -3.013 &
 $\epsilon^{\rm Mn}_{d_{eg},-}$ & 2.364 \\
 \hline
 \end{tabular}
 \label{table1}
\end{center}
\caption{Parameter of our tight-binding model for Mn in
GaAs. All energies are given in electronvolts and referred to the
top of GaAs valence
band parametrized in Ref. \onlinecite{Talwar:1982_a}.}
\end{table}

\subsection{Green's functions formalism for evaluating ${\rm Mn}$ $d$-orbital occupation numbers}

As mentioned above, $d$-orbitals are also included in the tight-binding description of the cation
sites occupied by the host Ga atoms. Atomic levels of Ga
d-orbitals are, nevertheless, chosen far from the valence band so that their
mixing with the band states is practically excluded. In this way we can use the established $sp^{3}$ second-nearest-neighbor tight-binding parameters of GaAs,
while having a tight-binding basis consisting of 26 orbitals per unit cell (18 $spd$ orbitals on the cation site and 8 $sp$ orbitals on the anion site). In this basis the introduction of the Mn impurity does not require to enlarge the Hilbert space but is only represented by shifted on-site energies of the $spd$ orbitals (Koster-Slater impurity \cite{Koster:1954_a}); recall that the hopping energies are considered to by the same for Ga and Mn.

The Hamiltonian of the GaAs host crystal in the
representation of the basis Bloch sums (Eq.~(\ref{bloch})) can be written as
\begin{equation}
H_{0}({\bm k}) = D_{0} + W({\bm k})\;,
\label{H0}
\end{equation}
where $D^{o}$ is a diagonal matrix (26$\times$26) with Ga and As on-site energies on
its diagonal, and the "kinetic energy" matrix $W(k)$ is constructed from the hopping
integrals Ga-As, Ga-As, and As-As hopping integrals.

The propagator (retarded Green's function) from one atom to another in the GaAs lattice is given by,
\begin{equation}
\langle\phi_{a^{\prime}\alpha^{\prime}}({\bm\rho}_{n^{\prime}a^{\prime}})|G_0(t)|
\phi_{a\alpha}({\bm\rho}_{na})\rangle
=\langle\phi_{a^{\prime}\alpha^{\prime}}({\bm\rho}_{n^{\prime}a^{\prime}})|e^{-\frac{i}{\hbar}H_0}|
\phi_{a\alpha}({\bm\rho}_{na})\rangle
\end{equation}
In Fourier transform,
\begin{equation}
G_0(z)=\frac{1}{i\hbar}\int_0^{\infty}dte^{\frac{i}{\hbar}z}G(t)=(z\mathds{1}-H_0)^{-1}\;,
\end{equation}
where $z=\varepsilon+i\eta$.  In the
representation of the basis Bloch sums (Eq.~(\ref{bloch})) we obtain,
\begin{equation}
G_0(z)=\sum_{{\bm k}(B.z.),a^{\prime}\alpha^{\prime},a\alpha}|\Phi_{{\bm k}a^{\prime}\alpha^{\prime}}\rangle
\langle\Phi_{{\bm k}a\alpha}|\left(z\mathds{1} - H_0({\bm k})\right)^{-1}_{a^{\prime}\alpha^{\prime},a\alpha}
\end{equation}
The orbital-diagonal on-site elements of the Green's function for the 0-th unit cell on the Ga sublattice then read,
\begin{equation}
G_{0,\alpha}(z)\equiv\langle\phi_{Ga\alpha}(0)|G_0(z)|\phi_{Ga\alpha}(0)\rangle=\frac{1}{N}\sum_{{\bm k}(B.z.)}\left(z\mathds{1} - H_0({\bm k})\right)^{-1}_{Ga\alpha}
=\Omega_{u.c.}\int_{B.z.}\frac{d{\bm k}}{(2\pi)^3}\left(z\mathds{1} - H_0({\bm k})\right)^{-1}_{Ga\alpha}\;,
\label{g0}
\end{equation}
where $\Omega_{u.c.}$ is the unit cell volume and the ${\bm k}$-sums and integral are over the 1st Brilloun zone.
We neglect the orbital-off-diagonal on-site elements of the Green's function. (Note that due to the tetrahedral symmetry, most of the orbital-off-diagonal on-site elements of $G(z)$ are zero.)

Neglecting the long-range Coulomb part of the impurity potential of the substitutional Mn$_{\rm Ga}$, we can describe Mn
by a Koster-Slater model
\cite{Koster:1954_a}. The Koster-Slater model assumes that the
perturbation is restricted to the impurity site and represented by
shifts of the atomic levels. For the Mn$_{\rm Ga}$ impurity we write
\begin{equation}
\delta_{\alpha,s} = \epsilon^{\rm Mn}_{\alpha,s} -\epsilon^{\rm Ga}_{\alpha}.
\end{equation}
The Hamiltonian of GaAs with a single Mn impurity at site "0" then reads,
\begin{equation}
H({\bm k})=H_0({\bm k})+H_1\;,
\end{equation}
where
\begin{eqnarray}
H_1=\sum_{\alpha,s}\delta_{\alpha,s}|\alpha\rangle\langle\alpha|\nonumber \\
|\alpha\rangle\equiv|\phi_{Ga\alpha}(0)\rangle
\end{eqnarray}

The Green's function for $H$ can be written as,
\begin{eqnarray}
G&=&(z\mathds{1}-H_0-H_1)^{-1}=\left((z\mathds{1}-H_0)(\mathds{1}-(z\mathds{1}-H_0)^{-1}H_1)\right)^{-1}
=G_0(\mathds{1}-G_0H_1)^{-1}\nonumber \\
&=&G_0\sum_j(G_0H_1)^j=G_0+G_0H_1G_0+G_0H_1G_0H_1G_0+\ldots
\end{eqnarray}
Because of the orbital-diagonal form of both the on-site Green's function matrices
and the perturbation, the calculation of the impurity Green's
function reduces to a set of scalar equations. We obtain,
\begin{eqnarray}
G_{\alpha,s}&=&G_{0,\alpha}+\sum_{\alpha^{\prime},s}\langle\alpha|G_0|\alpha^{\prime}
\rangle\delta_{\alpha^{\prime},s}\langle\alpha^{\prime}|G_0|\alpha\rangle+\ldots\nonumber \\
&=&G_{0,\alpha}+G_{0,\alpha}\delta_{\alpha,s}G_{0,\alpha}+
G_{0,\alpha}\delta_{\alpha,s}G_{0,\alpha}\delta_{\alpha,s}G_{0,\alpha}\ldots\nonumber\\
&=&G_{0,\alpha}+G_{0,\alpha}\delta_{\alpha,s}G_{0,\alpha}\sum_j\left(\delta_{\alpha,s}G_{0,\alpha}\right)^j\nonumber\\
&=&\frac{G_{0,\alpha}}{1-\delta_{\alpha,s}G_{0,\alpha}}\;.
\label{g-function}
\end{eqnarray}
Finally we rewrite the Green's functions in a physically and computationally convenient way,
\begin{equation}
G_{0,\alpha}(z)=\left(z-\epsilon^{\rm Ga}_{\alpha}-\Gamma_{\alpha}(z)\right)^{-1}\;,
\label{g0gamma}
\end{equation}
which, together with Eq,~(\ref{g0}), should be regarded as a definition of $\Gamma_{\alpha}(z)$. This definition useful because Eqs.~(\ref{g0gamma}) and (\ref{g-function})yield
\begin{equation}
G_{\alpha,s}(z)=\left(z-\epsilon^{\rm Mn}_{\alpha,s}-\Gamma_{\alpha}(z)\right)^{-1}\;,
\label{ggamma}
\end{equation}
where the "self-energy" $\Gamma_{\alpha}(z)$, calculated from the host GaAs Green's function, represents the effect of hopping energies and is the same in $G$ as in $G_0$, i.e., does not depend on the shifted energies on the impurity site.
In other words, Eq.~(\ref{ggamma}) shows how the local electronic structure of the
impurity depends both on the impurity potential ($\epsilon^{\rm Mn}_{\alpha,s}$) and on the
electronic structure of the surrounding crystal ($\Gamma_{\alpha}(z)$).

From the orbital-diagonal on-site Green's functions we obtain orbital decomposition of the local density of states (DOS),
\begin{equation}
g_{\alpha,s}(\varepsilon) = -\frac{1}{\pi} {\rm Im} G_{\alpha,s}(\varepsilon+i0)\;.
\label{gdos}
\end{equation}
The desired expression for occupation numbers of the Mn $d$-orbital states are then given by,
\begin{equation}
\langle n_{m,s}\rangle=\int d\varepsilon g_{m,s}(\varepsilon) f_{E_F}(\varepsilon)\;,
\label{occnum}
\end{equation}
where the Fermi function $f_{E_F}(\varepsilon)=\theta(E_F-\varepsilon)$. The self-consistent unrestricted Hatree-Fock calculation of the Mn $d$-orbital on-site energies, $\epsilon^{\rm Mn}_{m,s}$, start from assuming $\langle n_{m,+}\rangle=1-\delta_{m,+}$ for the five majority spin $d$-orbitals and $\langle n_{m,-}\rangle=\delta_{m,-}$ for the five minority spin $d$-orbitals. The initial small and $d$-orbital dependent  variations from 1 (or 0) are set to allow for spontaneous symmetry breaking which could result, e.g., in only four Mn $d$-electrons remaining strongly localized deep in the band and the fifth $d$ electron state shifted towards or above the top of the host semiconductor valence band. The self-consistency loop then proceeds by putting these initial occupation numbers in Eq.~(\ref{epsilonMn}) for the Mn on-site energies, then calculate the impurity Green's function from Eq.~(\ref{ggamma}) and obtain new occupation numbers from Eqs.~(\ref{gdos}) and (\ref{occnum}). These self-consistent calculations can be performed assuming pure GaAs environment, i.e., using the function $\Gamma_{\alpha}(z)$. The function $\Gamma_{\alpha}(z)$ can be also replaced with $\Gamma^{\Sigma}_{\alpha,s}(z)$ which describes the environment of (Ga,Mn)As in the coherent-potential approximation (CPA). The CPA method is described in the following subsection.

\subsection{Coherent-potential approximation}
At finite concentration of Mn the single-impurity picture breaks
down as soon as the typical distance of the impurities becomes
comparable with the extend of the related wave functions. We evaluate the
band structure of a mixed (Ga,Mn)As crystal using the (CPA), which is
particularly suitable for describing the system over the entire doping range from 0 to 100\% of Mn. The basic quantity in the
CPA is the configurationally averaged Green's function. The
averaging restores the translational invariance of the mixed
crystal. The procedure is based on replacing real but random atomic levels
by site-independent but
complex and energy-dependent selfenergies $\Sigma_{a\alpha,s}$
which are to be determined selfconsistently. Because of the  diagonal form of the Green's functions, $\Sigma_{a\alpha,s}$ will be also diagonal and since there's in disorder in (Ga,Mn)As on the As-sublattice,
\begin{equation}
\Sigma_{As\alpha} = \epsilon^{\rm As}_{\alpha}\;.
\end{equation}

The self-energy on the Ga-sublattice $\Sigma_{Ga\alpha,s}$ is obtained by from the CPA condition which states that the on-site Green's function for a periodic system in which the Ga-sublattice energies have been replaced by $\Sigma_{Ga\alpha,s}\equiv\Sigma_{\alpha,s}$ equals the composition-weighted sum of the single-impurity on-site Green's function assuming Ga or Mn energies on the site:
\begin{equation}
G^{\Sigma}_{0,\alpha,s}(z)=(1-x)G_{{\rm Ga},\alpha,s}(z)+xG_{{\rm Mn},\alpha,s}(z)\;.
\end{equation}
Here $x$ is the doping in the Ga$_{1-x}$Mn$_x$As mixed crystal.
Using Eqs.~(\ref{g0}) and (\ref{g0gamma}) we can write the Green's function of the effective periodic Hamiltonian, with the diagonal terms $D_0$ in  Eq.~(\ref{H0}) replaced with $\Sigma_{a\alpha,s}$, as
\begin{equation}
G^{\Sigma}_{0,\alpha,s}(z)=\left(z-\Sigma_{\alpha,s}(z)-\Gamma^{\Sigma}_{\alpha,s}(z)\right)^{-1}\;.
\label{g0cpa}
\end{equation}
Similarly the on-site impurity Green's functions read,
\begin{eqnarray}
G_{{\rm Ga},\alpha,s}(z)&=&\left(z-\epsilon^{\rm Ga}_{\alpha}-\Gamma^{\Sigma}_{\alpha,s}(z)\right)^{-1}\nonumber \\
G_{{\rm Mn},\alpha,s}(z)&=&\left(z-\epsilon^{\rm Mn}_{\alpha}-\Gamma^{\Sigma}_{\alpha,s}(z)\right)^{-1}\;.
\label{gcpa}
\end{eqnarray}
The convergence of the self-consistent solution of Eqs.~(\ref{g0cpa}) and (\ref{gcpa}) is guaranteed by starting
the calculations from the the virtual-crystal approximation
which assumes that $\Sigma_{\alpha,s}=\Sigma^{v.c.}_{\alpha,s}$ are the average atomic levels on the Ga sublattice for the given doping,
\begin{equation}
\Sigma^{v.c.}_{\alpha,s} = (1-x)\epsilon^{\rm Ga}_{\alpha}+x\epsilon^{\rm Mn}_{\alpha,s}\;.
\end{equation}

\section{Results: ${\rm\bf (Ga,Mn)As}$ band-structure calculated using the unmodified TBA parametrization}
\subsection{Cartoon representation of the expected band structure of Mn doped GaAs}

\begin{figure}[ht]
\vspace*{0cm}
\hspace*{-0.cm}\includegraphics[height=0.73\columnwidth]{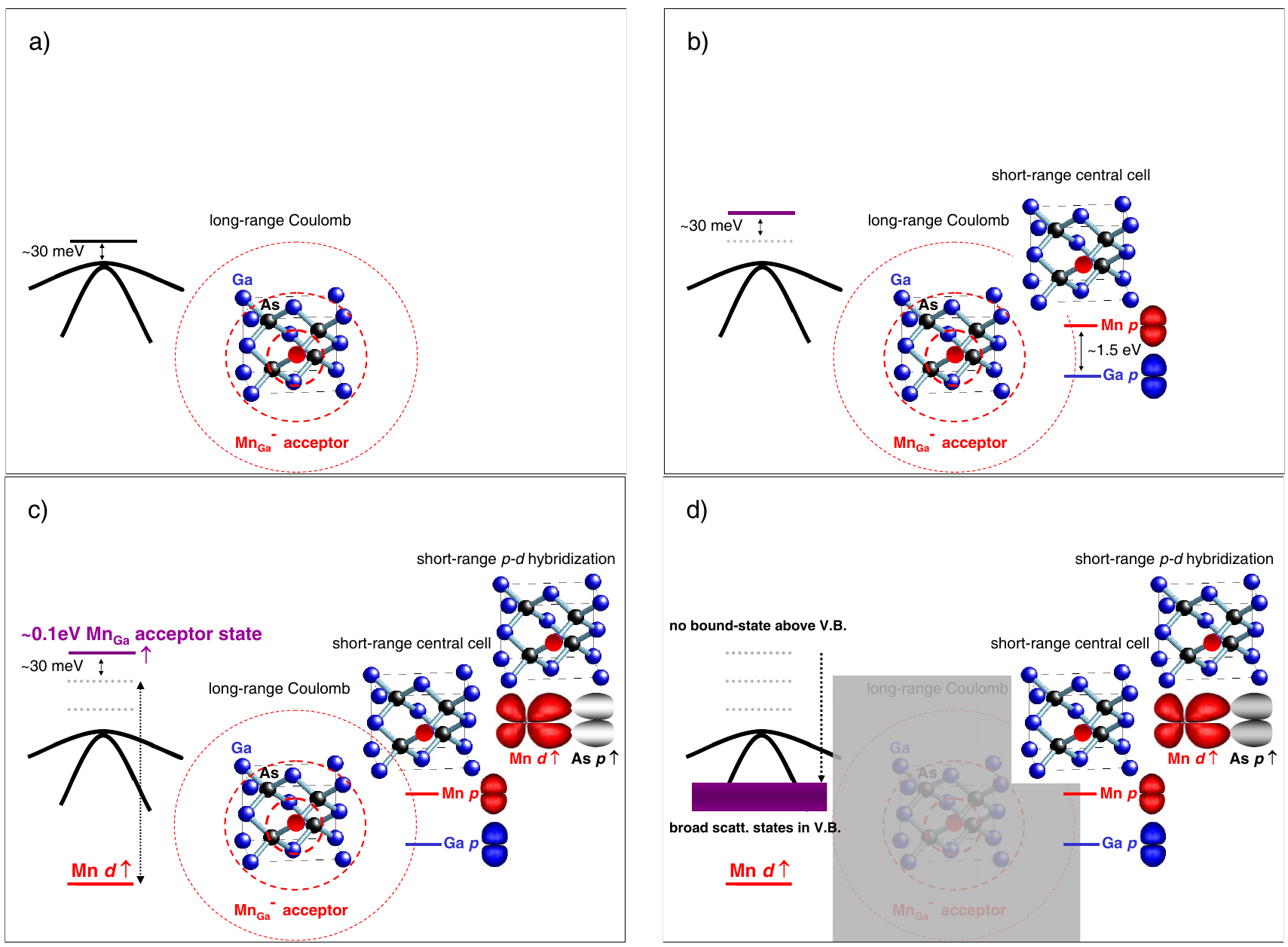}
\vspace*{-0cm} \caption{Schematic diagrams of the long-range (a) and the two short-range (b),(c) contributions to the single Mn acceptor level. (d) No bound state in the gap would form without the long-range Coulomb potential.}
\label{fig_cartoon}
\end{figure}

Fig.~\ref{fig_cartoon}(a) shows the expected $\sim 30$~meV binding
energy due to the long-range Coulomb potential of a single
acceptor in GaAs. Fig.~\ref{fig_cartoon}(b) shows the comparable
enhancement of the binding energy due to the central cell
correction whose strength is estimated form the conventionally
parametrized on-site energies on Mn and from comparison of the
Mn(Ga) substitution with the well established central cell
correction for the non-magnetic Zn(Mn) substitution (see main text
and references therein). Fig.~\ref{fig_cartoon}(c) shows the again
comparable enhancement of the binding energy due to the
hybridization of the Mn $d$-states with the host valence band
states (see main text for references). Fig.~\ref{fig_cartoon}(d)
illustrates that no bound state above the top of the valence band
is expected to form in the absence of the long-range Coulomb
potential of the charged Mn acceptor. When this potential is
screened in the high doped metallic ferromagnetic (Ga,Mn)As, it is
then plausible to expect that also no detached impurity band can
persist above the valence band and that the two must merge.

\subsection{Single Mn impurity: Local density of states and occupation of Mn d-orbitals}
\begin{figure}[th]
\vspace*{-0.5cm}
\includegraphics[height=0.7\columnwidth,angle=270]{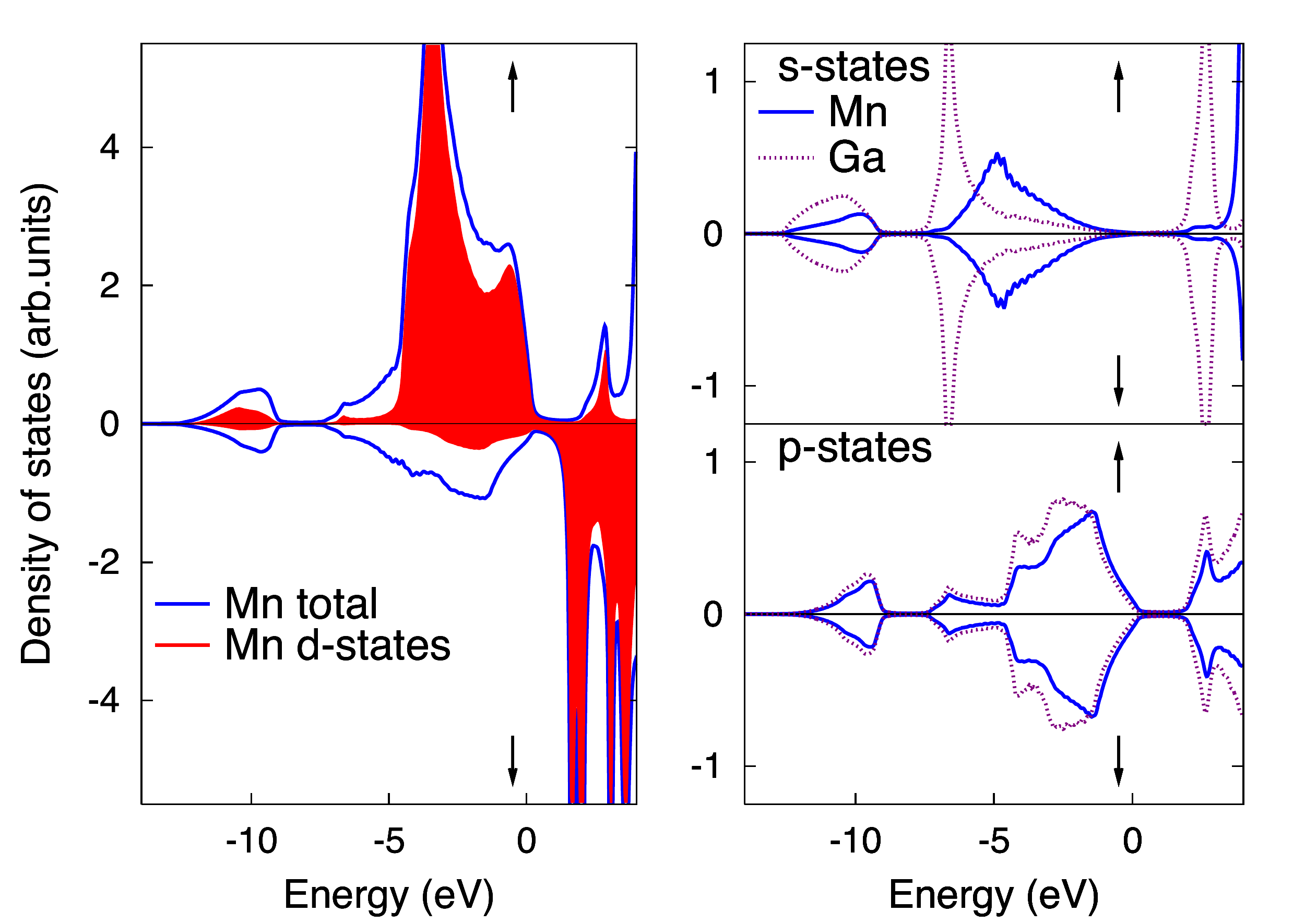}
\vspace*{-0cm} \caption{Local density of states at Mn$_{\rm Ga}$
and its orbital decomposition.}
\label{fig_imp-local}
\end{figure}
\begin{figure}[ht]
\vspace*{-0.5cm}
\includegraphics[height=0.5\columnwidth,angle=270]{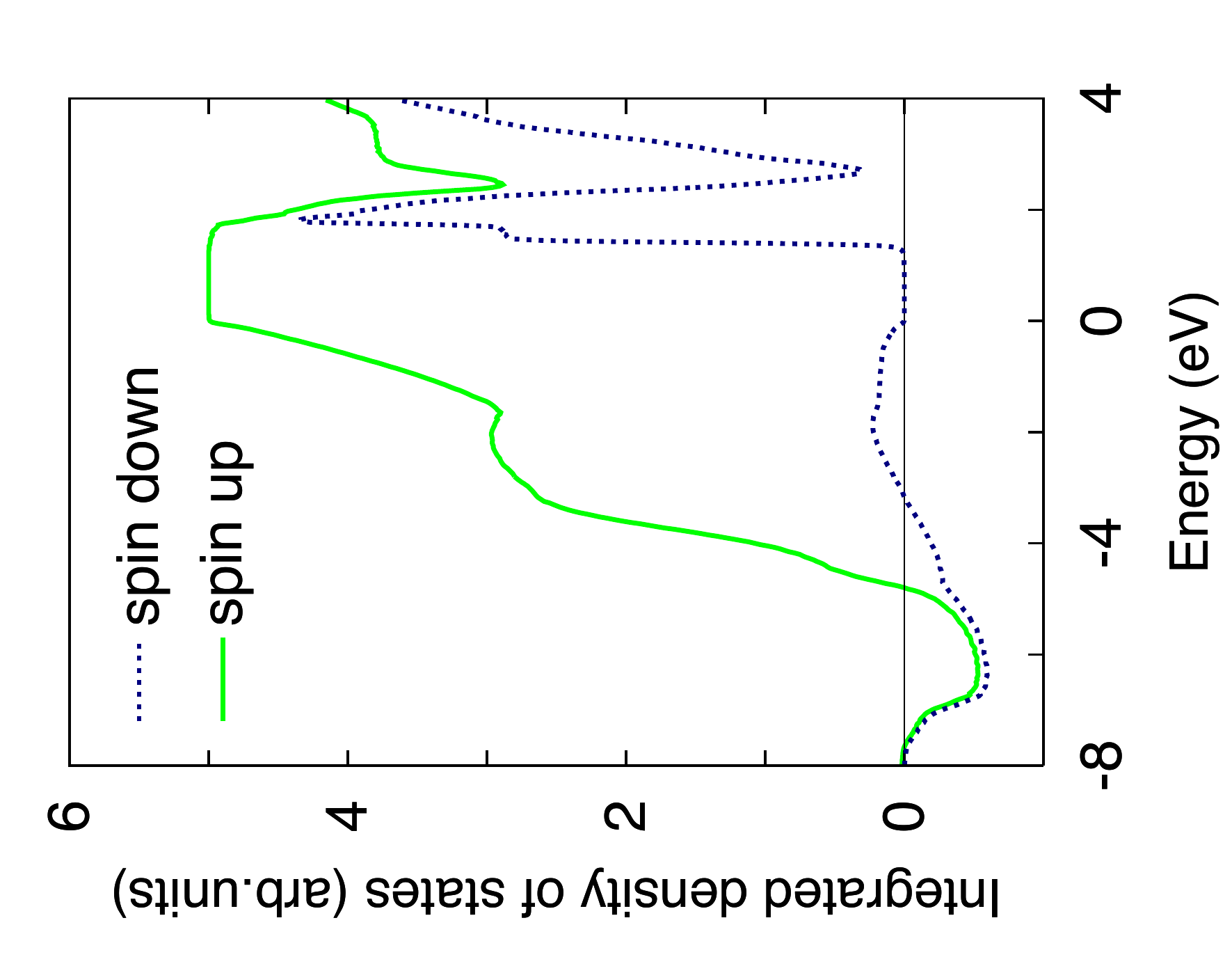}
\vspace*{-0cm} \caption{Integrated change $\delta N_{tot}(E)$ of
the total density of states induced by substitution of single Mn
for Ga in GaAs crystal. Electron states with spin parallel (up)
and antiparallel (down) to the local magnetic moment are treated
separately.}
\label{fig_imp-global}
\end{figure}
The self-consistent unrestricted Hartree-Fock occupation numbers
calculated for a Mn impurity in a pure GaAs environment
($\Gamma_{\alpha}(z)$) can be classified according to their spin
and to the tetrahedral symmetry of the impurity, i.e., there are
three degenerate $t_{2g}$-orbitals and two weakly hybridized
$e_g$-orbitals. The electron interactions within Mn $d$-shell are
strong enough to result into almost complete spin polarization
($\langle n_{t_{2g},+}\rangle=0.899$, $\langle
n_{e_{g},+}\rangle=0.979$, $\langle
n_{t_{2g},-}\rangle=0.108$ and $\langle
n_{e_{g},-}\rangle=0.038$) and to formation of a local
magnetic moment close to 5$\mu_{B}$. On the other hand, we did not
find any tendency to the quenching of the orbital moment, i.e.
symmetry breaking in these occupation numbers due to the electron
correlation.\cite{Anderson:1961_a} This is a key observation
allowing us to disregard a formation of Mn-related bound state due
to the correlation effects and stick to an effective one-particle 
picture of the impurity state.

Fig.~\ref{fig_imp-local} shows spin-polarized local density of
states (LDOS) at a single substitutional Mn impurity in GaAs in
the spectral range of valence and the lowest conduction bands. The
shaded area in the left panel represents the contribution of Mn
$d$-states. The change of local densities of $s$ and $p$-states  at the
Mn impurity, together with the corresponding local densities of
states at the host atom Ga, are shown in the right panel. Their
differences reflect the central-cell corrections to the atomic
levels $\varepsilon_{s}$ and $\varepsilon_{p}$, respectively.

An alternative picture of the impurity induced modification of
the electron spectrum is obtained from the change $\delta g_{tot}(E)$
of the total DOS in which the changes of the local densities
of states at all lattice sites are summed up. The integrated quantity
\begin{equation}
\delta N_{tot}(E)=\int_{-\infty}^{E}\delta g_{tot}(E')dE'.
\end{equation}
is particularly suitable to show spectral features with a small
weight at the impurity site. In our case of Mn$_{\rm Ga}$, $\delta
N_{tot}(E)$ combines two features, i.e., addition of the
d-orbitals and reconstruction of the band states due to both
central-cell corrections and hybridization.
Fig.~\ref{fig_imp-global} shows smooth, but non-monotonic increase
of $\delta N_{tot}(E)$ in the valence band. This is a signature of
mixing of Mn $d$-orbitals with the band states. In addition
to the changes in the valence band, a sharp dip in $\delta
N_{tot}(E)$ for both spin polarizations around 2 eV indicates that
Mn $d$-orbitals hybridize significantly also with GaAs conduction band states (away from the $\Gamma$-point).\cite{Masek:2006_a,Masek:2007_a}

On the other hand, as the Fermi energy remains pinned at the edge
of the valence band in the single impurity regime, 
$\delta N_{tot}(E_F)$ in the band gap define the number of electrons
accumulated at and around Mn impurity. These numbers are five and
zero for spin-up and spin-down electrons, respectively, as a
result of the fact that the impurity potential does not pull any
state from the valence band in the conventionally parametrized TBA with self-consistent Hatree-Fock Mn $d$-orbital on-site energies when the long-range
acceptor Coulomb potential is not included.

\protect\newpage

\subsection{DOSs of (Ga,Mn)As over the entire concentration range}

\begin{figure}[h!]
\vspace*{-0.cm}
\includegraphics[height=0.6\columnwidth,angle=270]{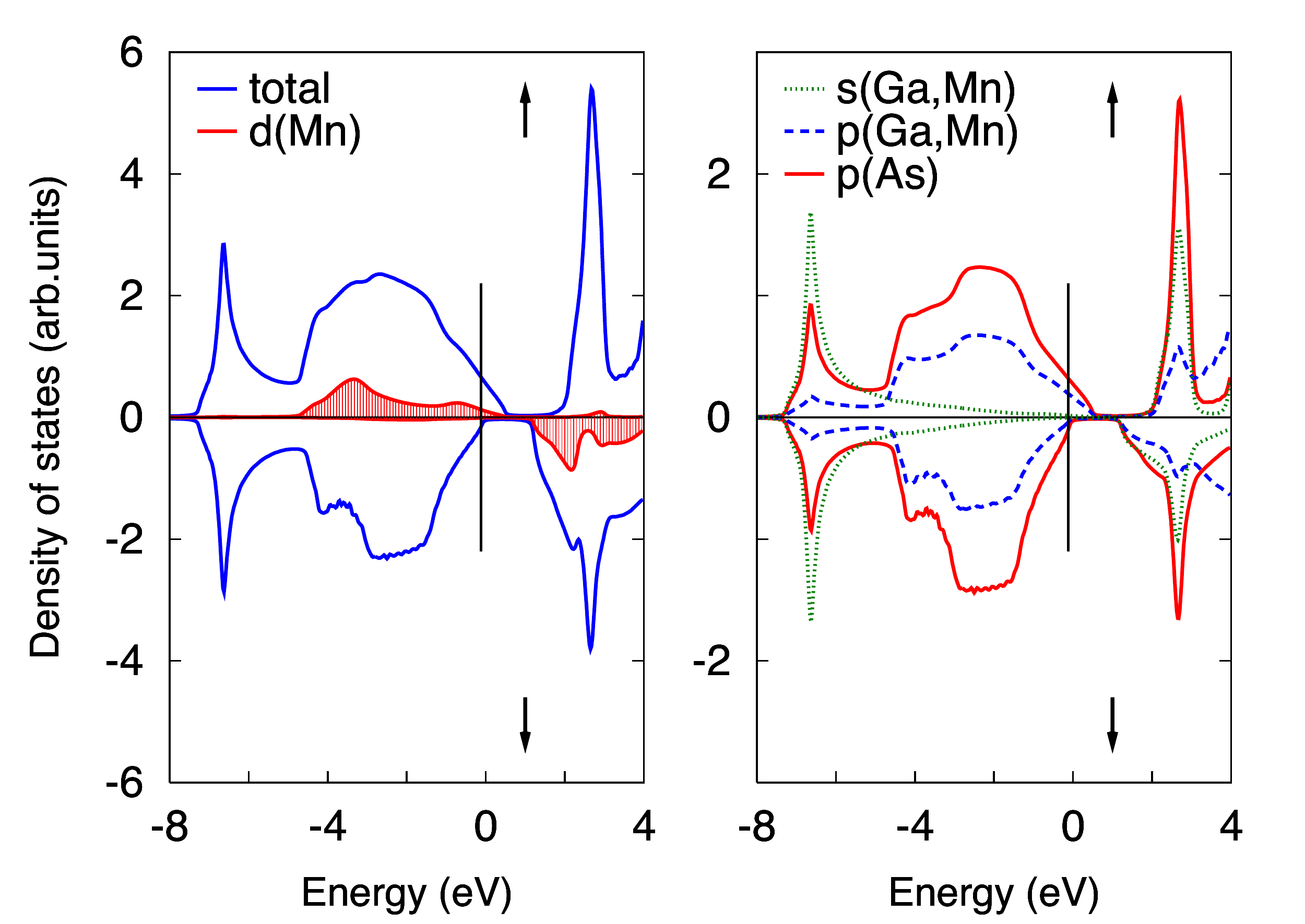}
\vspace*{-0cm} \caption{TBA density of states of
Ga$_{0.9}$Mn$_{0.1}$As and its orbital composition. Position of
Fermi energy is indicated by a vertical line.}
\label{fig_TBA-DOS10}
\end{figure}

We now fix the Hartree-Fock $d$-levels obtained for a single Mn
impurity and perform the CPA calculations of the DOS in Mn doped
GaAs as well as in (Ga,Mn)As mixed crystals with higher
concentrations of Mn. We show the atom and orbital resolved DOSs
for the entire doping range up to MnAs. Fig.~\ref{fig_TBA-DOS10} shows the result for GaAs doped with 10
percent of Mn. The admixture of Mn $d$-orbitals to the unoccupied
states at the top of the valence band is relatively small and the occupation
numbers of the $d$-orbitals do not differ significantly from the values
obtained in the single impurity case ($\langle
n_{t_{2g},+}\rangle=0.886$, $\langle
n_{e_{g},+}\rangle=0.984$, $\langle
n_{t_{2g},-}\rangle=0.110$ and $\langle
n_{e_{g},-}\rangle=0.035$). This confirms that also the
self-consistency condition for Mn atomic levels remains fulfilled
with a reasonable accuracy.

\begin{figure}[h!]
\vspace*{-0.cm}
\includegraphics[height=0.6\columnwidth
,angle=270]{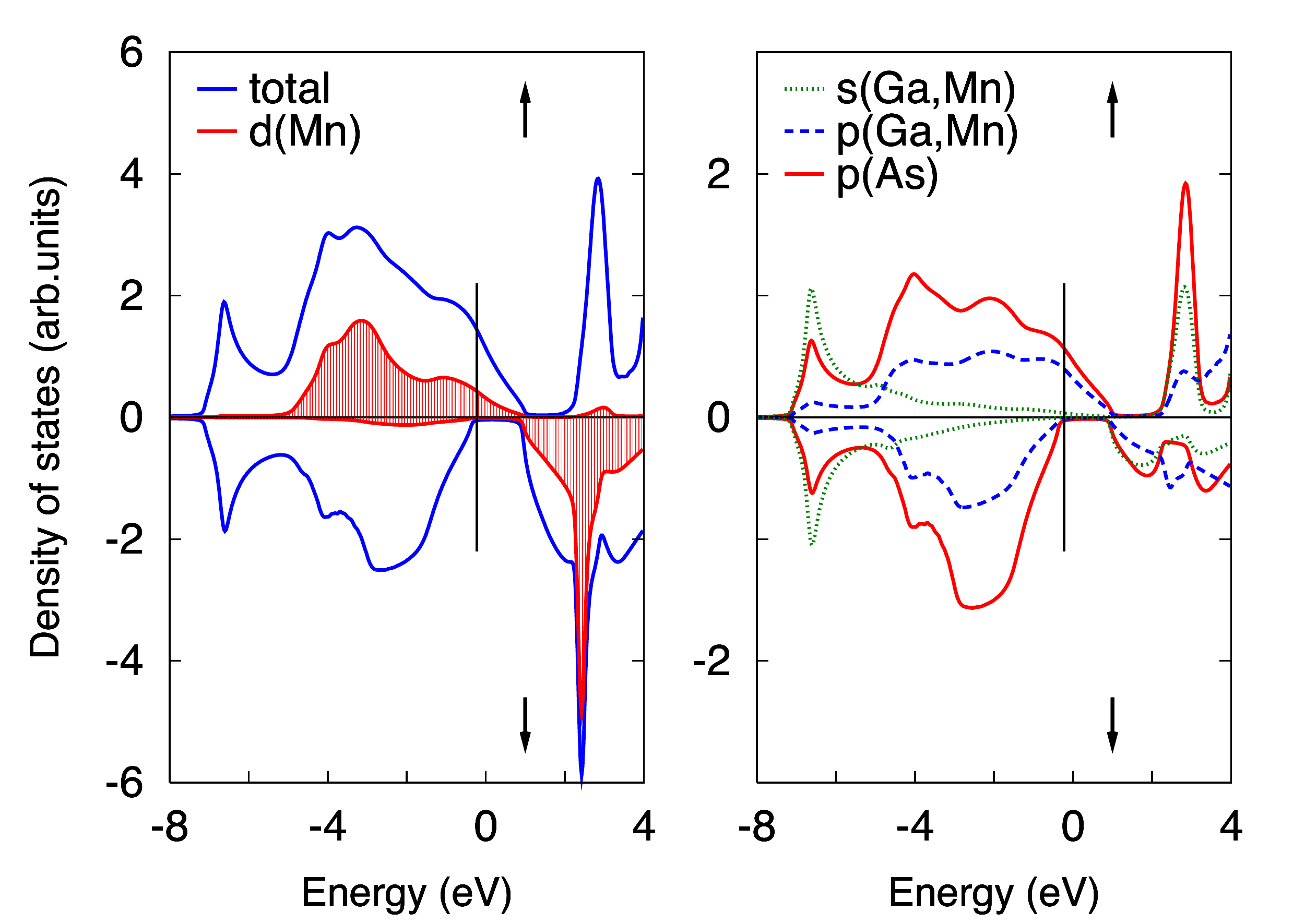}
\vspace*{-0cm} \caption{TBA density of states of
Ga$_{0.7}$Mn$_{0.3}$As and its orbital composition. Position of
Fermi energy is indicated.}
\label{fig_TBA-DOS30}
\end{figure}

\begin{figure}[h!]
\vspace*{-0.cm}
\includegraphics[height=0.6\columnwidth,angle=270]{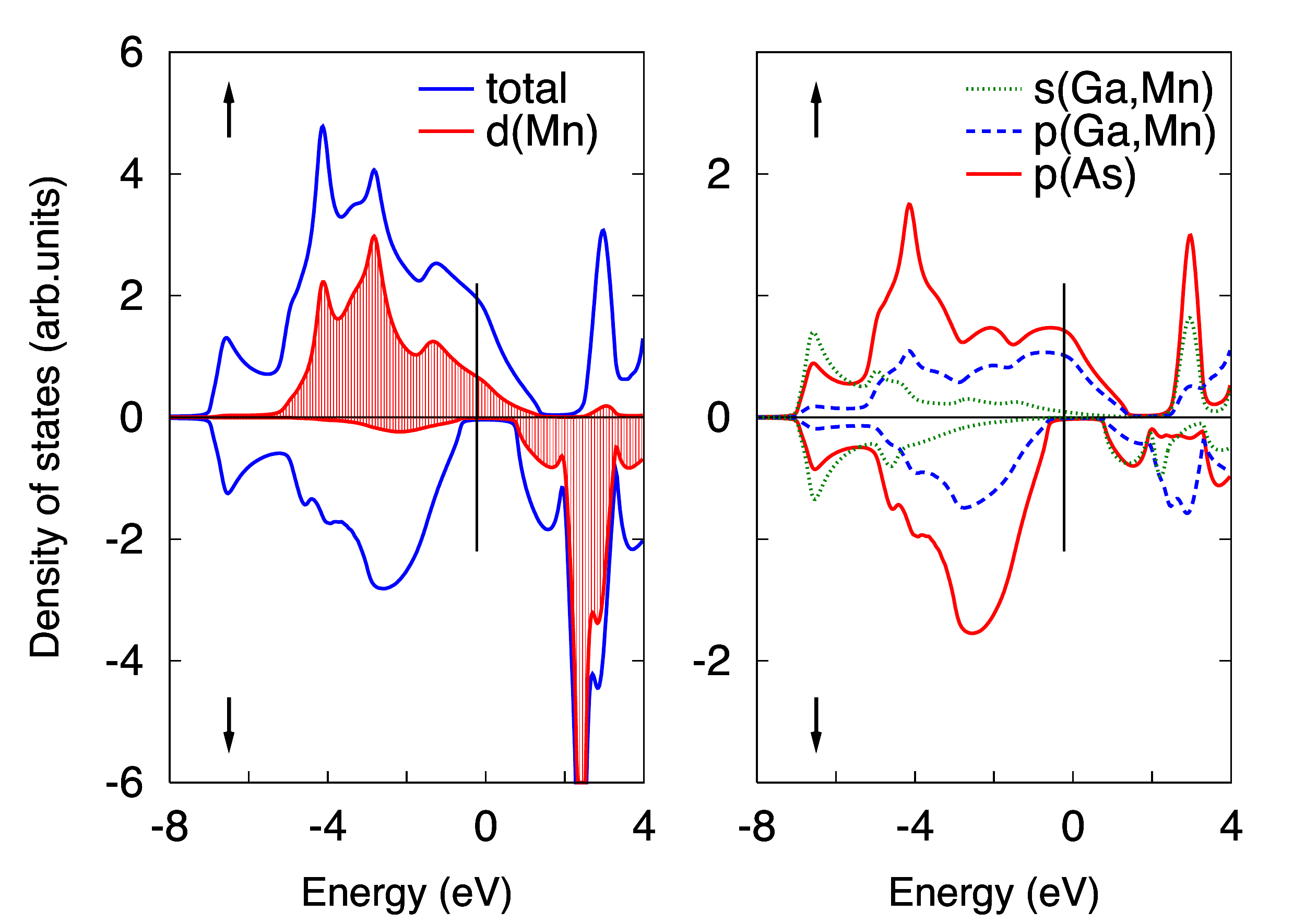}
\vspace*{-0cm} \caption{TBA density of states of
Ga$_{0.5}$Mn$_{0.5}$As and its orbital composition. Position of
Fermi energy is indicated}
\label{fig_TBA-DOS50}
\end{figure}

\begin{figure}[h!]
\vspace*{-0.cm}
\includegraphics[height=0.6\columnwidth,angle=270]{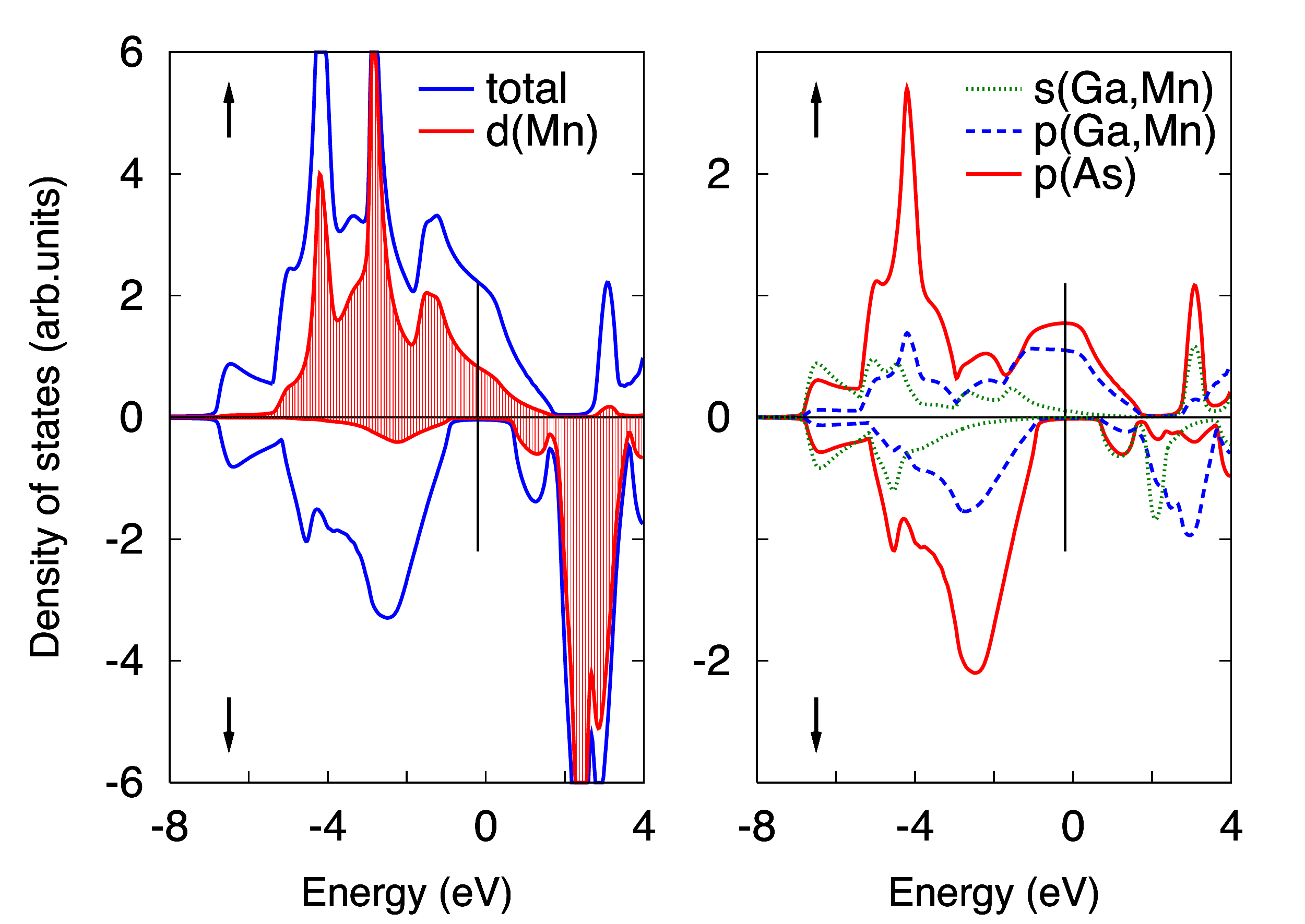}
\vspace*{-0cm} \caption{TBA density of states of
Ga$_{0.3}$Mn$_{0.7}$As and its orbital composition. Position of
Fermi energy is indicated}
\label{fig_TBA-DOS70}
\end{figure}

\begin{figure}[h!]
\vspace*{-0.cm}
\includegraphics[height=0.6\columnwidth,angle=270]{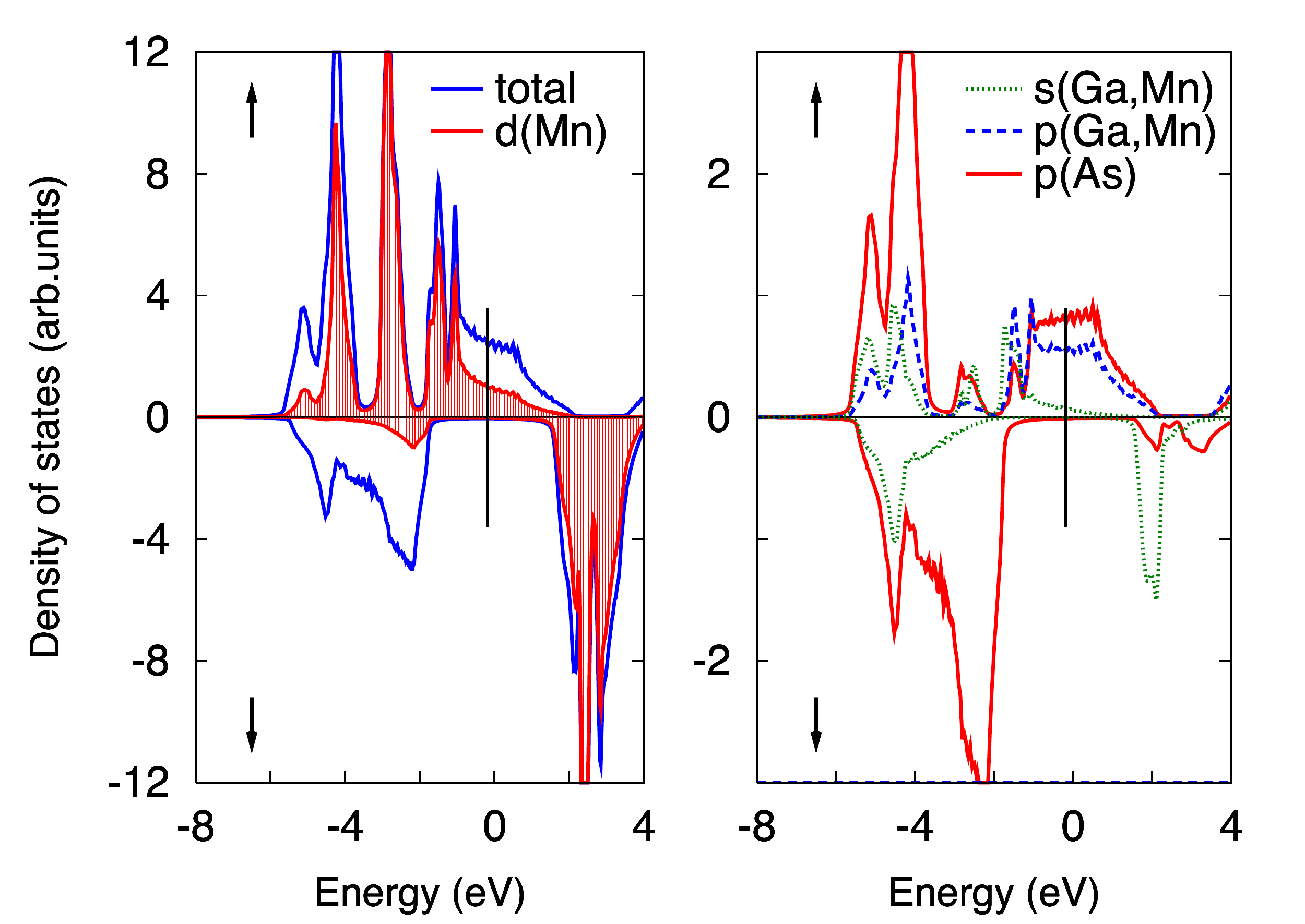}
\vspace*{-0cm} \caption{TBA density of states of a hypothetical
MnAs crystal with zinc blende structure. Orbital composition of
the bands and position of the Fermi energy are indicated.}
\label{fig_TBA-DOS100}
\end{figure}

For higher concentrations of Mn (see Figs.~\ref{fig_TBA-DOS30}-\ref{fig_TBA-DOS100}), the spectral weight near the
Fermi energy gradually evolves from being dominated by the As
$p$-orbitals to a comparable As $p$ and Mn $d$-orbital weight for
Mn content around 70 percent (see Fig.~\ref{fig_TBA-DOS70}) where
the DOS undergoes a transition from the shape characteristic to
Mn-doped GaAs to the form similar to the MnAs limit (see
Fig.~\ref{fig_TBA-DOS100}).

\protect\newpage

\vspace*{12cm}
\section{Results: ${\rm\bf (Ga,Mn)As}$ band-structure calculated using modified TBA parametrizations}
\subsection{Cartoon representation of the modified TBA parametrizations}

\begin{figure}[ht]
\vspace*{-0.5cm}
\hspace*{-0.cm}\includegraphics[height=0.73\columnwidth]{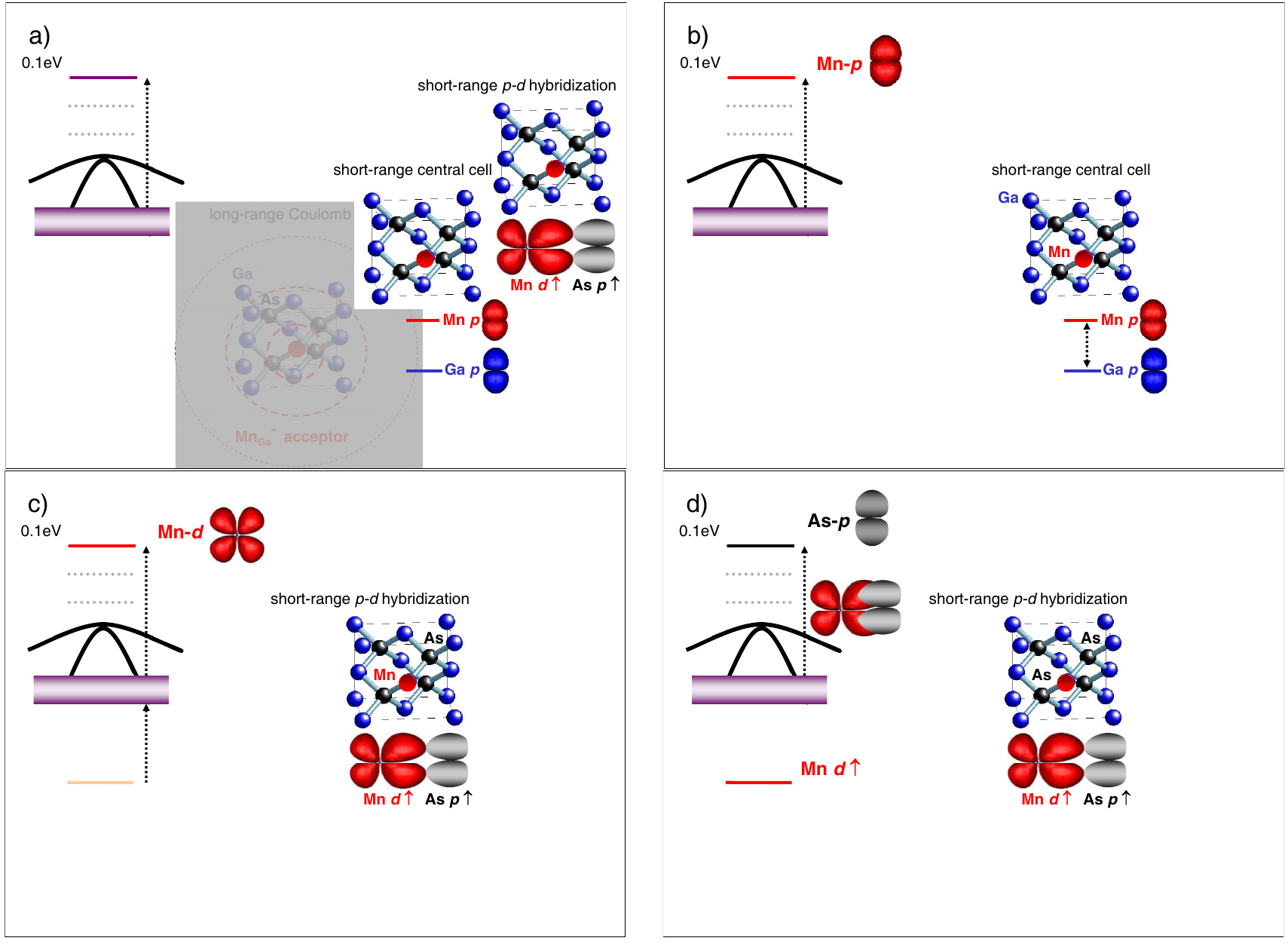}
\vspace*{-0cm} \caption{Cartoon representation of the modified TBA parametrizations.}
\label{fig_cartoon_ib}
\end{figure}
Fig.~\ref{fig_cartoon_ib}(a) illustrates the strategy of
calculations in this section in which we modify the conventional
TBA parametrization in order to obtain the 0.1~eV bound state of a
single Mn impurity in GaAs without the long-range Coulomb
potential. Fig.~\ref{fig_cartoon_ib}(b) shows the first,
central-cell approach in which we shift the Mn $p$-orbital on-site
energies. Figs.~\ref{fig_cartoon_ib}(c),(d) show the enhanced
$p-d$ hybridization approach in which we shift the Mn $d$-orbital
on-site energies or increase the values of the hopping energies
involving Mn $d$-orbitals.

\subsection{Enhanced central-cell correction by shifted Mn $p$-orbital on-site energies}
\begin{figure}[ht]
\vspace*{-0.5cm}
\includegraphics[height=0.7\columnwidth,angle=270]{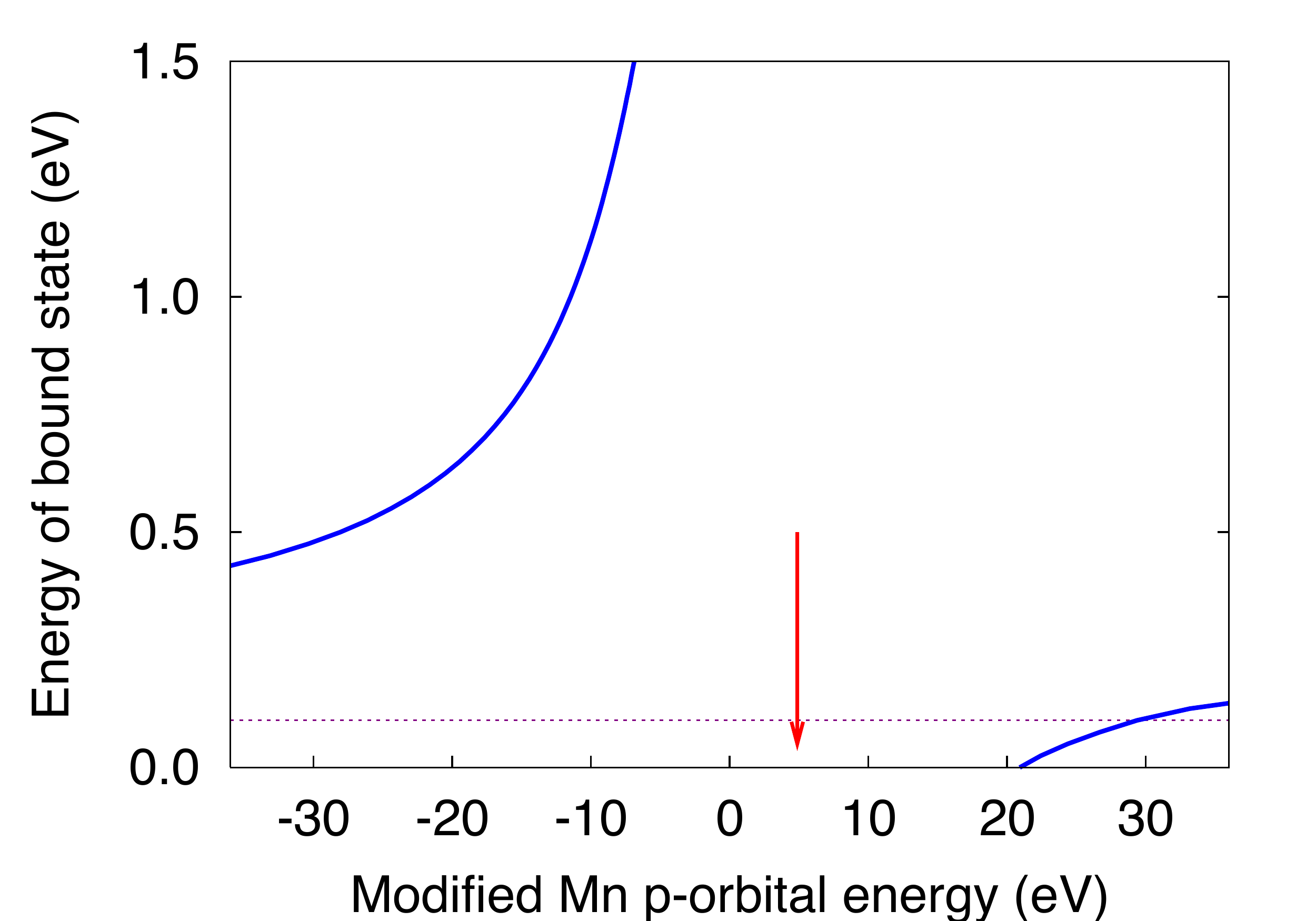}
\vspace*{-0cm} \caption{Relation between the p-orbital energy
$\bar{\epsilon}^{\rm Mn}_p$ at substitutional Mn impurity in GaAs
and position of related bound state in the band gap. Observed
Mn-related acceptor level is represented by horizontal dotted
line. The arrow indicates the conventionally parametrized TBA value of $\epsilon^{\rm Mn}_p$.}
\label{fig_cond-tbapc}
\end{figure}

Here we consider modified on-site Mn $p$-orbital energies,
\begin{equation}
\bar{\epsilon}^{\rm Mn}_p=\epsilon^{\rm Mn}_p+\Delta^{\rm Mn}_p\;.
\end{equation}
We examine a capability of such perturbation to create a localized
state in the band gap by using the impurity Green's function,
Eq.~\ref{ggamma}. A bound state is represented by a pole of the
Green's function, i.e., the energy $z$ of the bound state is
obtained from a condition
\begin{equation}
z-\bar{\epsilon}^{\rm Mn}_p-\Gamma_{p}(z)\;, \label{cond}
\end{equation}
Fig.~\ref{fig_cond-tbapc} shows that no bound state can be formed
in the band gap of GaAs for any physically meaningful value of $\Delta^{\rm
Mn}_p$.

\subsection{Enhanced $p-d$ hybridization by increased Mn $d$-orbital on-site energies -- TBA$^{\rm d}$}
\begin{figure}[ht]
\vspace*{-0.5cm}
\includegraphics[height=0.7\columnwidth,angle=270]{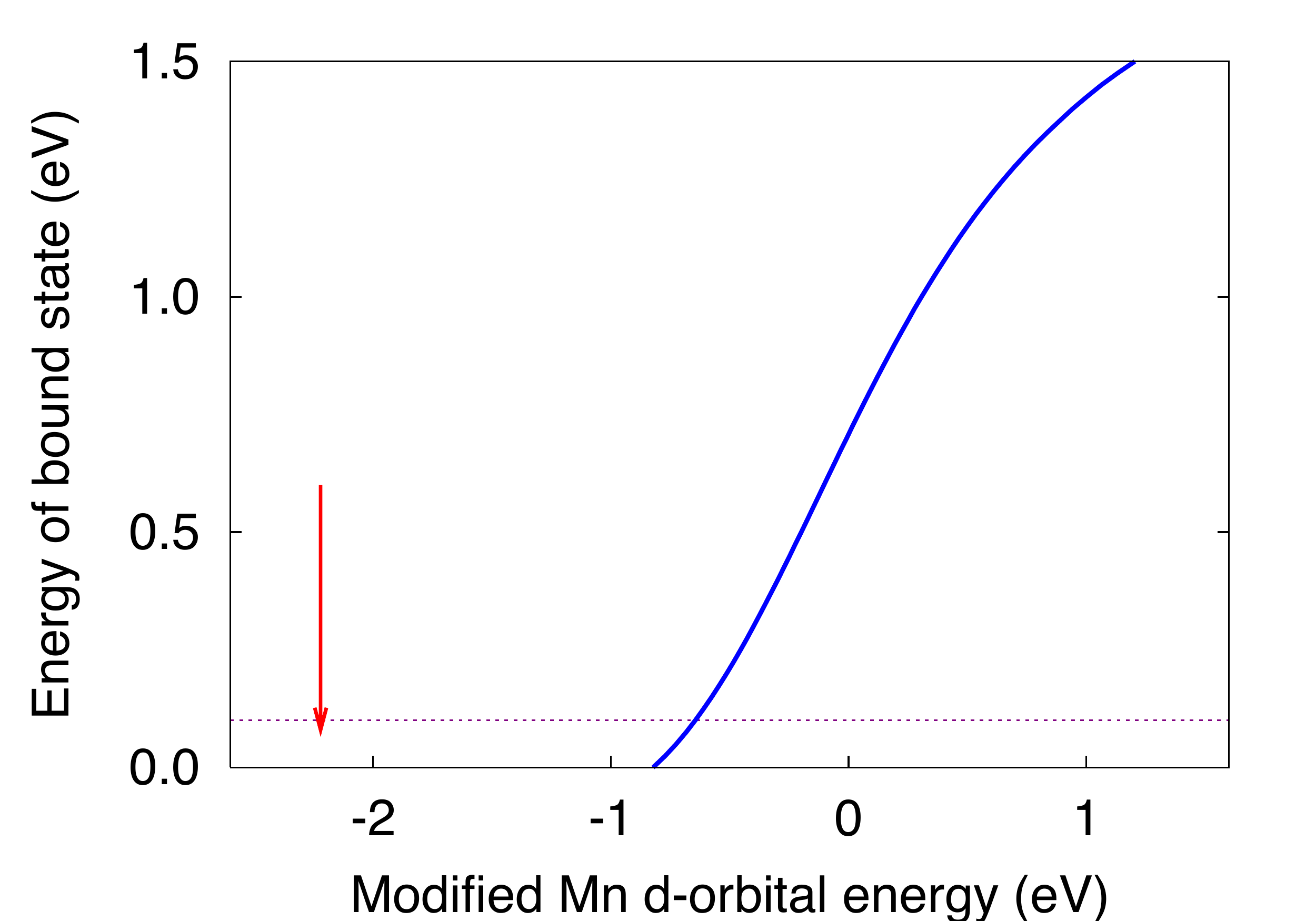}
\vspace*{-0cm} \caption{Relation between the energy $\bar{E}_d$ of
a t2g Mn d-orbital and position of related bound state in GaAs. Mn
acceptor level is represented by dotted line. The arrow indicates the conventionally parametrized TBA value of
$\epsilon^{\rm Mn}_{d_{t2g},+}$.}
\label{fig_cond-tbad}
\end{figure}

\begin{figure}[ht]
\vspace*{-0.5cm}
\includegraphics[height=0.7\columnwidth,angle=270]{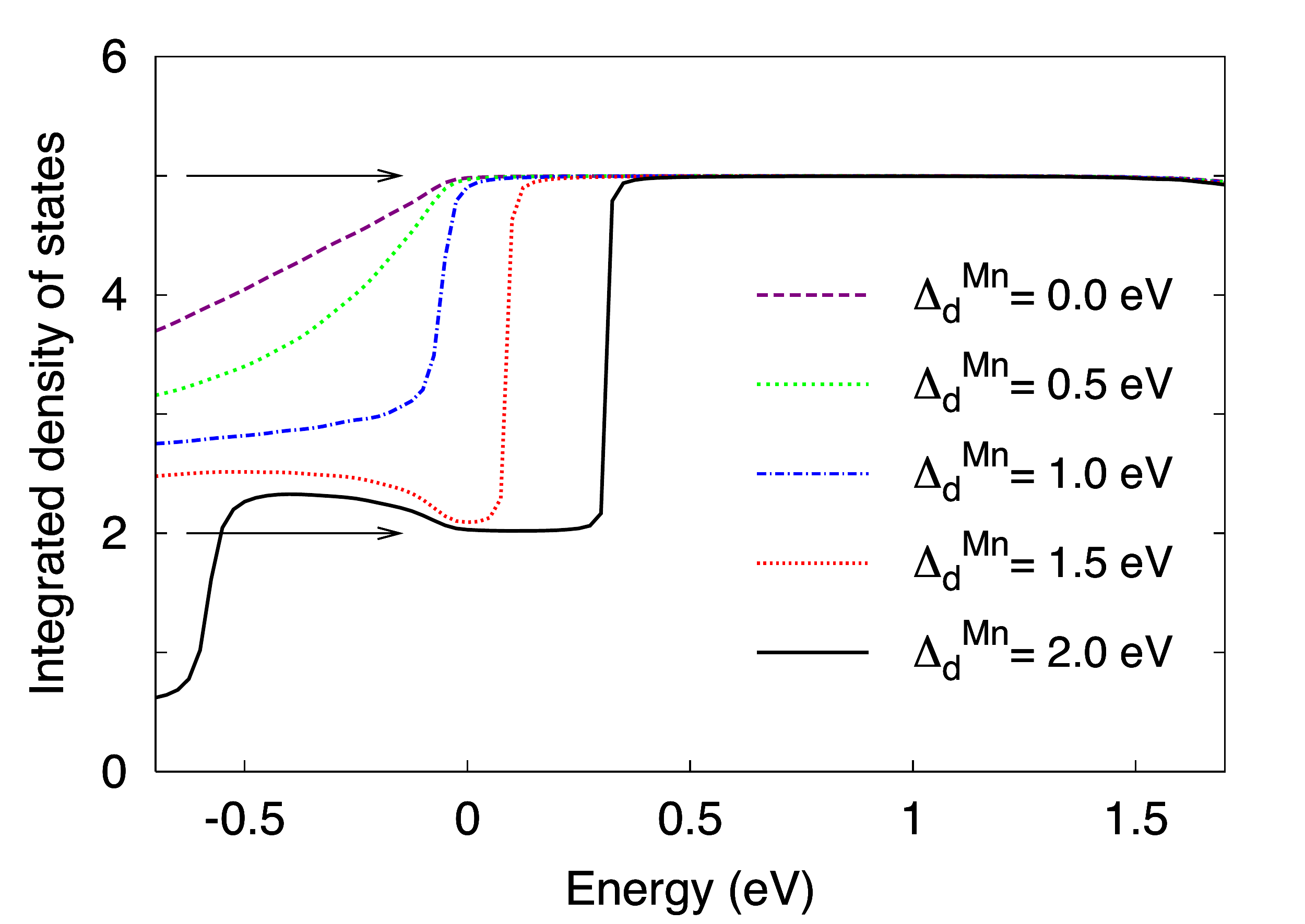}
\vspace*{-0cm} \caption{Integrated change $\delta N_{tot}(E)$ of
the total density of states for spin-up electrons visualizes
formation of a bound state for increasing $\Delta^{\rm Mn}_{d}$.
It is characterized by a step-like increase of $\delta N_{tot}(E)$
in the band gap and by a depletion of the DOS on the uppermost
part of the valence band. The arrows indicate complete occupations
of $e_{g}$ and $t_{2g}$ orbitals.}
\label{fig_serie-idos}
\end{figure}

Here we consider modified on-site Mn $d$-orbital energies obtained
by shifting the one-particle energies in Eq.~(\ref{epsilonMn}) as,
\begin{equation}
\bar{E}_d=E_d+\Delta^{\rm Mn}_{d}\;.
\end{equation}
Fig.~\ref{fig_cond-tbad} shows that $\Delta^{\rm Mn}_{d} \approx
1.6$ eV is necessary to create a bound state with a binding energy
0.1 eV above the edge of the valence band.
Fig.~\ref{fig_serie-idos} shows the
transformation of the impurity related resonance in the valence
band into a bound state in the band gap for increasing value of
$\Delta^{\rm Mn}_{d}$, indicating the same value of $\Delta^{\rm Mn}_{d}$ which gives the 0.1~eV bound state.

\begin{figure}[ht]
\vspace*{-0.cm}
\includegraphics[height=0.7\columnwidth,angle=270]{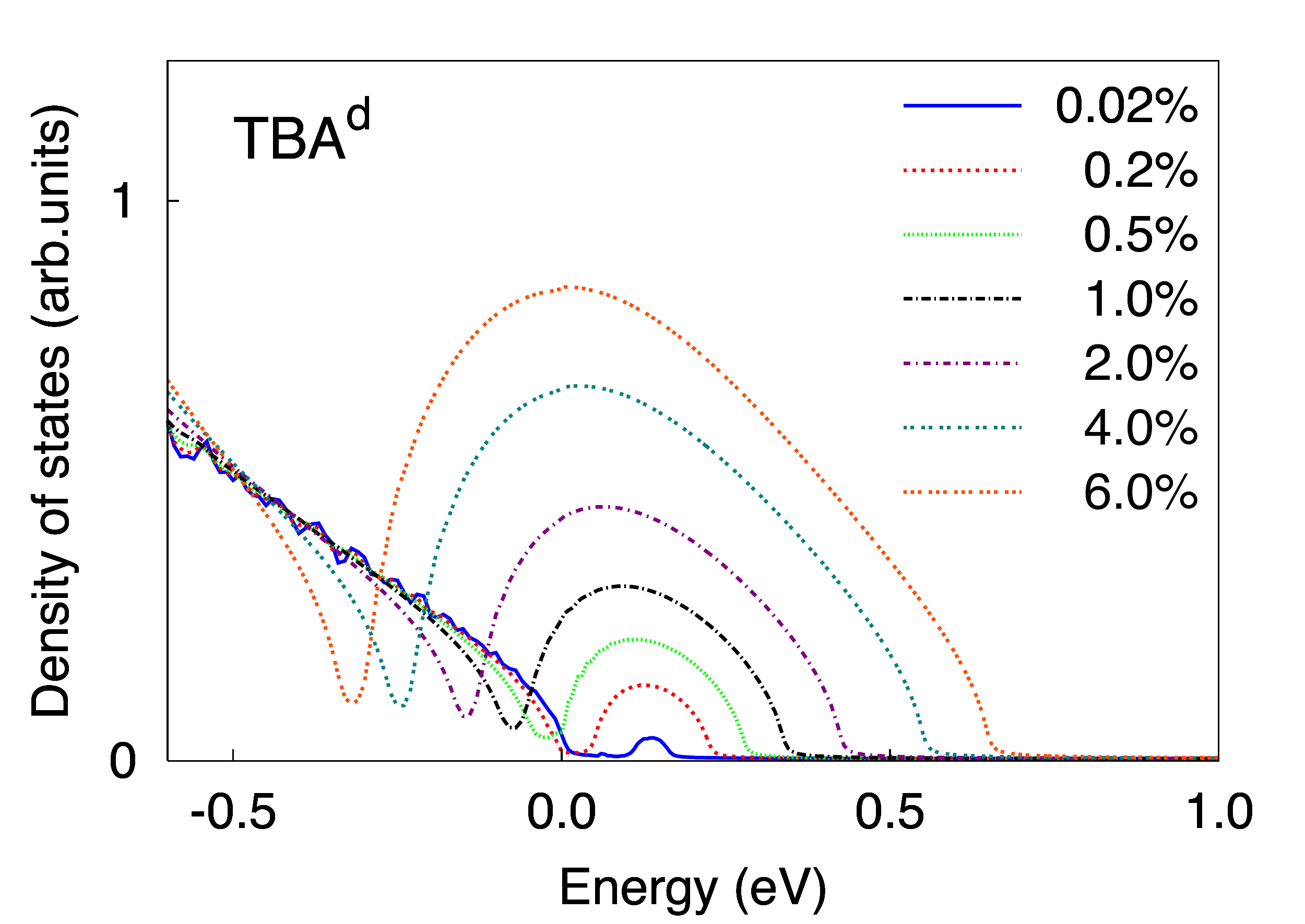}
\vspace*{-0cm} \caption{Total density of states of dilute
(Ga,Mn)As for $\Delta^{\rm Mn}_{d} = 1.59$ eV adjusted to the
position of the impurity level to the observed acceptor binding
energy 0.1 eV.}
\label{fig_dos-tbad}
\end{figure}

This value of $\Delta^{\rm Mn}_{d}$ is not as unphysical as in the above case of shifted Mn $p$-levels.
However, it represents a transfer of Mn $d$-orbital energies from
$\approx$ -2.5 eV to $\approx$ -1 eV and, as a result, a significant
enhancement of the hybridization effects leading, e.g.,
to a factor of 2 enhancement of the exchange coupling $J_{pd}$ between the local
moments and the holes. This value of 
$J_{pd}$ is much larger than any experimentally inferred value of this parameter.

\begin{figure}[ht]
\vspace*{-0.5cm}
\includegraphics[height=0.7\columnwidth,angle=270]{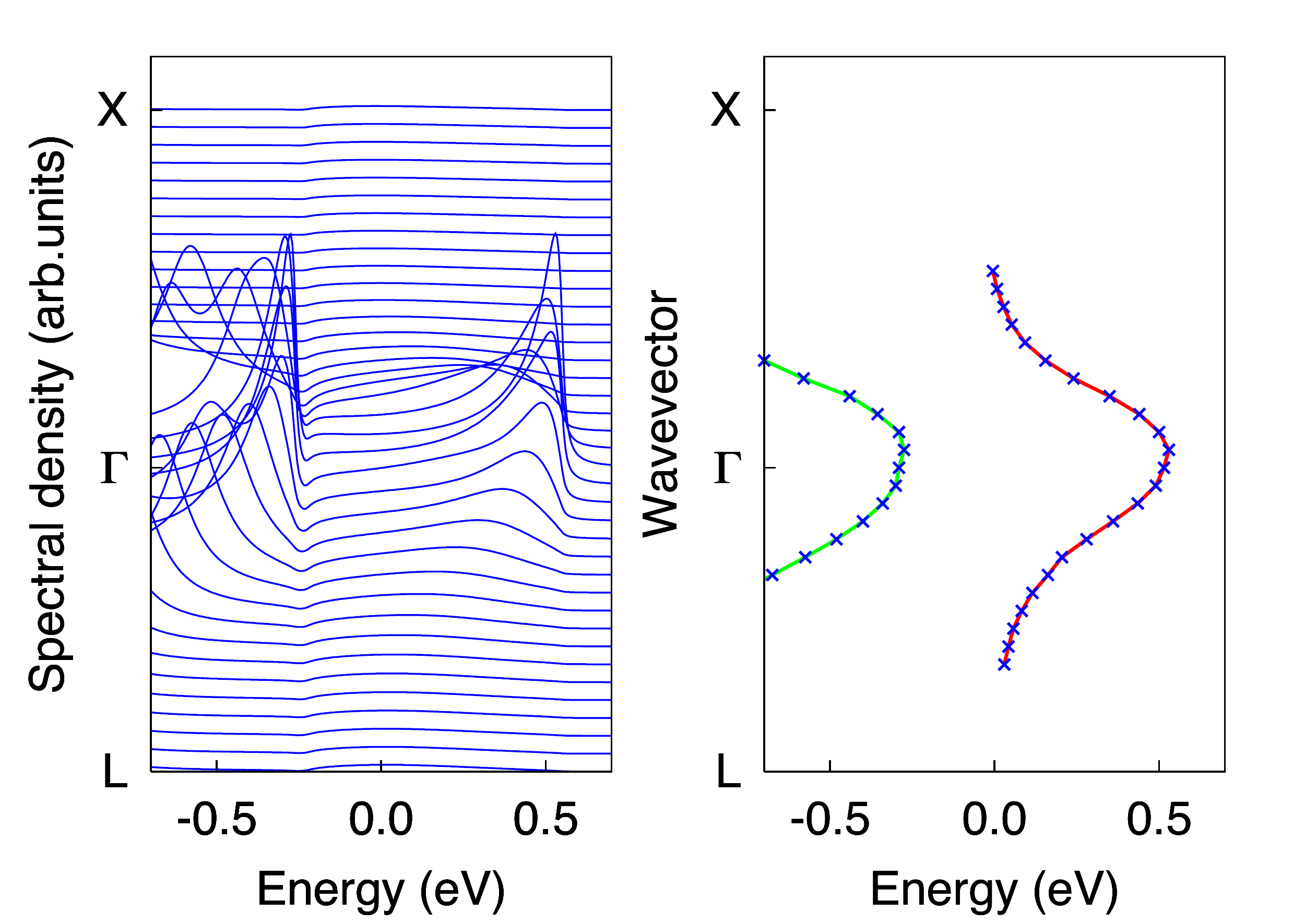}
\vspace*{-0cm} \caption{Spectral density of states of
Ga$_{0.96}$Mn$_{0.04}$As for $\Delta^{\rm Mn}_{d} = 1.59$ eV along
L-$\Gamma$-X path in the Brillouin zone. The maxima of the
spectral density are used to show effective dispersion relations
in the right panel.}
\label{fig_ak-tbad}
\end{figure}

We now use the modified TBA${\rm d}$ with $\Delta^{\rm Mn}_{d} = 1.59$ eV
that results in the localized state at 0.1 eV in the single impurity
case and use the CPA to visualize the evolution of the impurity
band with increasing concentrations $x$ of Mn.
Fig.~\ref{fig_dos-tbad} shows that the width of the impurity
band quickly increases with increasing $x$. The impurity band
merges with the valence band already for $x<0.1\%$ and at
$x\approx 6\%$ the width of the broad maximum of the deformed valence band extends over the lower half of the band gap.

Finally, the left panel of Fig.~\ref{fig_ak-tbad} shows the
spectral density for the mixed system with 4\% of Mn in the
spectral range close to the top of the valence band. We note that around the center of the Brillouin
zone, the spectral density has a double-maximum structure. The lower
maximum corresponds to the host valence band, the upper one to the
scattering states at the top of the valence band due to Mn impurities. This maximum broadens and finally disappears for
wave-vectors away from the center the Brillouin zone. 
The effective mass corresponding to the scattering and host parts of the band are similar as indicated in the right
panel of Fig.~\ref{fig_ak-tbad} .

\subsection{Enhanced $p-d$ hybridization by increased Mn $d$-orbital hopping energies -- TBA$^{\rm pd}$}
Here we consider modified Mn $d$-orbital hopping energies obtained
by multiplying the $\eta$ coefficients in Eq.~(\ref{eta}) by a factor $A>1$,
\begin{eqnarray}
\bar{\eta}_{sd\sigma}&=&A\eta_{sd\sigma}\nonumber\\
\bar{\eta}_{pd\sigma}&=&A\eta_{pd\sigma}\nonumber\\
\bar{\eta}_{pd\pi}&=&A\eta_{pd\pi}\;.
\end{eqnarray}

\begin{figure}[ht]
\vspace*{-0.cm}
\includegraphics[height=0.7\columnwidth,angle=270]{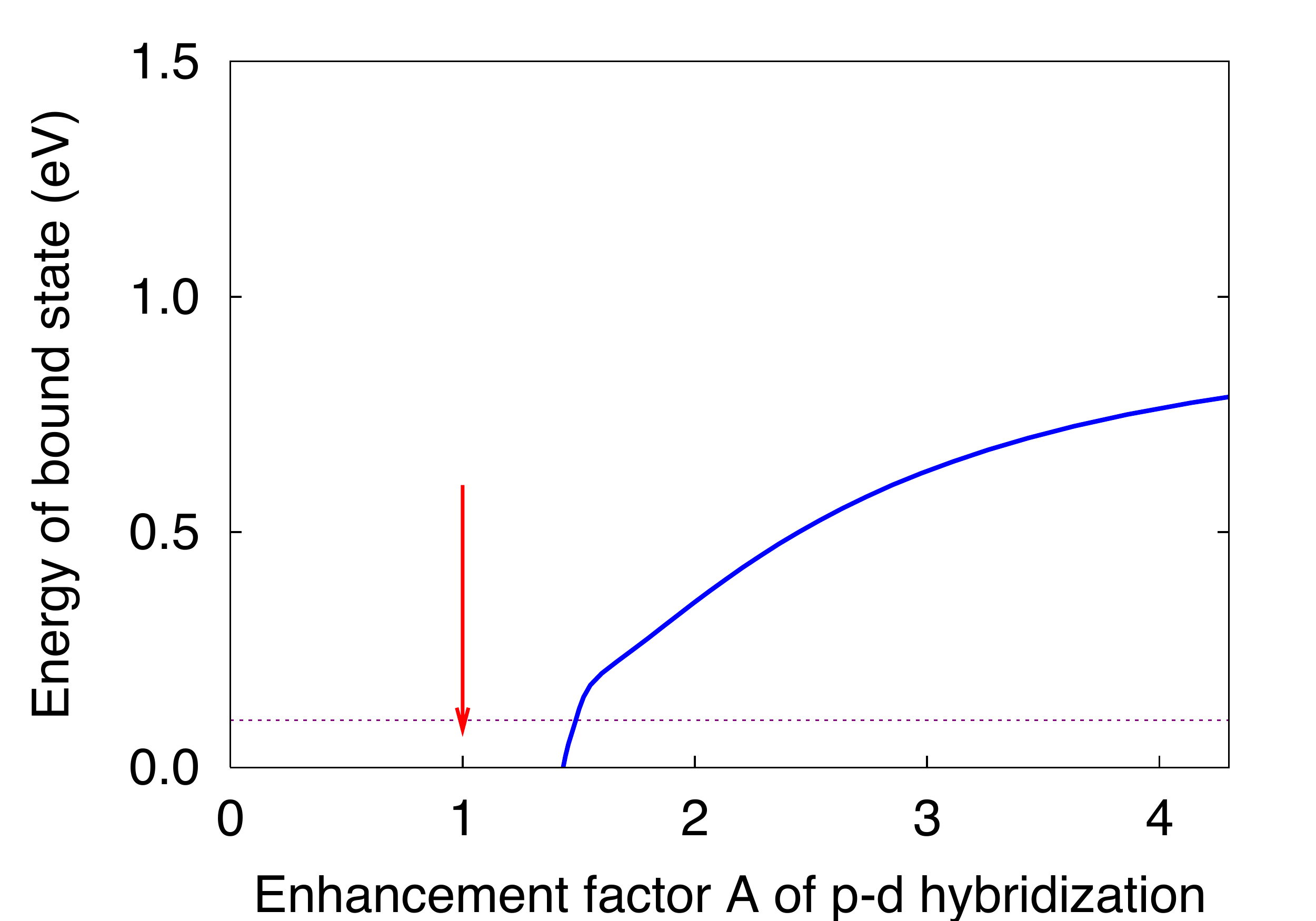}
\vspace*{-0cm} \caption{Relation between the energy of Mn-induced
related bound state in the band gap of GaAs and the enhancement
factor $A$ for sp-d hybridization. Dotted line represents Mn
acceptor level and the arrow corresponds to our basic
parametrization}
\label{fig_cond-tbapd}
\end{figure}

\begin{figure}[h!]
\vspace*{-0.cm}
\includegraphics[height=0.7\columnwidth,angle=270]{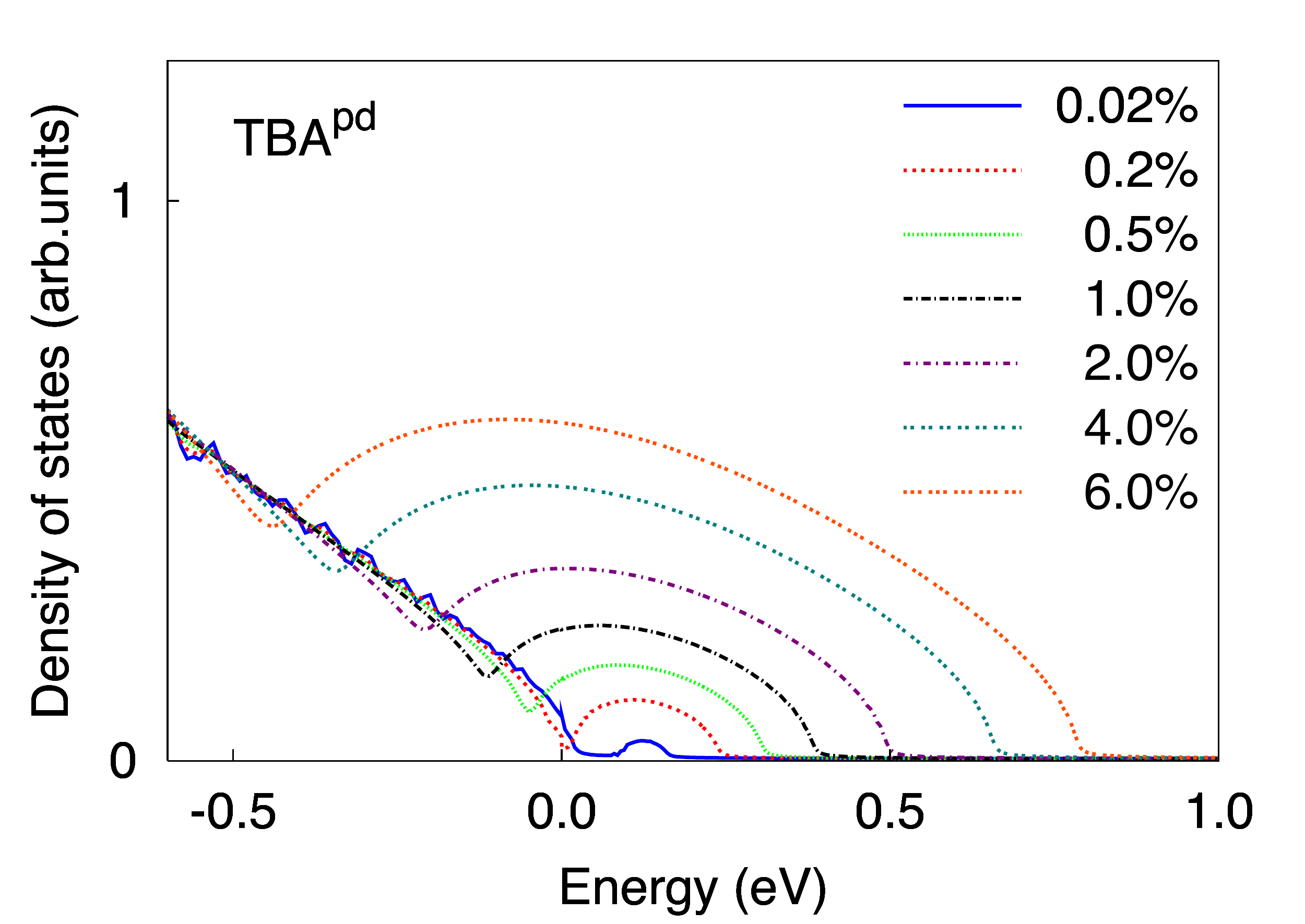}
\vspace*{-0cm} \caption{Total density of states of dilute
(Ga,Mn)As for the $sp-d$ hybridization factor $A$ adjusted to
obtain the observed acceptor level at 0.1 eV.}
\label{fig_dos-tbapd}
\end{figure}

Fig.~\ref{fig_cond-tbapd} shows that $A \approx 1.5$ eV is
necessary to create a bound state with a binding energy 0.1 eV
above the edge of the valence band. Similarly to the TBA$^{\rm p}$
case, 
$J_{pd}$ is enhanced by a factor of 2 in TBA$^{\rm pd}$. To show what can be expected at finite concentrations of
Mn, we adopted the modified hopping integrals with $A=1.5$ and
calculated the density of states shown in
Fig.~\ref{fig_dos-tbapd}. Even though the mechanism of formation
of the bound state was different, its transformation into a broad
spectral feature merged with the valence band is very similar to
the results from Fig.~\ref{fig_dos-tbad}.

\vspace*{5cm}
\section{Results: ${\rm\bf (Ga,Mn)As}$ band-structure calculated using the $sp$ tight-binding model with shifted ${\rm\bf As}$ $p$-levels -- TB$^{\rm p}$}
Here we show results obtained using the conventional tight-binding
parametrization of GaAs and an effective model of the Mn impurity based on 
shifting majority-spin As $p$-orbital on-site energies on the
neighbors of  the presumed Mn impurity.
\begin{eqnarray}
\bar{\epsilon}^{\rm As}_{p,+}&=&\epsilon^{\rm As}_p+\Delta^{\rm As}_p\nonumber \\
\bar{\epsilon}^{\rm As}_{p,-}&=&\epsilon^{\rm As}_p
\end{eqnarray}

\begin{figure}[h!]
\vspace*{-0.cm}
\includegraphics[height=0.7\columnwidth,angle=270]{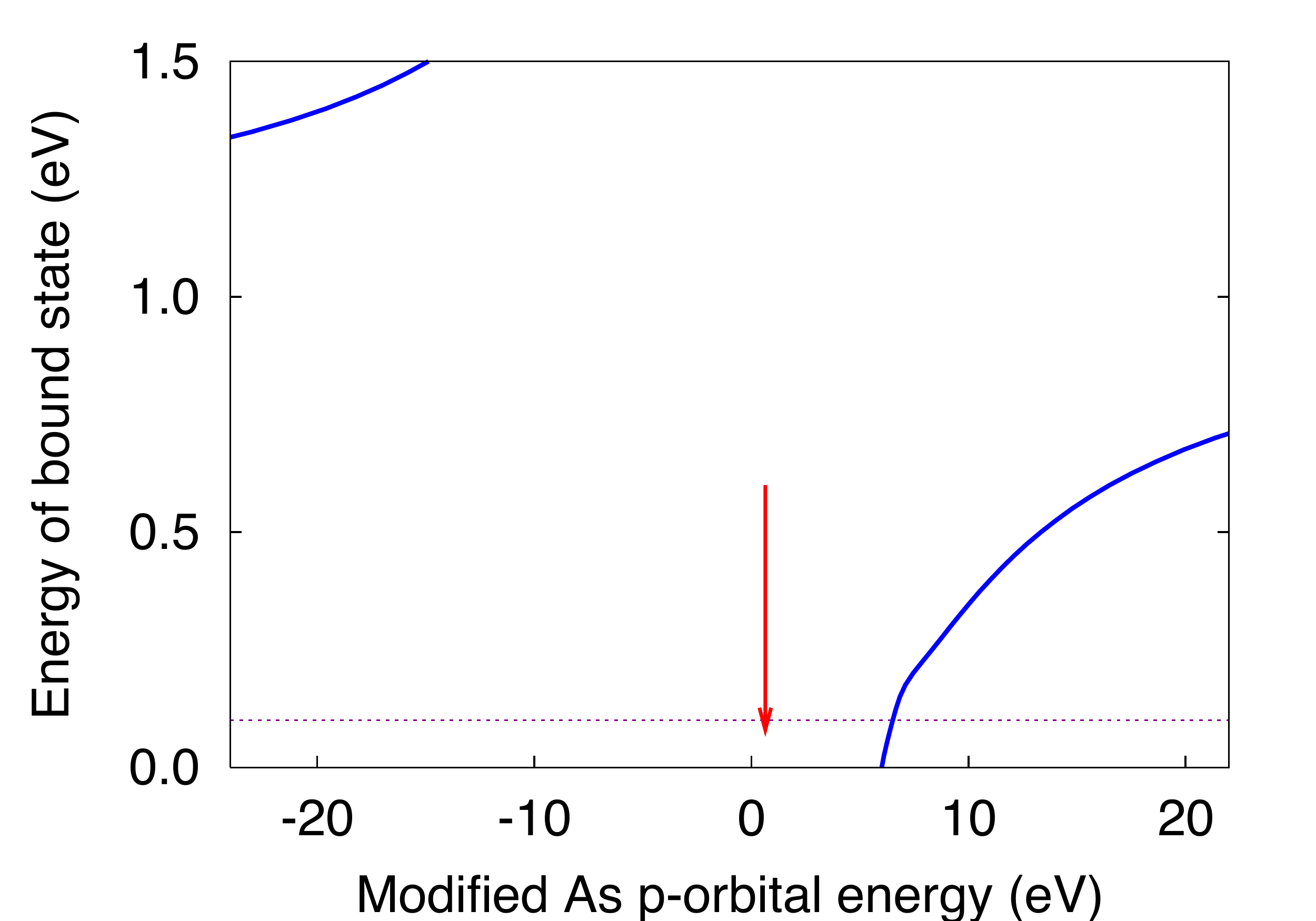}
\vspace*{-0cm} \caption{Relation between the modified As p-orbital
energy $\bar{\epsilon}^{\rm As}_{p,+}$, representing the exchange
field due to Mn impurity in GaAs, and a position of related bound
state. Dotted line represents Mn acceptor level and the arrow
corresponds to the unperturbed atomic level $\epsilon^{\rm
As}_{p}$.}
\label{fig_cond-tbapa}
\end{figure}

First we find and fix the value of  $\Delta^{\rm As}_p$ by
considering a single As impurity with $\bar{\epsilon}^{\rm As}_p$
and searching for the 0.1~eV bound state without the long-range
Coulomb potential. Fig.~\ref{fig_cond-tbapa} shows that
$\Delta^{\rm As}_{p} \approx 6$ eV must be used. Then we do the
CPA to calculate the band structure as a function of the presumed
Mn doping. Before writing down the CPA equations we need to
realize that the single As impurity problem with shifted As
$p$-orbital energies on a single As atom corresponds to 3/4 of one
Mn impurity. The factor 1/4 is because Mn interacts with 4 As
nearest-neighbors and the factor of 3 is because Mn is not
distributed symmetrically around the As but the bond is
directional. From the As $p_x$, $p_y$, and $p_z$ orbitals only one
of the three independent linear combinations is directed along the
bond with Mn; for example, for the
$\rho_{0As}=\frac{a_{lc}}{4}(0,0,0)$ bond, it is the combination
$p_x+p_y+p_z$. The CPA equation then reads,
\begin{equation}
G^{\Sigma}_{0,\alpha,s}(z)=(1-\bar{x})G_{{\rm
As},\alpha,s}(z)+\bar{x}G_{{\rm \bar{As}},\alpha,s}(z)\;,
\end{equation}
where the single-site Green's function $G_{{\rm As},\alpha,s}(z)$
corresponds to the unshifted As $p$-orbitals on the site,
$\epsilon^{\rm As}_p$, and $G_{{\rm \bar{As}},\alpha,s}(z)$ to the
shifted orbitals $\bar{\epsilon}^{\rm As}_{p,+}$, and the
effective Mn concentration $x=\frac34\,\bar{x}$ where $\bar{x}$ is
the concentration of  As atoms with shifted $p$-orbital energies
considered in the CPA calculations.

\begin{figure}[h!]
\vspace*{-0.cm}
\includegraphics[height=0.7\columnwidth,angle=270]{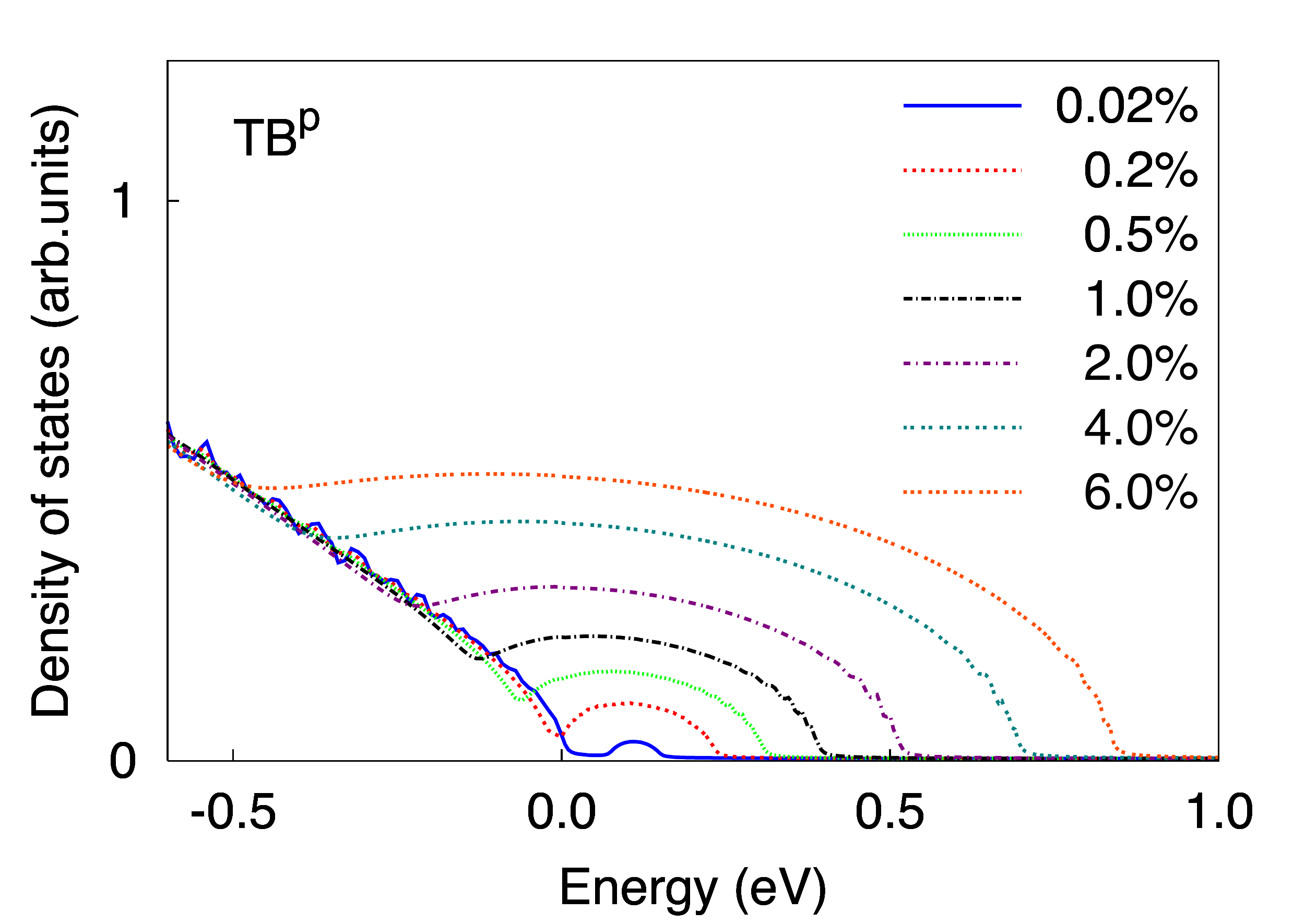}
\vspace*{-0cm} \caption{Total density of states of dilute
(Ga,Mn)As for the $\Delta^{\rm As}_p$ adjusted to obtain the
observed acceptor level at 0.1 eV.}
\label{fig_dos-tbapa}
\end{figure}

The results are summarized in Fig.\ref{fig_dos-tbapa}. Also in
this case, the width of the broad maximum corresponding to the Mn-induced scattering states depends on Mn
concentration in the same way as for the models studied in the
previous Section. 


\vspace*{1cm}
\section{Full-potential supercell ab initio calculations in LDA and LDA+U}

To support the tight-binding studies, we present the results of
the supercell calculations performed by using the full-potential
linearized-augmented-plane-wave method (WIEN
package\cite{WIEN97}). Besides the standard LDA
calculations (with an improved local-density approximation by the
generalized-gradient approximation, the GGA) we used also LDA+U (or GGA+U to be precise)
approximation with the same parameters $U$ and $J$ as used in the
tight-binding calculations. Because the LDA+U approximation is
known to shift the occupied $d$-states to lower energy and
unoccupied $d$-states to higher energies, the relation between LDA+U
and LDA seems analogous to the relation between conventional and
modified versions of the tight-binding method.

\begin{figure}[ht]
\vspace*{-0.cm}
\includegraphics[height=0.7\columnwidth,angle=270]{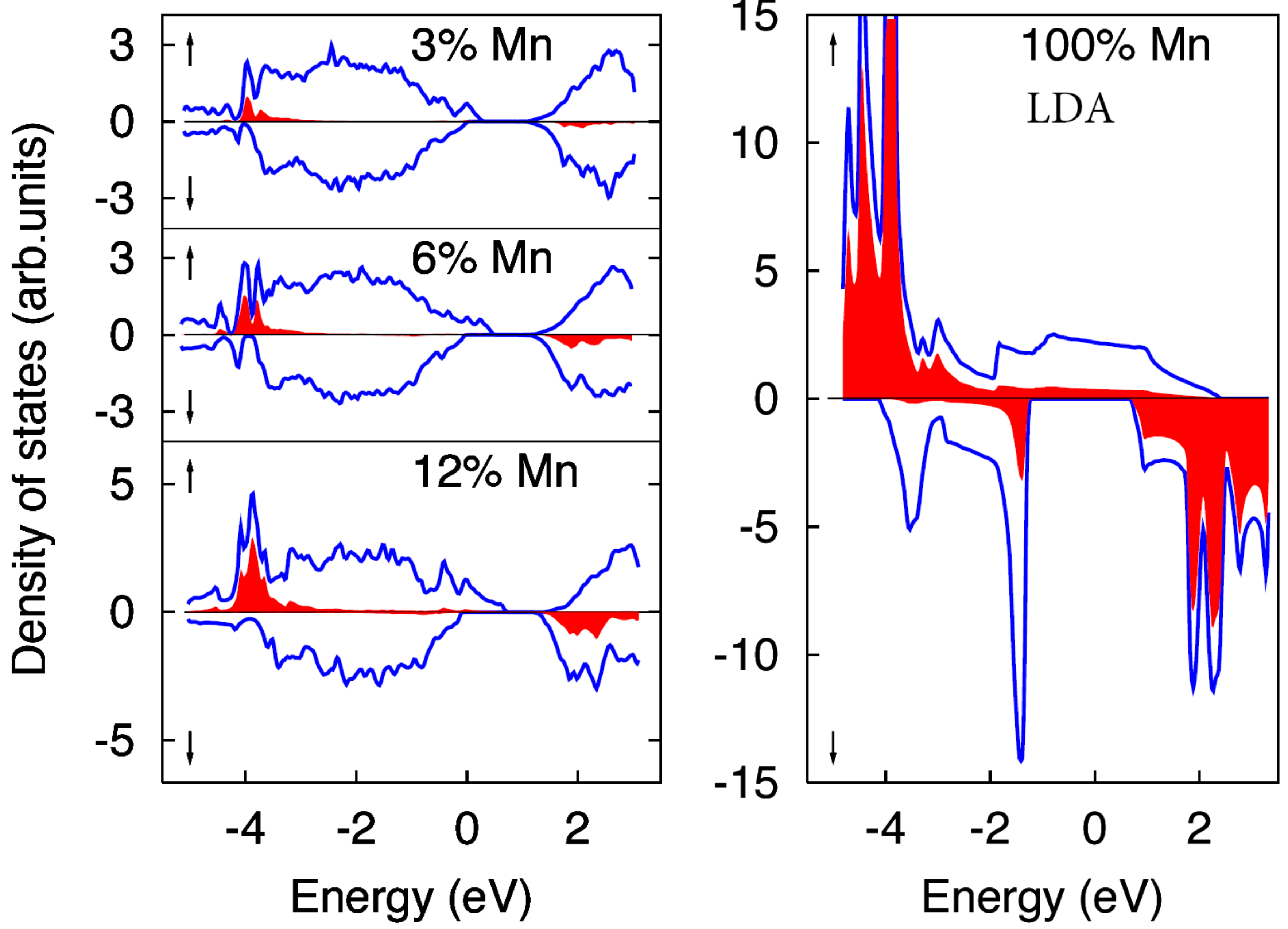}
\vspace*{-0cm} \caption{FP LAPW densities of states of (Ga,Mn)As
supercells representing three typical concentrations of Mn
together with the DOS of zinc-blende MnAs crystal obtained with LDA (GGA to by precise)
approximation. DOSs are normalized to the volume of conventional
unit cell.}
\label{fig_flapw-gga}
\end{figure}

\begin{figure}[ht]
\vspace*{-0.cm}
\includegraphics[height=0.7\columnwidth,angle=270]{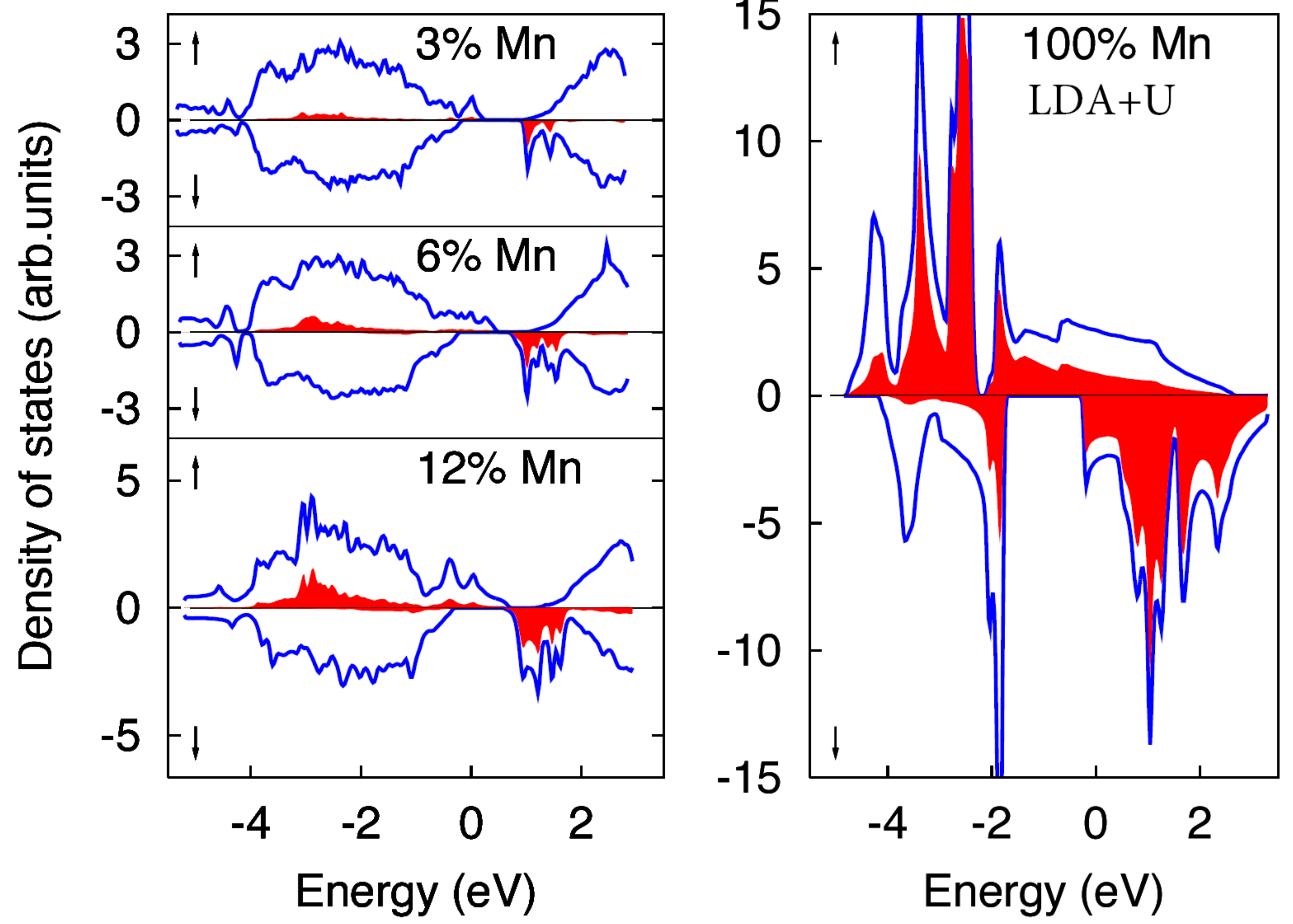}
\vspace*{-0cm} \caption{FP LAPW densities of states of (Ga,Mn)As
supercells representing three typical concentrations of Mn
together with the DOS of zinc-blende MnAs crystal obtained with LDA+U (GGA+U to be precise)
approximation. DOSs are normalized to the volume of conventional
unit cell.}
\label{fig_flapw-ggau}
\end{figure}

We present the densities of states calculated for the
superlattices Ga$_{7}$MnAs$_{8}$, Ga$_{15}$MnAs$_{16}$, and
Ga$_{31}$MnAs$_{32}$, representing the diluted systems with
roughly 12\%, 6\%, and 3\% Mn, respectively, and also for a
hypothetical MnAs crystal with a zinc-blende structure as a
limiting case. The results are summarized in
Fig.~\ref{fig_flapw-gga} and Fig.~\ref{fig_flapw-ggau}.

\vspace*{8cm}
\section{TB-LMTO CPA LDA and LDA+U calculations}
\begin{figure}[ht]
\vspace*{-0.5cm}
\includegraphics[height=0.7\columnwidth,angle=270]{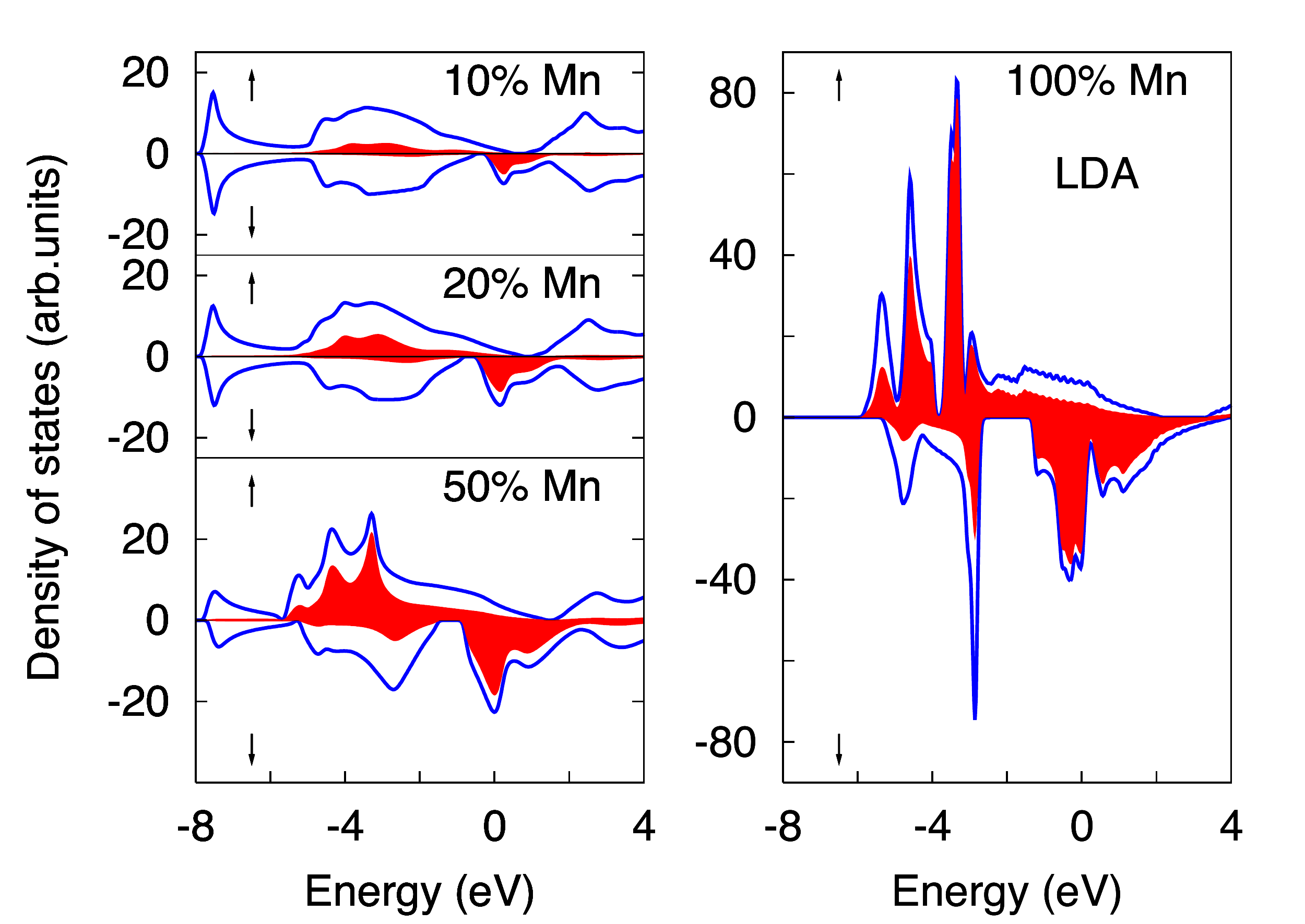}
\vspace*{-0cm} \caption{Partial density of Mn d-states in dilute
(Ga,Mn)As magnetic semiconductors. }
\label{fig_lmto-imp}
\end{figure}

\begin{figure}[ht]
\vspace*{-0.cm}
\includegraphics[height=0.7\columnwidth,angle=270]{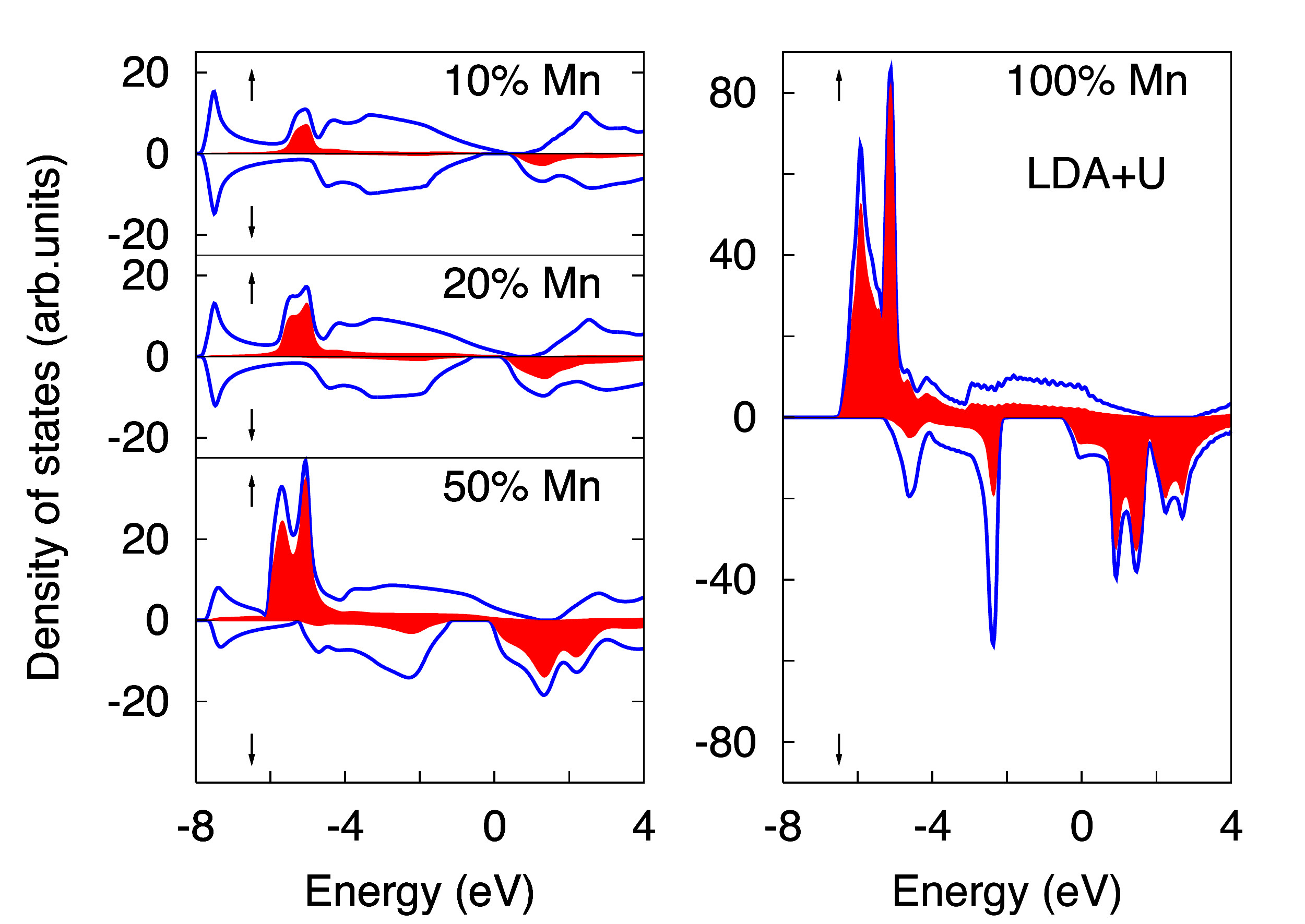}
\vspace*{-0cm} \caption{Density of states for (Ga,Mn)As mixed
crystals with various content of Mn obtained in the LDA
approximation. Shaded area shows the partial density of Mn
d-states.}
\label{fig_lmto-lda}
\end{figure}

\begin{figure}[ht]
\vspace*{-0.cm}
\includegraphics[height=0.7\columnwidth,angle=270]{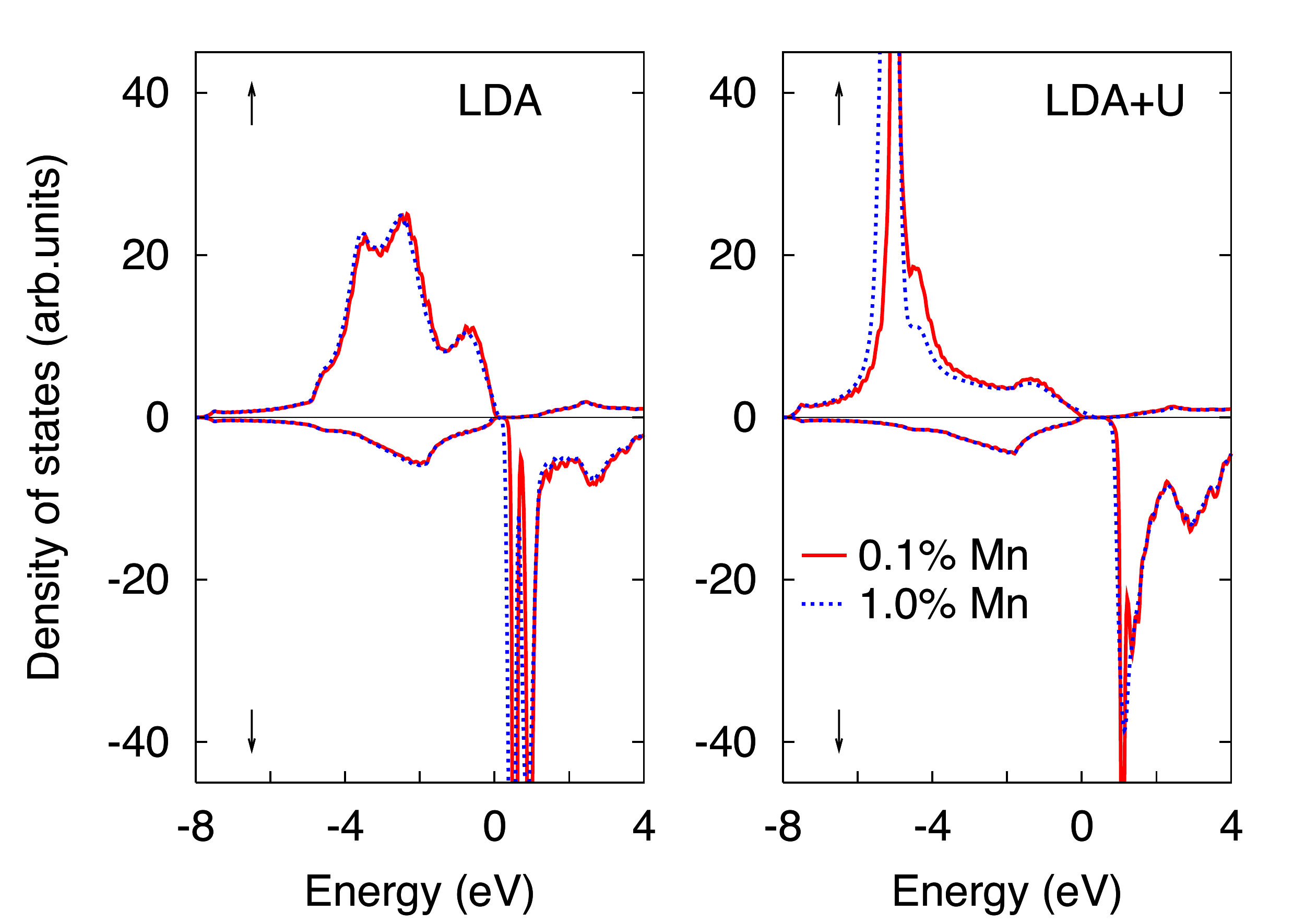}
\vspace*{-0cm} \caption{Density of states for (Ga,Mn)As mixed
crystals with various content of Mn obtained in the LDA+U
approximation. Shaded area shows the partial density of Mn
d-states.}
\label{fig_lmto-ldau}
\end{figure}

The supercell calculations are restricted to a limited range of
chemical compositions. In particular, it is difficult to represent
very diluted system with Mn concentrations below one percent. We
took an advantage of the available tight-binding linear muffin-tin
approximation (TB-LMTO) version of the CPA \cite{Turek:1997_a,Kudrnovsky:2004_a,Sato:2003_b} and performed a series
of calculations for a complementary set of Mn concentrations.
Again, we are interested in differences induced by accounting the
correlation effects and compare the DOSs obtained with LDA and with LDA+U. The results for
very diluted (Ga,Mn)As magnetic semiconductors are summarized in
Fig.~\ref{fig_lmto-imp}, densities of states for concentrated
mixed crystals are shown in Fig.~\ref{fig_lmto-lda} and
Fig.~\ref{fig_lmto-ldau}.

\vspace*{7cm}

\begin{thebibliography}{25}
\expandafter\ifx\csname natexlab\endcsname\relax\def\natexlab#1{#1}\fi
\expandafter\ifx\csname bibnamefont\endcsname\relax
  \def\bibnamefont#1{#1}\fi
\expandafter\ifx\csname bibfnamefont\endcsname\relax
  \def\bibfnamefont#1{#1}\fi
\expandafter\ifx\csname citenamefont\endcsname\relax
  \def\citenamefont#1{#1}\fi
\expandafter\ifx\csname url\endcsname\relax
  \def\url#1{\texttt{#1}}\fi
\expandafter\ifx\csname urlprefix\endcsname\relax\def\urlprefix{URL }\fi
\providecommand{\bibinfo}[2]{#2}
\providecommand{\eprint}[2][]{\url{#2}}

\bibitem[{\citenamefont{Chapman and Hutchinson}(1967)}]{Chapman:1967_a}
\bibinfo{author}{\bibfnamefont{R.~A.} \bibnamefont{Chapman}} \bibnamefont{and}
  \bibinfo{author}{\bibfnamefont{W.~G.} \bibnamefont{Hutchinson}},
  \bibinfo{journal}{Phys. Rev. Lett.} \textbf{\bibinfo{volume}{18}},
  \bibinfo{pages}{443} (\bibinfo{year}{1967}).

\bibitem[{\citenamefont{Ohno}(1998)}]{Ohno:1998_a}
\bibinfo{author}{\bibfnamefont{H.}~\bibnamefont{Ohno}},
  \bibinfo{journal}{Science} \textbf{\bibinfo{volume}{281}},
  \bibinfo{pages}{951} (\bibinfo{year}{1998}).

\bibitem[{\citenamefont{Jungwirth et~al.}(2006)\citenamefont{Jungwirth, Sinova,
  {Ma\v{s}ek}, {Ku\v{c}era}, and MacDonald}}]{Jungwirth:2006_a}
\bibinfo{author}{\bibfnamefont{T.}~\bibnamefont{Jungwirth}}
{\it et al}, \bibinfo{journal}{Rev. Mod. Phys.}
  \textbf{\bibinfo{volume}{78}}, \bibinfo{pages}{809} (\bibinfo{year}{2006}).

\bibitem[{\citenamefont{eds. T.~Dietl et~al.}(2008)\citenamefont{eds. T.~Dietl,
  Awschalom, Kaminska, and Ohmo}}]{Dietl:2008_b}
\bibinfo{author}{\bibnamefont{eds. T.~Dietl}} {\it et al},
in
  \emph{\bibinfo{booktitle}{Spintronics}} (\bibinfo{publisher}{Elsevier},
  \bibinfo{year}{2008}), vol.~\bibinfo{volume}{82} of
  \emph{\bibinfo{series}{Semiconductors and Semimetals}}.

\bibitem[{\citenamefont{Woodbury and Blakemore}(1973)}]{Woodbury:1973_a}
\bibinfo{author}{\bibfnamefont{D.~A.} \bibnamefont{Woodbury}} \bibnamefont{and}
  \bibinfo{author}{\bibfnamefont{J.~S.} \bibnamefont{Blakemore}},
  \bibinfo{journal}{Phys. Rev.} \textbf{\bibinfo{volume}{B 8}},
  \bibinfo{pages}{3803} (\bibinfo{year}{1973}).


\bibitem[{\citenamefont{Marder}(2000)}]{Marder:1999_a}
\bibinfo{author}{\bibfnamefont{M.~P.} \bibnamefont{Marder}},
  \emph{\bibinfo{title}{Condensed Matter Physics}} (\bibinfo{publisher}{Wiley,
  New York}, \bibinfo{year}{2000}), \eprint{Supplementary material by author}.


\bibitem[{\citenamefont{Jungwirth et~al.}(2007)\citenamefont{Jungwirth, Sinova,
  MacDonald, Gallagher, {Nov\'{a}k}, Edmonds, Rushforth, Campion, Foxon, Eaves
  et~al.}}]{Jungwirth:2007_a}
\bibinfo{author}{\bibfnamefont{T.}~\bibnamefont{Jungwirth}} {\it et al},
\bibinfo{journal}{Phys. Rev.}
  \textbf{\bibinfo{volume}{B 76}}, \bibinfo{pages}{125206}
  (\bibinfo{year}{2007}).

\bibitem[{\citenamefont{{Nov\'{a}k} et~al.}(2008)\citenamefont{{Nov\'{a}k},
  {Olejn\'{i}k}, Wunderlich, Cukr, {V\'{y}born\'{y}}, Rushforth, Edmonds,
  Campion, Gallagher, Sinova et~al.}}]{Novak:2008_a}
\bibinfo{author}{\bibfnamefont{V.}~\bibnamefont{{Nov\'{a}k}}} {\it et al},
   \bibinfo{journal}{Phys. Rev. Lett.}
  \textbf{\bibinfo{volume}{101}}, \bibinfo{pages}{077201}
  (\bibinfo{year}{2008}).

\bibitem[{\citenamefont{Wang et~al.}(2008)\citenamefont{Wang, Campion,
  Rushforth, Edmonds, Foxon, and Gallagher}}]{Wang:2008_e}
\bibinfo{author}{\bibfnamefont{M.}~\bibnamefont{Wang}} {\it et al},
  \bibinfo{journal}{Appl. Phys. Lett.} \textbf{\bibinfo{volume}{93}},
  \bibinfo{pages}{132103} (\bibinfo{year}{2008}).

\bibitem[{\citenamefont{Burch et~al.}(2006)\citenamefont{Burch, Shrekenhamer,
  Singley, Stephens, Sheu, Kawakami, Schiffer, Samarth, Awschalom, and
  Basov}}]{Burch:2006_a}
\bibinfo{author}{\bibfnamefont{K.~S.} \bibnamefont{Burch}} {\it et al},
  \bibinfo{journal}{Phys. Rev. Lett.} \textbf{\bibinfo{volume}{97}},
  \bibinfo{pages}{087208} (\bibinfo{year}{2006}).

\bibitem[{\citenamefont{Stone et~al.}(2008)\citenamefont{Stone, Alberi, Tardif,
  Beeman, Yu, Walukiewicz, and Dubon}}]{Stone:2008_a}
\bibinfo{author}{\bibfnamefont{P.~R.} \bibnamefont{Stone}} {\it et al},
  \bibinfo{journal}{Phys. Rev. Lett.} \textbf{\bibinfo{volume}{101}},
  \bibinfo{pages}{087203} (\bibinfo{year}{2008}).

\bibitem[{\citenamefont{Ando et~al.}(2008)\citenamefont{Ando, Saito, Agarwal,
  Debnath, and Zayets}}]{Ando:2008_a}
\bibinfo{author}{\bibfnamefont{K.}~\bibnamefont{Ando}} {\it et al},
  \bibinfo{journal}{Phys. Rev. Lett.} \textbf{\bibinfo{volume}{100}},
  \bibinfo{pages}{067204} (\bibinfo{year}{2008}).

\bibitem[{\citenamefont{Tang and {Flatt\'{e}}}(2008)}]{Tang:2008_a}
\bibinfo{author}{\bibfnamefont{J.-M.} \bibnamefont{Tang}} \bibnamefont{and}
  \bibinfo{author}{\bibfnamefont{M.~E.} \bibnamefont{{Flatt\'{e}}}},
  \bibinfo{journal}{Phys. Rev. Lett.} \textbf{\bibinfo{volume}{101}},
  \bibinfo{pages}{157203} (\bibinfo{year}{2008}).

\bibitem[{\citenamefont{Burch et~al.}(2008)\citenamefont{Burch, Awschalom, and
  Basov}}]{Burch:2008_a}
\bibinfo{author}{\bibfnamefont{K.~S.} \bibnamefont{Burch}},
  \bibinfo{author}{\bibfnamefont{D.~D.} \bibnamefont{Awschalom}},
  \bibnamefont{and} \bibinfo{author}{\bibfnamefont{D.~N.} \bibnamefont{Basov}},
  \bibinfo{journal}{J. Magn. Magn. Mater.} \textbf{\bibinfo{volume}{320}},
  \bibinfo{pages}{3207} (\bibinfo{year}{2008}).

\bibitem[{sup()}]{suppl}
\bibinfo{note}{Supplementary material.}

\bibitem[{\citenamefont{Harrison}(1980)}]{Harrison:1980_a}
\bibinfo{author}{\bibfnamefont{W.~A.} \bibnamefont{Harrison}},
  \emph{\bibinfo{title}{Electronic Structure and the Properties of Solids}}
  (\bibinfo{publisher}{Freeman, San Francisco}, \bibinfo{year}{1980}).

\bibitem[{\citenamefont{Bhattacharjee and {\`{a}}~la
  Guillaume}(2000)}]{Bhattacharjee:2000_a}
\bibinfo{author}{\bibfnamefont{A.~K.} \bibnamefont{Bhattacharjee}}
  \bibnamefont{and} \bibinfo{author}{\bibfnamefont{C.~B.}
  \bibnamefont{{\`{a}}~la Guillaume}}, \bibinfo{journal}{Solid State Commun.}
  \textbf{\bibinfo{volume}{113}}, \bibinfo{pages}{17} (\bibinfo{year}{2000}).

\bibitem[{\citenamefont{Linnarsson et~al.}(1997)\citenamefont{Linnarsson,
  {Janz\'{e}n}, Monemar, Kleverman, and Thilderkvist}}]{Linnarsson:1997_a}
\bibinfo{author}{\bibfnamefont{M.}~\bibnamefont{Linnarsson}} {\it et al},
  \bibinfo{journal}{Phys. Rev.} \textbf{\bibinfo{volume}{B 55}},
  \bibinfo{pages}{6938} (\bibinfo{year}{1997}).

\bibitem[{\citenamefont{Turek et~al.}(2008)\citenamefont{Turek, Siewert, and
  Fabian}}]{Turek:2008_a}
\bibinfo{author}{\bibfnamefont{M.}~\bibnamefont{Turek}},
  \bibinfo{author}{\bibfnamefont{J.}~\bibnamefont{Siewert}}, \bibnamefont{and}
  \bibinfo{author}{\bibfnamefont{J.}~\bibnamefont{Fabian}},
  \bibinfo{journal}{Phys. Rev.} \textbf{\bibinfo{volume}{B 78}},
  \bibinfo{pages}{085211} (\bibinfo{year}{2008}), \eprint{arXiv:0805.4350}.

\bibitem[{\citenamefont{Okabayashi et~al.}(1998)\citenamefont{Okabayashi,
  Kimura, Rader, Mizokawa, Fujimori, Hayashi, and Tanaka}}]{Okabayashi:1998_a}
\bibinfo{author}{\bibfnamefont{J.}~\bibnamefont{Okabayashi}} {\it et al},
  \bibinfo{journal}{Phys. Rev.} \textbf{\bibinfo{volume}{B 58}},
  \bibinfo{pages}{R4211} (\bibinfo{year}{1998}).

\bibitem{note2} From the technical viewpoint, the tight-binding approach is not well suited to account for long-range potentials.

\bibitem[{\citenamefont{Matsukura et~al.}(1998)\citenamefont{Matsukura, Ohno,
  Shen, and Sugawara}}]{Matsukura:1998_a}
\bibinfo{author}{\bibfnamefont{F.}~\bibnamefont{Matsukura}} {\it et al},
  \bibinfo{journal}{Phys. Rev.} \textbf{\bibinfo{volume}{B 57}},
  \bibinfo{pages}{R2037} (\bibinfo{year}{1998}).

\bibitem[{\citenamefont{Szczytko et~al.}(1999)\citenamefont{Szczytko, Mac,
  Twardowski, Matsukura, and Ohno}}]{Szczytko:1999_a}
\bibinfo{author}{\bibfnamefont{J.}~\bibnamefont{Szczytko}} {\it et al},
  \bibinfo{journal}{Phys. Rev.} \textbf{\bibinfo{volume}{B 59}},
  \bibinfo{pages}{12935} (\bibinfo{year}{1999}).

\bibitem[{\citenamefont{Omiya et~al.}(2000)\citenamefont{Omiya, Matsukura,
  Dietl, Ohno, Sakon, Motokawa, and Ohno}}]{Omiya:2000_a}
\bibinfo{author}{\bibfnamefont{T.}~\bibnamefont{Omiya}} {\it et al},
  \bibinfo{journal}{Physica} \textbf{\bibinfo{volume}{E 7}},
  \bibinfo{pages}{976} (\bibinfo{year}{2000}).

\bibitem[{\citenamefont{Tang and {Flatt\'{e}}}(2004)}]{Tang:2004_a}
\bibinfo{author}{\bibfnamefont{J.-M.} \bibnamefont{Tang}} \bibnamefont{and}
  \bibinfo{author}{\bibfnamefont{M.~E.} \bibnamefont{{Flatt\'{e}}}},
  \bibinfo{journal}{Phys. Rev. Lett.} \textbf{\bibinfo{volume}{92}},
  \bibinfo{pages}{047201} (\bibinfo{year}{2004}).

\bibitem{LDA}
A.B.~Shick, J.~Kudrnovsk\'y,  V.~Drchal, Phys. Rev. B {\bf 69}, 125207 (2004) L.M.~Sandratskii, P.~Bruno, J.~Kudrnovsk\'y, {\em ibid} {\bf 69}, 195203; M.~Wierzbowska, D.~Sanchez-Portal, S.~Sanvito, {\em ibid} 70, 235209.

\bibitem[{sup()}]{note_lda}
Note that the only marked difference between the TBA and LDA+U results is in the semiconductor band-gap  underestimated in the density-functional theories.
\end{thebibliography}

\begin{thebibliography}{16}
\expandafter\ifx\csname natexlab\endcsname\relax\def\natexlab#1{#1}\fi
\expandafter\ifx\csname bibnamefont\endcsname\relax
  \def\bibnamefont#1{#1}\fi
\expandafter\ifx\csname bibfnamefont\endcsname\relax
  \def\bibfnamefont#1{#1}\fi
\expandafter\ifx\csname citenamefont\endcsname\relax
  \def\citenamefont#1{#1}\fi
\expandafter\ifx\csname url\endcsname\relax
  \def\url#1{\texttt{#1}}\fi
\expandafter\ifx\csname urlprefix\endcsname\relax\def\urlprefix{URL }\fi
\providecommand{\bibinfo}[2]{#2}
\providecommand{\eprint}[2][]{\url{#2}}

\bibitem[{\citenamefont{Slater and Koster}(1954)}]{Slater:1954_a}
\bibinfo{author}{\bibfnamefont{J.~C.} \bibnamefont{Slater}} \bibnamefont{and}
  \bibinfo{author}{\bibfnamefont{G.~F.} \bibnamefont{Koster}},
  \bibinfo{journal}{Phys. Rev.} \textbf{\bibinfo{volume}{94}},
  \bibinfo{pages}{1498} (\bibinfo{year}{1954}).

\bibitem[{\citenamefont{Harrison}(1980)}]{Harrison:1980_a2}
\bibinfo{author}{\bibfnamefont{W.~A.} \bibnamefont{Harrison}},
  \emph{\bibinfo{title}{Electronic Structure and the Properties of Solids}}
  (\bibinfo{publisher}{Freeman, San Francisco}, \bibinfo{year}{1980}).

\bibitem[{\citenamefont{Talwar and Ting}(1982)}]{Talwar:1982_a}
\bibinfo{author}{\bibfnamefont{D.~N.} \bibnamefont{Talwar}} \bibnamefont{and}
  \bibinfo{author}{\bibfnamefont{C.~S.} \bibnamefont{Ting}},
  \bibinfo{journal}{Phys. Rev.} \textbf{\bibinfo{volume}{B 25}},
  \bibinfo{pages}{2660} (\bibinfo{year}{1982}).

\bibitem[{\citenamefont{Chadi}(1977)}]{Chadi:1977_a}
\bibinfo{author}{\bibfnamefont{D.~J.} \bibnamefont{Chadi}},
  \bibinfo{journal}{Phys. Rev.} \textbf{\bibinfo{volume}{B 16}},
  \bibinfo{pages}{790} (\bibinfo{year}{1977}).

\bibitem[{\citenamefont{Vogl et~al.}(1983)\citenamefont{Vogl, Hjalmarson, and
  Dow}}]{Vogl:1983_a}
\bibinfo{author}{\bibfnamefont{P.}~\bibnamefont{Vogl}},
  \bibinfo{author}{\bibfnamefont{H.~P.} \bibnamefont{Hjalmarson}},
  \bibnamefont{and} \bibinfo{author}{\bibfnamefont{J.~D.} \bibnamefont{Dow}},
  \bibinfo{journal}{J. Phys. Chem. Solids} \textbf{\bibinfo{volume}{44}},
  \bibinfo{pages}{365} (\bibinfo{year}{1983}).

\bibitem[{\citenamefont{Koster and Slater}(1954)}]{Koster:1954_a}
\bibinfo{author}{\bibfnamefont{G.~F.} \bibnamefont{Koster}} \bibnamefont{and}
  \bibinfo{author}{\bibfnamefont{J.~C.} \bibnamefont{Slater}},
  \bibinfo{journal}{Phys. Rev.} \textbf{\bibinfo{volume}{95}},
  \bibinfo{pages}{1167} (\bibinfo{year}{1954}).

\bibitem[{\citenamefont{Ralph et~al.}(1975)\citenamefont{Ralph, Simpson, and
  Elliott}}]{Ralph:1975_a}
\bibinfo{author}{\bibfnamefont{H.~I.} \bibnamefont{Ralph}},
  \bibinfo{author}{\bibfnamefont{G.}~\bibnamefont{Simpson}}, \bibnamefont{and}
  \bibinfo{author}{\bibfnamefont{R.~J.} \bibnamefont{Elliott}},
  \bibinfo{journal}{Phys. Rev.} \textbf{\bibinfo{volume}{B 11}},
  \bibinfo{pages}{2948} (\bibinfo{year}{1975}).

\bibitem[{\citenamefont{Anderson}(1961)}]{Anderson:1961_a}
\bibinfo{author}{\bibfnamefont{P.~W.} \bibnamefont{Anderson}},
  \bibinfo{journal}{Phys. Rev.} \textbf{\bibinfo{volume}{124}},
  \bibinfo{pages}{41} (\bibinfo{year}{1961}).

\bibitem[{\citenamefont{Dworin and Narath}(1970)}]{Dworin:1970_a}
\bibinfo{author}{\bibfnamefont{L.}~\bibnamefont{Dworin}} \bibnamefont{and}
  \bibinfo{author}{\bibfnamefont{A.}~\bibnamefont{Narath}},
  \bibinfo{journal}{Phys. Rev. Lett.} \textbf{\bibinfo{volume}{25}},
  \bibinfo{pages}{1287} (\bibinfo{year}{1970}).

\bibitem[{\citenamefont{Parmenter}(1973)}]{Parmenter:1973_a}
\bibinfo{author}{\bibfnamefont{R.~H.} \bibnamefont{Parmenter}},
  \bibinfo{journal}{Phys. Rev.} \textbf{\bibinfo{volume}{B 8}},
  \bibinfo{pages}{1273} (\bibinfo{year}{1973}).

\bibitem[{\citenamefont{{Ma\v{s}ek}
  et~al.}(2007{\natexlab{a}})\citenamefont{{Ma\v{s}ek}, {Kudrnovsk\'{y}},
  {M\'{a}ca}, Sinova, MacDonald, Campion, Gallagher, and
  Jungwirth}}]{Masek:2006_a}
\bibinfo{author}{\bibfnamefont{J.}~\bibnamefont{{Ma\v{s}ek}}},
  \bibinfo{author}{\bibfnamefont{J.}~\bibnamefont{{Kudrnovsk\'{y}}}},
  \bibinfo{author}{\bibfnamefont{F.}~\bibnamefont{{M\'{a}ca}}},
  \bibinfo{author}{\bibfnamefont{J.}~\bibnamefont{Sinova}},
  \bibinfo{author}{\bibfnamefont{A.~H.} \bibnamefont{MacDonald}},
  \bibinfo{author}{\bibfnamefont{R.~P.} \bibnamefont{Campion}},
  \bibinfo{author}{\bibfnamefont{B.~L.} \bibnamefont{Gallagher}},
  \bibnamefont{and}
  \bibinfo{author}{\bibfnamefont{T.}~\bibnamefont{Jungwirth}},
  \bibinfo{journal}{Phys. Rev.} \textbf{\bibinfo{volume}{B 75}},
  \bibinfo{pages}{045202} (\bibinfo{year}{2007}{\natexlab{a}}),
  \eprint{arXiv:cond-mat/0609158}.

\bibitem[{\citenamefont{{Ma\v{s}ek}
  et~al.}(2007{\natexlab{b}})\citenamefont{{Ma\v{s}ek}, {Kudrnovsk\'{y}},
  {M\'{a}ca}, and Jungwirth}}]{Masek:2007_a}
\bibinfo{author}{\bibfnamefont{J.}~\bibnamefont{{Ma\v{s}ek}}},
  \bibinfo{author}{\bibfnamefont{J.}~\bibnamefont{{Kudrnovsk\'{y}}}},
  \bibinfo{author}{\bibfnamefont{F.}~\bibnamefont{{M\'{a}ca}}},
  \bibnamefont{and}
  \bibinfo{author}{\bibfnamefont{T.}~\bibnamefont{Jungwirth}},
  \bibinfo{journal}{Acta Phys. Polon. A} \textbf{\bibinfo{volume}{112}},
  \bibinfo{pages}{215} (\bibinfo{year}{2007}{\natexlab{b}}).

\bibitem[{WIE()}]{WIEN97}
\bibinfo{note}{P.~Blaha, K.~Schwarz, and J.~Luitz, WEIN97, FPLAPW package for
  calculating crystal properties, TU Vienna.}

\bibitem[{\citenamefont{Turek et~al.}(1997)\citenamefont{Turek, Drchal,
  {Kudrnovsk\'{y}}, {\v{S}ob}, and Weinberger}}]{Turek:1997_a}
\bibinfo{author}{\bibfnamefont{I.}~\bibnamefont{Turek}},
  \bibinfo{author}{\bibfnamefont{V.}~\bibnamefont{Drchal}},
  \bibinfo{author}{\bibfnamefont{J.}~\bibnamefont{{Kudrnovsk\'{y}}}},
  \bibinfo{author}{\bibfnamefont{M.}~\bibnamefont{{\v{S}ob}}},
  \bibnamefont{and}
  \bibinfo{author}{\bibfnamefont{P.}~\bibnamefont{Weinberger}},
  \emph{\bibinfo{title}{Electronic Structure of Disordered Alloys, Surfaces,
  and Interfaces}} (\bibinfo{publisher}{Kluwer Academic, Boston},
  \bibinfo{year}{1997}).

\bibitem[{\citenamefont{{Kudrnovsk\'{y}}
  et~al.}(2004)\citenamefont{{Kudrnovsk\'{y}}, Turek, Drchal, {M\'{a}ca},
  Weinberger, and Bruno}}]{Kudrnovsky:2004_a}
\bibinfo{author}{\bibfnamefont{J.}~\bibnamefont{{Kudrnovsk\'{y}}}},
  \bibinfo{author}{\bibfnamefont{I.}~\bibnamefont{Turek}},
  \bibinfo{author}{\bibfnamefont{V.}~\bibnamefont{Drchal}},
  \bibinfo{author}{\bibfnamefont{F.}~\bibnamefont{{M\'{a}ca}}},
  \bibinfo{author}{\bibfnamefont{P.}~\bibnamefont{Weinberger}},
  \bibnamefont{and} \bibinfo{author}{\bibfnamefont{P.}~\bibnamefont{Bruno}},
  \bibinfo{journal}{Phys. Rev.} \textbf{\bibinfo{volume}{B 69}},
  \bibinfo{pages}{115208} (\bibinfo{year}{2004}).

\bibitem[{\citenamefont{Sato et~al.}(2003)\citenamefont{Sato, Dederichs,
  Katayama-Yoshida, and Kudrnovsky}}]{Sato:2003_b}
\bibinfo{author}{\bibfnamefont{K.}~\bibnamefont{Sato}},
  \bibinfo{author}{\bibfnamefont{P.~H.} \bibnamefont{Dederichs}},
  \bibinfo{author}{\bibfnamefont{H.}~\bibnamefont{Katayama-Yoshida}},
  \bibnamefont{and}
  \bibinfo{author}{\bibfnamefont{J.}~\bibnamefont{Kudrnovsky}},
  \bibinfo{journal}{Physica B} \textbf{\bibinfo{volume}{340-342}},
  \bibinfo{pages}{863} (\bibinfo{year}{2003}).

\end{thebibliography}

\end{document}